\documentclass[twoside,chap,11pt]{usthesis} \usepackage[hang]{caption2}
\usepackage{epsfig} 
\usepackage{amsmath,epsf} 
\singlespace

\newfont{\Bbb}{bbmss12}
\newcommand{\R}{\mbox{\Bbb R}}
\newcommand{\cR}{\mbox{$\cal R$}}
\newcommand{\cM}{\mbox{$\cal M$}}
\newcommand{\be}{\begin{equation}}
\newcommand{\ee}{\end{equation}}
\newcounter{EX}[chapter]
\newenvironment{example}[1]
     {\stepcounter{EX}\hfill\break {\bf Example
         \thechapter.\arabic{EX}} {\em #1} \newline} 
     {\hskip 1in \hfill$\bullet$\newline}
\newcounter{THM}[chapter]
\newenvironment{thm}[1]
     {\stepcounter{THM}\hfill\break {\bf Theorem
         \thechapter.\arabic{THM}} {(#1)} \newline\em} 
     {\hskip 1in \newline\rm}
\newcommand{\Tr}[1]{\mbox{Tr}_{#1}\;}
\def\unit{\leavevmode\hbox{\small1\kern-3.8pt\normalsize1}}
\newtheorem{lemma}{Lemma}

\newcommand{\lan}{\langle}
\newcommand{\ran}{\rangle}
\newcommand{\UP}{\uparrow}
\newcommand{\DN}{\downarrow}
\newcommand{\spinupvec}{|\!\UP\ran}
\newcommand{\spindownvec}{|\!\DN\ran}

\begin{document}

 
 
\thesistitle{\bf Quantum Information Theory } 
 
\author{Robert Helmut Schumann}         
\degree{Master~of~Science} 
\supervisor{Professor H.B. Geyer} 
\submitdate{December 2000}         
\dedicateto{my mother}
\dedicatequote{}
\copyrightyear{2000}

\titlepage 
\specialhead{Abstract}

{\em What are the information processing capabilities of physical
systems?}

As recently as the first half of the $20^{\rm th}$ century this
question did not even have a definite meaning.  What is information,
and how would one process it?  It took the
development of theories of computing (in the 1930s) and information
(late in the 1940s) for us to formulate mathematically what it means to 
compute or communicate.

Yet these theories were abstract, based on axiomatic mathematics: what 
did physical systems have to do with these axioms?  Rolf Landauer had
the essential insight --- ``Information is physical'' --- that
information is always encoded in the state of a physical system, whose 
dynamics on a microscopic level are well-described by quantum physics.
This means that we cannot discuss information without discussing how
it is represented, and how nature dictates it should behave.  

Wigner considered the situation from another perspective when he wrote
about ``the unreasonable effectiveness of mathematics in the natural
sciences''. Why are the computational techniques of mathematics so
astonishingly useful in describing the physical world
\cite{Deutsch:1999}?  One might begin to suspect foul play in the
universe's operating principles.

Interesting insights into the physics of information accumulated
through the 1970s and 1980s --- most sensationally in the proposal for a
``quantum computer''.  If we were to mark a particular year in which
an explosion of interest took place in information physics, that year
would have to be 1994, when Shor showed that a problem of
practical interest (factorisation of integers) could be solved easily
on a quantum computer.  But the applications of information in physics
--- and vice versa --- have been far more widespread than this popular
discovery.  These applications range from improved experimental
technology, more sophisticated measurement techniques, methods for
characterising the quantum/classical boundary, tools for quantum chaos, 
and deeper insight into quantum theory and nature.

In this thesis I present a short review of ideas in quantum
information theory.  The first chapter contains introductory material, 
sketching the central ideas of probability and information theory.
Quantum mechanics is presented at the level of advanced undergraduate
knowledge, together with some useful tools for quantum mechanics of
open systems.  In the second chapter I outline how classical
information is represented in quantum systems and what this means for
agents trying to extract information from these systems.  The final
chapter presents a new resource: quantum information.  This resource
has some bewildering applications which have been discovered in the
last ten years, and continually presents us with unexpected insights
into quantum theory and the universe.


\clearpage 
\clearpage 

\dedication              
\specialhead{Acknowledgements}

This thesis was begun as an intrepid adventure.  It caught me quite
unexpectedly: a suggestion from a friend who heard quantum computing
was the next big thing; some enquiries as to whether anybody in South
Africa would be interested in supervising this ``fringe'' science; and 
then suddenly I moved town, university and field within one month.

I owe most of the success of this venture to three people: Professor
Hendrik Geyer, Chris Fuchs and Marcelle Olivier.

Professor Geyer --- who had dabbled in the quantum computing
literature --- agreed to supervise a thesis in quantum computing which 
subsequently evolved into the present work.  He has provided an
excellent example to me of a scientist, a leader in science and as an
intellectual.

My correspondence with Chris Fuchs only began a few months before the
completion of this thesis, but I have felt his presence since meeting
him in Turin and delving through his contributions to quantum
information theory.  His support and input have been rewarding and
enlightening.

Marcelle, as a perfectionist herself, indulged me in my fits of
writing, reading and not doing enough work.  She is my muse.

Thanks are also due to my various colleagues at the Instituut vir
Teoretiese Fisika: Andrew van Biljon (who's always been here), Leandro 
Boonzaaier, Lucian Anton, Professor Frikkie Scholtz and Jacques Kotze.
I gratefully acknowledge the financial support of the National
Research Foundation, in particular for the funds made available for my 
attendance at the TMR Network summer school on quantum computing and
quantum information theory in Turin, Italy in 1999.  Thanks also to my 
parents, who are a little bewildered at what I do (and would prefer it 
if I were making money) but love me anyway.  Many more people
have helped and encouraged me during my time at Stellenbosch --- and
before --- and their contribution to making me is also appreciated.

And to Debbie, who watched me suddenly develop into a Quantum Computer
Scientist after her brother mentioned it.


\tableofcontents         
\listoffigures           
 
\clearpage 
\clearpage

\chapter{Prolegomenon}

Information theory and quantum mechanics form two cornerstones of an
immense construction of technology achieved in the $20^{\rm th}$
century.  And apart from being highly successful in their respective
realms -- and indeed in the communal realm of computing -- they have
recently interacted in a way their discoverers hadn't dreamed of.  By
stretching the envelope of their specifications, the field of quantum 
information theory was born: quantum mechanics by asking questions
about how small computers could be made, and how much energy is
required for their operation; and information theory by asking how the 
abstract notion of a logical bit is implemented in the nuts-and-bolts
world of physics.

Of course, before we begin our exploration of quantum information, we
require knowledge of the two theories on which it is based.  This
first chapter provides a basis for the definitions of information
theory and some demonstration as to why these notions are appropriate, 
and then lays the groundwork for the style of quantum mechanics required
for later developments.

\section{Information Theory}

Information is such a rich concept that trying to pin it
down with a definition amputates some of its usefulness.  We therefore 
adopt a more pragmatic approach, and ask questions which (hopefully)
have well-defined quantitative answers, such as ``By what factor can
given information be compressed?'' and ``How much redundancy must be
incorporated into this message to ensure correct decoding when shouted 
over a bad telephone connection?''.  The answers to these questions
are given by a small number of measures of information, and the
methods of answering often yield valuable insights into the
fundamentals of communication.

But first: what are the basic objects of communication, 
with which the theory deals?  For this we consider a simple canonical
example.  Suppose we have to set up a communication link from the
President to Defence Underground Military Bombers (DUMB) from where
nuclear weapons are launched.  When the President wakes up in the
morning, he either presses a button labelled {\bf Y} (to launch) or a
button labelled {\bf N} (to tell DUMB to relax).  This communication
channel requires two symbols which constitute an alphabet, but in
general we could envisage any number of symbols, such as 256 in
conventional ASCII.  As 21$^{\rm st}$ century historians, we may be
interested in the series of buttons pushed by the President in 1992.
We hope that he pushed {\bf N} significantly more times than he pushed 
{\bf Y}, so it may be more economical to store in our archive
 the sequence of integers 
$N_1, \ldots, N_k$ where $N_i$ is the number of {\bf N}'s between the 
$(i-1)^{\rm st}$ and $i\,^{\rm th}$ destructive 
{\bf Y}'s.  From this toy model we learn that
our mathematical notion of information should involve an 
{\em alphabet} and a {\em
probability distribution} of the alphabet symbols.  However
it also seems desirable that the information be invariant under
changes in the alphabet or representation - we don't want the amount of
information to change simply by translating from the set $\{{\bf Y},
{\bf N}\}$ to the set $\{0,1,\ldots,365\}$. The probability
distribution seems to be a good handle onto the amount of information,
with the proviso that our measure be invariant under these ``translations''.

From this example we also note a more subtle point about information:
it quantifies our {\em ignorance}.  A sequence that is completely
predictable, about which there is complete knowledge and no ignorance, 
contains no information.  If the president's actions were entirely a
function of his childhood there would be no point in storing all the
{\bf Y}'s and {\bf N}'s, or indeed for DUMB to pay attention to
incoming signals - we could calculate them from publicly
available knowledge.  For a sequence to contain information there
must be a degree of ignorance about future signals, so in a sense a
probability is the firmest grip we can get on this ignorance.

Hence we interrupt our development of information for a sojourn into
probability theory.

\subsection{Notions of Probability}

We will consider our alphabet to be a set $A = \{a_1, a_2,\ldots, a_n\}$ 
of $n$ symbols.   The notion of probability we
employ here, the Bayesian approach, is closely tied to the concept 
of information.  Informally a probability $p(a_i)$ is a measure of how 
much we'd be willing to bet on the outcome of a trial (which perhaps
tells us which symbol to transmit) being $a_i$ \cite{Rice:1994}.  
Clearly this will depend on our
subjective knowledge of how a system was prepared (or perhaps which
horse has been doped), and explains the popularity of card counting in 
Blackjack.  We begin with an event $H$ of interest, which for the sake 
of definiteness we specify as
``The next card dealt will be an Ace'' and prior knowledge that the 
deck is complete and well-shuffled with no wildcards.  In this case,
our understanding of the situation tells us that
\be
p(H) = {1 / 13}.
\end{equation}
Our prior knowledge in this case is implicit and is usually clear
from the context.  There are situations in which our prior knowledge
might change and must be explicit, as for example when we know that
a card has been removed from our shuffled pack.  How much we'd be
willing to bet depends on the value of that card; we then use the
notation
\begin{equation} \label{conditionalaces}
p(H \mid \mbox{Ace removed}) = {3 \over 51} \mbox{ and } 
p(H \mid \mbox{Other card removed}) = {4 \over 51}
\end{equation}
to demonstrate the dependence, assuming that all other
prior knowledge remains the same.

To make these ideas more formal, we consider a set $A$ (of signals,
symbols or events), and we define a probability measure as a function 
$p$ from the subsets\footnote{In 
more generality, a probability measure is defined on a
$\sigma$-algebra of subsets of $A$ \cite{Taylor:1997}.  This allows us
to extend this 
description of a probability to continuous spaces.} of $A$
to the real numbers satisfying the following axioms\footnote{These
axioms can be derived from some intuitive principles of inductive 
reasoning; see e.g. \cite{Smith:1988} and \cite{Cox:1946}.}:
\begin{enumerate}
\item $p(\phi) = 0$ (probability of null event)
\item \label{certain}$p(A) = 1$ (probability of certain event)
\item For any $M \subseteq A$, $p(M) \geq 0$
\item For $a, b\in A$, $p({a})+p({b}) = p({a,b})$ (probability of 
disjoint events is additive)\footnote{When $p$ is
defined on a $\sigma$-algebra, we demand that $p$ be additive over
countable sequences of pairwise disjoint subsets from the
$\sigma$-algebra.}.
\end{enumerate}
This formalism gives a mathematical structure to the ``plausibility of 
a hypothesis'' in the presence of (unstated) prior knowledge.  Happily 
this machinery also coincides in cases of importance
with the frequency interpretation of
probability, which allows us to employ counting arguments in
calculating probabilities in many situations.  Because of the last
requirement above, we can specify a probability measure by giving its
value on all the singleton subsets ${a}$ of $A$.  In this case we
typically write $p(\{a\}) = p(a)$ where no confusion can arise, and
even this is occasionally shrunk to $p_a$.  We also use the terms
``measure'' and ``distribution'' interchangeably - the former simply
being more abstractly mathematical in origin than the latter.

We will frequently be interested in a {\em random variable}, defined as
a function from $A$ to the real numbers.  For example, if the sample
space $A$ is a full pack of cards then the function $X$ which takes
``Spades'' to 1 and other suits to 0 counts the number of Spades; we
write
\be
\mbox{Prob}(X=1)=p(1)=1/4\hspace{10 mm} \mbox{Prob}(X=0)=p(0)=3/4
\end{equation}
A function of interest might then be $F = \sum_i X_i$ where each $X_i$ 
is one of these ``Spade-counting'' random variables; in this case $F$
is defined on the space $A\times\ldots\times A$ and $X_i$ is a
random variable on the $i^{\mbox{th}}$ space.  The {\em expectation
value} of a random variable $X$ is then defined as 
\be
E_pX = \sum_{a\in A}p(a)X(a)
\end{equation}
where the subscript makes explicit the distribution governing the
random variable.  This is just the first of a host of quantities of
statistical interest, such as mean and variance, defined on random
variables.

If we have two sample spaces, $A$ and $B$, we can amalgamate them and
consider {\em joint} probability distributions on $A\times B$.  The
probability measure is then specified by the singleton probabilities
$p(a,b)$ where $a\in A$ and $b\in B$.  By axiom \ref{certain} above,
we have that
\begin{equation}
\sum_{a\in A,\, b\in B} p(a,b) = 1.
\end{equation}
If for each $a\in A$ we define $p_A(a)=\sum_{b\in B}p(a,b)$ and 
similarly $p_B(b)=\sum_{a\in A}p(a,b)$ then $p_A$ and
$p_B$ are also probability measures, on the spaces $A$ and $B$
respectively; these measures are called the {\em marginal}
distributions.  Conventionally we drop the subscripts on the marginal
distributions where confusion cannot arise.  But notice that these
measures are not necessarily 
``factors'' of the joint probability, in that $p(a,b)\neq
p_A(a)p_B(b)$ for all $a\in A, b\in B$\footnote{Those events for which 
it is true that $p(a,b)=p_A(a)p_B(b)$ are called independent, and
if true for all events the distribution is called independent.}. This
prompts us to define the {\em conditional} probability distributions
\begin{equation}
p(a|b) = {p(a,b) \over p(b)} = {p(a,b) \over \sum_{x\in A}p(x,b)}
\hbox{\; and\; } p(b|a) = {p(a,b) \over p(a)}.
\end{equation}
These definitions lend rigour to the game of guessing Aces described
by Eqn \ref{conditionalaces}.  Note in this definition that if we
choose a fixed member $b$ from the set $B$, then the distribution
$p_b(a)=p(a|b)$ is also a well-defined distribution on $A$.  In
effect, learning which signal from the set $B$ has occurred gives us
partial knowledge of which signal from $A$ will occur --- we have
updated our knowledge, conditional on $b$.

This definition can quite easily be extended to more than two sample
spaces.  If $A_1,\ldots,A_n$ are our spaces and $p$ is the joint
probability distribution on $A=\prod A_i$, then, for example,
\be
p(a_na_{n-1}|a_{n-2},\ldots,a_1) = {p(a_1,\ldots,a_n) \over
p(a_1,\ldots, a_{n-2})} = {p(a_1,\ldots,a_n) \over \sum_{x\in A_n,
y\in A_{n-1}} p(a_1,\ldots,a_{n-2},y,x)}
\end{equation}
is the probability of sampling the two symbols $a_{n-1}$ and $a_n$
given the sampled sequence $a_1,\ldots,a_{n-2}$.

Conditional probabilities give us a handle on the ``inverse
probability'' problem.  In this problem, we are told the outcome of a
sampling from the set $A$ (or perhaps of several identically
distributed samplings from the set $A$ and asked to ``retrodict'' the
preparation.  We might perhaps be told that one suit is missing from a 
pack of cards and, given three cards drawn from the smaller pack, asked
to guess which suit this is.  The tool we should use is {\em Bayes'
Theorem},
\begin{equation}\label{bayesthm}
p(b|a) = {p(a|b)p(b) \over p(a)} = {p(a|b)p(b) \over
\sum_{y\in B}p(a|y)p(y)}
\end{equation}
which is a simple consequence of $p(a|b)p(b) = p(a,b) = p(b|a)
p(a)$.  In applying Eqn \ref{bayesthm}, we typically have knowledge 
of $p(b)$ --- the probability that each suit was removed --- or if we
don't, we apply {\em Bayes' postulate}, or the ``principle of
insufficient reason'' \cite{Peres:1993}, which says in the absence of
such knowledge we assume a uniform distribution\footnote{There is some 
ambiguity here since a uniform distribution over $x^2$ is not uniform
over $x$; see \cite{Lane:1980}.  This is a source of much confusion
but is not a major obstacle to retrodiction.} over $B$.  A knowledge 
of $p(a|b)$ comes from our analysis of the situation: If all the
Clubs have been  removed, what is the probability of drawing the three
cards represented by $a$?  Once we have calculated $p(a|b)$ from
our knowledge of the situation, and obtained the ``prior''
probabilities $p(b)$ from some assumption, we can use Eqn
\ref{bayesthm} to calculate the
``posterior'' probability $p(b|a)$ of each preparation $b$ based on
our sampling result $a$. 

\paragraph{Stochastic processes}
We will later characterise an information source as a {\em stochastic
process} and this is an opportune place to introduce the definition.
A stochastic process is defined as an indexed sequence of random
variables \cite{Cover:1991} from the same symbol set $A$, where we may
imagine the index to refer to consecutive time steps or individual
characters in a sequence produced by an information source.  There may
be an arbitrary dependence between the random variables, so the
sequence may be described by a distribution
$\mbox{Prob}(X_1=x_1,\ldots,X_n=x_n) = p(x_1,\ldots,x_n)$.  A
stochastic source is described as {\em stationary} if the joint
distribution of any subset of symbols is invariant with respect to
translations in the index variable, i.e. for any index $s$
\be
p(x_1,\ldots,x_n) = p(x_{s+1},\ldots,x_{s+n}).
\end{equation}

\begin{example}{The weather}
The assumption behind most forms of weather forecasting is that the
weather operates as a stochastic process.  Thus if we know the vector
of variables like temperature, wind speed, air pressure and date for a 
series of days, we can use past experience to develop a probability
distribution for these quantities tomorrow (except for the date, which 
we hope is deterministic).

On the other hand, over a much longer time scale, the Earth's climate
does not appear to be stochastic.  It shows some sort of dynamical
behaviour which is not obviously repetitive, and so a probability 
description is less appropriate.
\end{example}

\subsection{Information Entropy}

Our aim in this section is to motivate the choice of the Shannon
entropy\footnote{Throughout this thesis, logarithms are assumed to be
to base 2 unless explicitly indicated.},
\be\label{Shannonentropy}
H(p) = - \sum_i p(X=x_i)\log p(X=x_i)
\end{equation}
as our measure of the information contained in a random variable $X$
governed by probability distribution $p$\ \footnote{$H$ is a functional
of the function $p$, so the notation in Eq \ref{Shannonentropy} is
correct. However, we frequently employ the random variable $X$ as the
argument to $H$; where confusion can arise, the distribution will be
explicitly noted.}.  There are in fact dozens of
ways of motivating this choice; we shall mention a few.

\begin{figure}
\begin{center}
\epsfig{file=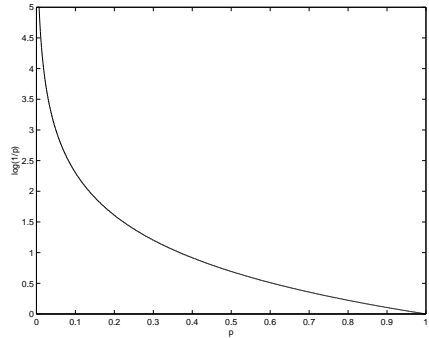, height=45mm}
\end{center}
\caption{The unexpectedness of an event of probability
$p$.}\label{unexpectedness} 
\end{figure}

As a first approach to Shannon's entropy function, we may consider the
random variable $u(x) = -\log p(x)$, which is sketched in Figure
\ref{unexpectedness}.
Intuitively we may justify calling $u$ the {\em unexpectedness} of the 
event $x$; this event is highly unexpected if it is almost sure not to 
happen, and has low unexpectedness if its probability is almost 1.
The information in the random variable $X$ is thus the ``eerily
self-referential'' \cite{Cover:1991} expectation value of $X$'s
unexpectedness.

Shannon \cite{Shannon:1949} formalised the requirements for an
information measure $H(p_1,\ldots,p_n)$ with the following criteria:
\begin{enumerate}
\item $H$ should be continuous in the $p_i$.
\item If the $p_i$ are all equal, $p_i=1/n$, then $H$ should be a
monotonic increasing function of $n$.
\item $H$ should be objective:
\be
H(p_1,\ldots,p_n) = H(p_1+p_2,p_3,\ldots,p_n) + (p_1+p_2)H\!\left(
\frac{p_1}{p_1+p_2}, \frac{p_2}{p_1+p_2}\right)
\end{equation}
\end{enumerate}
The last requirement here means that if we lump some of the outcomes
together, and consider the information of the lumped probability
distribution plus the weighted information contained in the individual 
``lumps'', we should have the same amount of information as in the
original probability distribution.  In the mathematical terminology of
Acz\'el and Dar\'oczy \cite{Aczel:1975}, the entropy is strongly
additive. Shannon proved that the unique (up
to an arbitrary factor) information measure satisfying these
conditions is the entropy defined in Eqn \ref{Shannonentropy}. Several
other authors, notably R\'{e}nyi, and Acz\'el and Dar\'oczy have
proposed other criteria which uniquely specify Shannon entropy as a
measure of information.

The arbitrary factor may be removed by choosing an appropriate
logarithmic base.  The most convenient bases are 2 and $e$, the unit
of information in these  cases being called the {\em bit} or the {\em
nat} respectively; if in this discussion the base of the logarithm is
significant it will be mentioned.

Some useful features to note about $H$ are:
\begin{itemize}
\item $H=0$ if and only if some $p_i=1$ and all others are zero; the
information is zero only if we are certain of the outcome.
\item If the probabilities are {\em equalised}, i.e.\ any two
probabilities are changed to more equal values, then $H$ increases.
\item If $p(a,b)=p_A(a)p_B(b)$ for all $a\in A, b\in B$ then
\begin{eqnarray*}
H(A,B)&=&- \sum_{x\in A, y\in B}p(x,y)\log p_A(x)p_B(y) \\
&=&- \sum_{x\in A, y\in B}p(x,y)\log p_A(x) - \sum_{x\in A, y\in
B}p(x,y)\log p_B(y) \\
&=&- \sum_{x\in A} p_A(x)\log p_A(x) - \sum_{y\in B} p_B(y)\log p_B(y) \\
&=&H(A) + H(B)
\end{eqnarray*}
\end{itemize}

We will be interested later in the information entropy of a more
general distribution on a product sample space.  So consider the
information contained in the distribution $p(a,b)$ on random variables 
$X$ and $Y$, where $p$ is not a product distribution:
\begin{eqnarray}
H(A,B) &=&- \sum_{x\in A, y\in B}p(x,y)\log p(x,y) \nonumber\\
&=& - \sum_{x\in A, y\in B}p(x,y)\log p(y|x)p(x) \nonumber\\
&=& - \sum_{x\in A, y\in B}p(x,y)\log p(y|x) - \sum_{x\in A}p_A(x)
\log p_A(x) \nonumber\\
&=& H(B|A) + H(A) \label{totentropy}
\end{eqnarray}
where we have defined the {\em conditional} entropy as
\be
H(B|A) = - \sum_{x\in A, y\in B}p(x,y)\log p(y|x) = \sum_{x\in A} p(x) 
\sum_{y\in B}p(y|x)\log p(y|x).
\end{equation}
Note that, for fixed $x\in A$, $p(y|x)$ is a probability distribution; 
so we could describe $H(Y|x) = \sum_{y\in B}p(y|x)\log p(y|x)$ as the
$x$-based entropy.  The conditional entropy is then the expectation
value of the $x$-based entropy.

Using the concavity of the $\log$ function, it can be proved that
$H(X,Y) \leq H(X) + H(Y)$: the entropy of a joint event is bounded by
the sum of entropies of the individual events.  Equality is achieved
only if the distributions are independent, as shown above.  From this
inequality and Eqn \ref{totentropy} we find
\be
H(Y|X) \leq H(Y)
\end{equation}
with equality only if the distributions of $Y$ and $X$ are independent.
In the case where they are not independent, learning about which value
of $x$ was sampled from the set $A$ allows us to update our knowledge
about what will be sampled from $B$, so our conditioned uncertainty
(entropy) is less than the unconditioned.  We could even extend our
idea of conditioned information to many more than just two sample
spaces; if we consider the random variables $X_1,\ldots,X_n$ defined
on spaces $A_1,\ldots,A_n$, we can define 
\begin{eqnarray*}
H(X_n|X_{n-1},\ldots,X_1) &= -\sum_{x_1\ldots x_n}&
p(x_1,\ldots,x_{n-1})\times \\
& & \sum_{x_n\in A_n} p(x_n|x_1,\ldots,x_{n-1})\log
p(x_n|x_1,\ldots,x_{n-1}) 
\end{eqnarray*}
to be the entropy of the next sampling given the sampled sequence once 
we know the preceding $n-1$ samples.  By repeated application of the
inequality above, we can show that
\be
H(X_n|X_{n-1},\ldots,X_1)\leq H(X_n|X_{n-1},\ldots,X_2) \leq\ldots\leq 
H(X_n|X_{n-1}) \leq H(X_n).
\end{equation}
In general, conditioning reduces our entropy and uncertainty.

We will now employ the characterisation of an information source as a
{\em stochastic process} as mentioned earlier.  Consider a stationary
stochastic source producing a sequence of random variables
$X_1,X_2,\ldots$ and the ``next-symbol'' entropy $H(X_{n+1}|X_n,
\ldots,X_1)$.  Note that
\begin{eqnarray*}
H(X_{n+1}|X_n,\ldots,X_1)&\leq& H(X_{n+1}|X_n, \ldots,X_2) \\
&=& H(X_n|X_{n-1},\ldots,X_1)
\end{eqnarray*}
where the equality follows from the stationarity of the process.  Thus 
next-symbol entropy is a decreasing sequence of non-negative
quantities and so has a limit.  We call this limit the entropy rate of 
the stochastic process, $H(X)$.  For a stationary stochastic process
this is equal to the limit of the average entropy per symbol,
\be
\frac{H(X_1,\ldots,X_n)}{n},
\end{equation}
which is a further justification for calling this limit the unique
entropy rate of the stochastic process.

\begin{example}{Entropy rate of a binary channel}\label{binarysource}
Consider a stochastic source producing a random variable from the set
$\{0,1\}$, with $p(0)=p, p(1)=1-p$.  Then the entropy is a function of 
the real number $p$, given by $H(p) = -p\log p - (1-p)\log(1-p)$.
This function is plotted in Figure~\ref{binaryH}.
\begin{figure}
\begin{center}
\epsfig{file=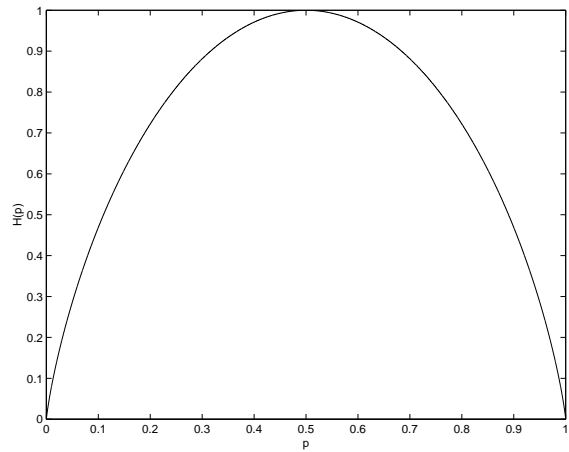, height=60mm}
\end{center}
\caption{The entropy of a binary random variable.}\label{binaryH}
\end{figure}
Notice that the function is concave and achieves its maximum value of
unity when $p=1/2$.
\end{example}

It was mentioned previously that entropy is a measure of our
ignorance, and since we interpret probabilities as subjective belief
in a proposition this ``ignorance'' must also be subjective.  The
following example, due to Uffink and quoted in \cite{Peres:1993},
illustrates this.  Suppose my key is in my pocket with probability 0.9 
and if it is not there it could equally likely be in a hundred other
places, each with probability $0.001$.  The entropy is then
$-0.9\log0.9-100(0.001\log0.001)=0.7856$.  If I 
put my hand in my pocket and the key is {\em not} there, the entropy
jumps to $-\log0.01=4.605$: I am now extremely uncertain where it is!
However, after checking my pocket there will on average be less
uncertainty; in fact the weighted average is $-0.9\log1-0.1\log0.01 =
0.4605$.

\subsection{Data Compression: Shannon's Noiseless Coding Theorem}
\label{largenumbers}

What is the use of the ``entropy rate'' of a stochastic source?  If
our approach to information is to be pragmatic, as mentioned
previously, then we need to find a useful interpretation of this
quantity.

We will first consider a stochastic source which produces independent, 
identically-distributed random variables $X_1, X_2,\ldots,X_n$.  A
typical sequence drawn from this source might be $x_1x_2\ldots x_n$
with probability $p(x_1,\ldots,x_n) = p(x_1)\ldots p(x_n)$.  Taking
logarithms of both sides and dividing by $n$, we find
\be
\frac{1}{n}\log p(x_1,\ldots,x_n) = \frac{1}{n}\sum \log p(x_i)
\end{equation}
and by the law of large numbers\footnote{The law of large numbers
states that if $x_1,\ldots,x_N$ are independent
identically-distributed random variables with mean $\bar{x}$ and
finite variance, then $P(|\frac{1}{N}\sum x_i - \bar{x}|>\delta)<\epsilon$ for
any $\delta,\epsilon>0$ \cite{Rice:1994}.} the quantity on the
right approaches the expectation value of the random variable $\log
p(X)$, which is just the entropy $H(X)$.  More precisely, if we
consider the set 
\be\label{palmer}
A_\epsilon^{(n)} = \{(x_1,\ldots,x_n)\in A|2^{-n(H(X)+\epsilon)}\leq
p(x_1, \ldots,x_n) \leq 2^{-n(H(X)-\epsilon)}\}
\end{equation}
then the following statements are consequences of the law of large
numbers:
\begin{enumerate}
\item $\mbox{Prob}(A_\epsilon^{(n)}) > 1-\epsilon$ for $n$ sufficiently large.
\item $|A_\epsilon^{(n)}| > (1-\epsilon)2^{n(H(X)-\epsilon)}$ for $n$
sufficiently  large, and $|\cdot|$ denotes the number of elements in
the set.
\item $|A_\epsilon^{(n)}| < 2^{n(H(X)+\epsilon)}$.
\end{enumerate}
Thus we can find a set of roughly $2^{nH(X)}$ strings (out of a
possible $|A|^n$, where $A$ is the set of possible symbols) which all
have approximately the same probability, and the probability of
producing a string {\em not} in the set is arbitrarily small.  

The practical importance of this is evident once we associate, with
each string in this ``most-likely'' set, a unique string from the set
$\{0,1\}^{nH(X)}$.  We thus have a code for representing strings
produced by  the stochastic source $\{X_i\}$ which makes arbitrarily
small errors\footnote{For the rare occasions when an unlikely string
$x_1\ldots x_n$ occurs, it does not matter how we deal with it: if our
system can tolerate errors we can associate it with the string
$0\ldots 0$; if not, we can code it into any unique string of length
$<<1/p(x_1,\ldots, x_n)$.}, and which uses $nH(X)$ binary digits to
represent $n$ source symbols.  We thus have the extremely useful
interpretation of $H(X)$ as the expected number of bits required to
represent long strings from the stochastic source producing random
variables $\{X_i\}$, in the case where each random variable in the set 
is independently distributed.

We have not quite done all we claimed in the previous paragraph, since 
we have ignored the possibility of a more compact coding strategy.
Let the strings from $A^n$ be arranged in order of decreasing
probability, and let $N(q)$ be the number of strings, taken in this
order, required to form a set of total probability $q$.  Then Shannon
proved the remarkable result that for $q\neq 0,1$
\be
\lim_{n\rightarrow\infty} \frac{\log N(q)}{n} = H.
\end{equation}
For large $n$ it makes no difference how we define ``probable'':
all probable sets contain about $2^{nH}$ elements!

Note also that if our strings are produced by an independent binary
source with $p(0)\neq 1/2$ then $H(p)<1$ (see Ex~\ref{binarysource}).
Thus our code strings are shorter than the source strings --- we have
compressed the information.  The resulting code strings will have
$p(0)\approx p(1)\approx 1/2$ and entropy rate close to unity.

We will now look at what changes when a fully stochastic source is
considered.  Instead of the random variables $X_1,\ldots,X_n$ being
independent, they are described by a probability distribution $p(x_1,
\ldots,x_n)$.  From the considerations above, we can design a code
with string lengths $l(x_1\ldots x_n)$ such that the expected length
per symbol $L = \frac{1}{n}\sum p(x_1,\ldots,x_n)l(x_1\ldots x_n)$
satisfies 
\be
\left|\frac{H(X_1,\ldots,X_n)}{n}-L\right| <\epsilon.
\end{equation}
If our stochastic source is stationary, then $H(X_1,\ldots,X_n)/n$
approaches the entropy rate $H(X)$.  Thus in this case too, the
entropy rate describes the shortest code available for a particular
source.  A provably optimal --- that is, shortest --- code can be 
found using an algorithm discovered by Huffman \cite{Cover:1991}, and
the communication theorist now has an enormous variety of codes to
choose from to suit his application.

For reference, Shannon's first theorem in full generality is given
below.

\begin{thm}{Shannon I}
Let ${\cal C}: A^{n} \rightarrow B$ be a code with binary codewords,
and suppose ${\cal C}$ has the property that no code word is the prefix
of another code word.  For each ${\bf x}\in A^n$ we denote the length
of the codeword ${\cal C}({\bf x})$ by $l_{\cal C}({\bf x})$ and
define $L_{\cal C} = \frac{1}{n}\sum p({\bf x}) l_{\cal C}({\bf x})$, the 
expected code word length per source symbol.  Suppose the source is
stochastic.  Then there exists a code ${\cal C}$ such that
\begin{equation}
\frac{H(X_1,\ldots,X_n)}{n}\leq L_{\cal C} \leq
\frac{H(X_1,\ldots,X_n)}{n} + \frac{1}{n}.
\end{equation}
An arbitrarily small error rate can be achieved if and only if the
first inequality is satisfied.
\end{thm}

We can also use Shannon's theorem to interpret the conditional entropy 
\be
H(X|Y) = H(X,Y) - H(Y).
\end{equation}
The right-hand side of this equation may be viewed as the number of
bits required to code $X$ and $Y$ together, less the number of bits
required to specify $Y$ alone.  The difference must surely be the
average number of bits required to code $X$ once $Y$ is known - one
could perhaps envisage using a different code for each random variable
$Y$ already in our possession.  This interpretation will be important
for considerations in the next section.

\subsection{Information Channels and Mutual Information}

Suppose we have a microphone on a stage, and the President is speaking 
into it.  The microphone will be hooked up to a loudspeaker system so
that the assembled throng will be able to hear him.  The President in
this situation is an information source, stochastically producing
words (or more generally sounds) near the microphone.  Considered
entirely separately, the loudspeaker is also a stochastic information
source.  If the technical crew have done their job there should not
only be a correlation between the output of the loudspeaker, there
should be a one-to-one correspondence.  Another way of saying this is
that if we know the sounds produced by the President, we should be
absolutely certain of what sounds will be produced by the
loudspeaker.  A mathematical way of expressing this is
\be
H(\mbox{loudspeaker}|\mbox{President})=0:
\end{equation}
the uncertainty (entropy) once we know the President's words should be 
zero.  Of course a real microphone-amplifier-loudspeaker (MAL) system
does introduce some errors, so in general the conditional entropy above
may  be some small positive amount.  A measure of the fidelity of the
MAL system could then be the reduction in entropy once we know the
original information:
\be\label{presidentinfo}
I(\mbox{loudspeaker}:\mbox{President}) = H(\mbox{loudspeaker}) -
H(\mbox{loudspeaker}|\mbox{President}).
\end{equation}
If somebody accidentally unplugged the microphone from the amplifier,
then there would be no correlation between the President's speech and
the sound produced by the loudspeaker and these could be considered to 
be independent sources, in which case $H(\mbox{loudspeaker}|
\mbox{President}) = H(\mbox{loudspeaker})$ and
\be
I(\mbox{loudspeaker}:\mbox{President}) = 0.
\end{equation}

The quantity defined in Eqn \ref{presidentinfo} is called the {\em
mutual information} between the President and the loudspeaker.  In a
general setting we would have two stochastic sources $X$ and $Y$ with
a given joint distribution $p(x,y)$ (which could in fact be a
distribution over n-tuples of symbols from $X$ and $Y$). The mutual
information would then be
\begin{eqnarray}
I(X:Y) &=& H(X) - H(X|Y) \label{pizza}\\
       &=& H(X) - H(X,Y) + H(Y) \nonumber\\
       &=& H(Y) - H(Y|X)
\end{eqnarray}
where we have used Eqn~\ref{totentropy} to obtain the second
equality. In this context, $H(X|Y)$ is sometimes referred to as the
equivocation of the channel.

Before continuing to the application and interpretation of the mutual
information, we note a few mathematical features of this function.  We 
note first the pleasing symmetry $I(X:Y)=I(Y:X)$; the
amount of information we gain about $X$ when we learn $Y$ is the same
as the amount of information gained about $Y$ on learning $X$.
Notice that, because $0\leq H(X|Y)\leq H(X)$, the mutual
information is always non-negative and always less than the entropies
$H(X)$ and $H(Y)$; the mutual information is zero if and only if the
distributions on $X$ and $Y$ are independent.  Finally observe that
$H(X|X)=0$, whence $I(X:X)=H(X)$ --- the self-information of a source
is equal to its information entropy.

The mutual information is in fact a special case of another function,
the {\em relative information} (otherwise known as Kullback-Leibler
distance\footnote{The Kullback-Leibler ``distance'' is not in fact a
metric: it is clearly not symmetric and doesn't satisfy a triangle 
inequality.}) between two distributions, which
we mention here for completeness.  The relative information between
two probability distributions $p(x)$ and $q(x)$ is defined to be
\begin{eqnarray}\label{crow}
D(p||q) &=& \sum_{x\in X} p(x) \log \frac{p(x)}{q(x)} \\
        &=& E_p \log \frac{p(x)}{q(x)}. \nonumber
\end{eqnarray}
Note that the relative information is {\em not} symmetric in $p$ and
$q$.  In fact $D$ has several very useful interpretations, notably as
the expected number of bits over and above the $H(p)$ required by
Shannon's theorem if the code you are using is optimised for the
non-occurring distribution $q$.  It is also easy to see that the
probability of observing a string $x_1x_2\ldots x_n$ from a source
producing independent, identically-distributed symbols with
distribution $p(x_i)$ is related to the distance between the observed
distribution and the real distribution.  If we let
$q(a_i)=\frac{n_i}{n}$ be the empirical distribution drawn from the
alphabet $\{a_1,\ldots,a_N\}$ then 
\begin{eqnarray}
p(x_1x_2\ldots x_n) &=& \prod_{i=1}^n p(x_i) = \prod_{j=1}^N p(a_j)^{n
q(a_i)} \nonumber \\
&=& \prod_{j=1}^N \exp[nq(a_j)\log p(a_j)] \nonumber\\
&=& \exp\left(n\sum_{j=1}^N[q(a_j)\log p(a_j) - q(a_j)\log q(a_j) +
q(a_j)\log q(a_j)]\right) \nonumber\\
&=& \exp\left(-n[H(q) + D(q||p)]\right). \nonumber
\end{eqnarray}
The mutual information is seen to be the relative information between
the product distribution of $X$ and $Y$ and the true joint
distribution:
\begin{eqnarray}
I(X:Y) &=& \sum_{x\in X}\sum_{y\in Y} p(x,y)\log\frac{p(x,y)}
{p(x)p(y)}\\
&=& D(p(x,y)||p(x)p(y)) \nonumber
\end{eqnarray}
and so is a measure of the correlation between $X$ and $Y$, that is,
the extent to which they differ from being independent.

As discussed in the previous section, the conditional entropy $H(X|Y)$ 
is a measure of the expected number of bits required to code $X$ once
we know the value of $Y$; the mutual information thus quantifies the
information about $X$ conveyed by $Y$ and vice-versa, and lends a more 
rigorous interpretation to the mutual information than the heuristic
explanation above.  It also leads us directly to the idea of an {\em
information channel}.

We characterise an information channel by the mistakes it makes, and
this knowledge is generally derived from an understanding of the
physical system used to convey the information.  For example in the
MAL system described above, the response of the amplifier will be
frequency dependent and perhaps have an upper and lower cutoff; the
loudspeaker in turn will also have a characteristic response, all of
which will lead to hissing, random noise, feedback and other such
unwanted effects.  If we assume that the alphabet under consideration
is a sequence of phonemes spoken by the President (we could
alternatively analyse the spectrum of his voice), then what his PR 
people will be interested in is the probability that a given phoneme
is produced by the loudspeaker when the President utters another given 
phoneme.  The intermediate steps don't interest them; what they care
about is $p(\mbox{lspk=``ch''}|\mbox{Pres=``sh''})=0.1$, because this
substitution could be damaging.

More rigorously, an information channel is characterised by an input
alphabet $X$, an output alphabet $Y$ and the transition probabilities
$p(y|x)$ that describe the probability of the input symbol $x$ being
turned into output symbol $y$.  For a given probability distribution
over the input symbols $X$, we can calculate the mutual information
(per symbol) between the input and output --- and if we further assume
that the channel can accept $r$ input symbols per unit time, then we
begin to see an interesting problem before us:  What is the fastest
rate at which information can be conveyed across this channel?  And
can we transmit information with arbitrarily few errors {\em despite}
the introduction of probabilistic errors by the channel?

These questions will take us to the heart of classical information
theory.  But to jump the gun a bit: The answer to the second question
is Yes, and transmission without errors can take place at the rate
\be\label{channelcap}
C = \max_{p(x)} I(X:Y)
\end{equation}
bits per symbol.  This is surprising, since one would imagine that we
could {\em either} transmit rapidly {\em or} transmit faithfully, but
not both at the same time.  That this is so is the content of
Shannon's Second Theorem, the Noisy Coding Theorem, which will be the
subject of the next section.

The quantity defined in Eqn~\ref{channelcap} is called the {\em
capacity} of the channel\footnote{The mutual information is
continuous over the probability simplex, and this simplex is compact,
so we are justified in calling this the maximum in place of supremum.
This also implies that the maximum is attained.}.  In most cases of
interest this can not be calculated explicitly, but some examples
serve to illustrate the idea of capacity.

\begin{example}{Noiseless and useless channels}
If we have a noiseless binary channel, so that $p(0|0)=1$ and
$p(1|1)=1$ then the maximum possible output entropy is 1 bit per
symbol and this capacity is achieved if we simply ensure that the
source probabilities are $p(0)=p(1)=1/2$.  On the other hand, if all
the transition probabilities are equal to $1/2$ then we can never hope 
to transmit any information.
\end{example}
\begin{example}{Binary symmetric channel}
Suppose a channel transmitting binary signals has probability $p$ of
flipping each bit, independently of other bits or of the particular
value of this bit.  By the symmetry of the errors, we observe that to
maximise the mutual information we should set $p(0)=p(1)=1/2$, so that 
$H(\mbox{transmitted})=1$.  If the channel output is a 1, then Bob
knows $p(1|0)=p$ and $p(1|1)=1-p$, so he calculates
\be
C = 1 - H(p)
\end{equation}
where $H$ is the binary entropy function plotted in
Figure~\ref{binaryH}.
\end{example}
\begin{example}{Ternary channel \cite{Shannon:1949}}
Suppose we have a channel with three symbols as depicted in
Figure~\ref{ternchann}, 
\begin{figure}
\begin{center}
\epsfig{file=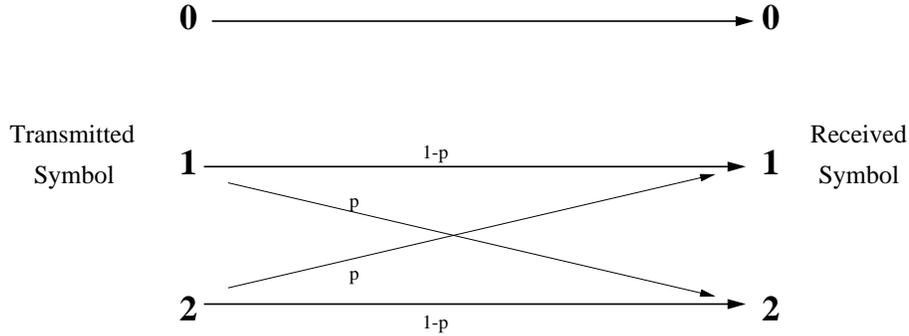, width=12cm}
\end{center}
\caption{The action of an uncertain ternary channel.}\label{ternchann}
\end{figure}
where one symbol 0 is transmitted without error, 
and the other two symbols 1 and 2 are interchanged with probability
$p$.  By symmetry, the capacity-achieving input source should have
$p(0)=P$, $p(1)=p(2)=Q$.  The mutual information will then be
\be
I = -P\log P -2Q\log Q -2Q\alpha
\end{equation}
where $\alpha = H(p) = -p\log p - (1-p)\log(1-p)$ is the noise due to
the channel.  We incorporate the constraint $P+2Q=1$ with a Lagrange
multiplier; we must maximise $U=-P\log P -2Q\log Q -2Q\alpha +\lambda
(P+2Q)$, whence\footnote{Here we assume the logarithm is to the base
$e$.}
\begin{eqnarray*}
\frac{\partial U}{\partial P} = -1 -\log P + \lambda = 0 \\
\frac{\partial U}{\partial Q} = -2 -2\log Q -2\alpha + 2\lambda = 0.
\end{eqnarray*}
Eliminating $\lambda$ we find $\log P=\log Q +\alpha$ or
$P=Qe^\alpha$.  Thus 
\be
P = \frac{e^\alpha}{e^\alpha+2}\hspace{10mm}Q=\frac{1}{e^\alpha+2}
\hspace{10mm} C = \log\frac{e^\alpha+2}{e^\alpha}.
\end{equation}
\end{example}

The channels we have considered here are described as memoryless.  For 
the brave-hearted and strong-willed out there, one can also consider
sources with memory i.e. where the error process can be considered as
a stochastic process depending on arbitrarily many previous input and
output symbols.  Memoryless channels are, fortunately, the rule in
situations of interest; and most examples of stochastic noise can be
approximated by memoryless channels transmitting large symbol blocks.

\subsection{Channel Capacity: Shannon's Noisy Coding Theorem}

The fundamental idea Shannon employed in showing that information can
be transmitted reliably over a noisy channel was to allow a small
probability of error, which goes to zero in some limit --- in
particular, in the limit when we code large blocks of symbols.

\begin{figure}
\epsfig{file=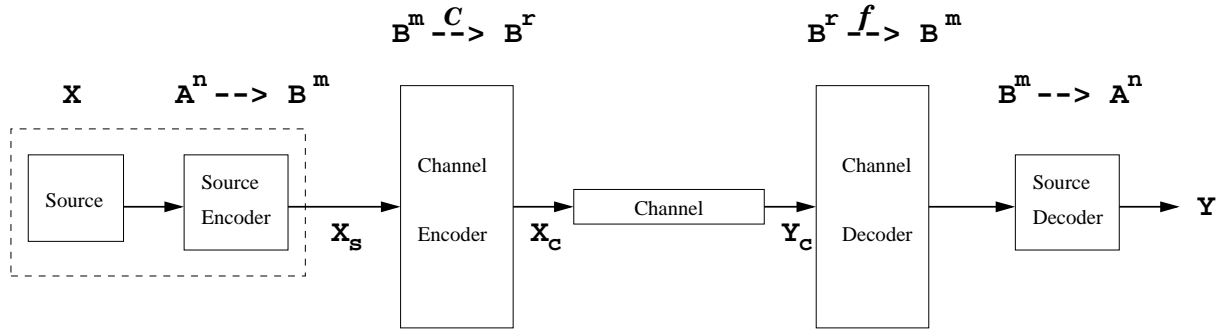, width=\textwidth}
\caption{Sketch of a communication system.}\label{commchann}
\end{figure}

Figure~\ref{commchann} is a sketch of the communication system
considered here.  The stochastic source produces symbols (represented
by the random variable $X$) drawn from the alphabet $A$, and a source
encoder optimally codes strings of source symbols into strings of
channel symbols drawn from the set $B$; this ``combined source'' is
represented by $X_s$.  The channel encoder introduces
some redundancy into the message by mapping $m$ channel symbols into
$r$ channel symbols, with $r>m$; this gives an effective source
$X_c$.  The output $Y_c$ of the channel is not
necessarily a string in the range of the code $C$, and we use a
decoding function $f$ to correct these changes.  Finally a source
decoder is applied; an error occurs whenever the random variable
$X\neq Y$.  With an optimal source encoder, we have 
$H(X_s)=\log |B|$ --- no redundancy --- and $H(X) = \frac{m}{n}\log
|B|$, where $|B|$ 
denotes the number of elements in B.  The input to the channel then
has entropy $H(X_c)=\frac{m}{r}H(X_s)$.  The rate of the
channel is defined to be $R =H(X_c)$, that is, the number of useful
bits conveyed per symbol sent.  The channel capacity $C$ in this case
is defined to be the mutual information of the {\em channel} symbols,
$X_s$ and $Y_c$, maximised over all codes,
\be\label{joubert}
C = \max \{ I(X_s:Y_c) | C:B^m\rightarrow B^r \},
\end{equation}
since the code will imply a probability distribution over the output
symbols.

We are now ready to state Shannon's
Noisy Channel Coding Theorem\footnote{The proof outline given here was
inspired by John Preskill's proof in {\it Lecture Notes for Physics 229:
Quantum Information and Computation}, available at
http://www.theory.caltech.edu/\~{}preskill/ph229.}.

\begin{thm}{Shannon II}
If $R\leq C$, there exists a code $C:B^m\rightarrow B^r$ such that the
probability of a decoding error for any message (string from $B^m$) is
arbitrarily small.
Conversely, if the probability of error is arbitrarily small for a
given code, then $R\leq C$.
\end{thm}

Before we sketch a proof of this, we return to the idea of conditional entropy
between the sent and received messages, $H(X_c|Y_c)$.  Suppose we have 
received an output $y_1\ldots y_N$; then if we had a noiseless side channel
which could convey the ``missing'' information $NH(X_c|Y_c)$, we could
perfectly reconstruct the sent message.  This means that there were on 
average $2^{NH(X_c|Y_c)}$ errors which this side channel would allow
us to correct, or equivalently that the received string, on average,
could have originated from one of $2^{NH(X_c|Y_c)}$
possible input strings.  And this last fact is crucial to coding,
since the optimal code will produce codewords which aren't likely to
diffuse to the same output string.

We begin by looking at the first statement in the theorem above.  
The strategy used by Shannon was
to consider random codes (i.e. random one-to-one functions from the
messages $B^m$ to the code words $B^r$) and average the probability of 
error over all these codes.  The decoding technique will be to
determine, 
for the received string $y_1\ldots y_r$, the set of
$2^{r(H(X_c|Y_c)+\delta)}$ most likely inputs, which we will call the
{\em decoding sphere} $D_\delta$. We then decode this received string
by associating it with a code word in its decoding sphere.  Suppose
without loss of generality that the input code word was $x^{(1)}$;
then there are two cases in which an error could occur: 
\begin{enumerate}
\item $x^{(1)}$ may not be in the decoding sphere; 
\item There may be other code words apart from $x^{(1)}$ in the
decoding sphere.
\end{enumerate}
Given arbitrary $\epsilon>0$ and $\delta>0$, the probability that
$x^{(1)}\in D_\delta$ can be made greater than $1-\epsilon$ by choosing
$r$ large.  So the probability of an error is
\be
P_e \leq \epsilon + p\left(\mbox{$x^{(i)}\in D_\delta$ for
at least one $i=2,3,\ldots, |B|^m$}\right).
\end{equation}

There are $2^{rR}$ code words distributed among $|B|^r=2^{rH(X_s)}$
strings from $B^r$; by the assumption of random coding, the
probability that an arbitrary string is a code word is
\be
p(\mbox{$X$ is a codeword}) = \frac{2^{rR}}{2^{rH(X_s)}} = 2^{-r(H(X_s) 
-R)}
\end{equation}
independently of other code word assignments.
We can now calculate the probability that $D_\delta$ contains a code
word (apart from $x^{(1)}$):
\begin{eqnarray}
p\left(\mbox{code word in $D_\delta$}\right)&=& \sum_{X\in D_\delta, 
X\neq x^{(1)}} p(\mbox{$X$ is a codeword}) \nonumber\\
&=& \left(|D_\delta|-1\right)2^{-r(H(X_s)-R)} = 2^{-r\left(H(X_s)-R -
H(X_c|Y_c)-\delta\right)}.
\end{eqnarray}
Now $H(X_c|Y_c)=H(X_s,X_c|Y_c)=H(X_c|X_s,Y_c)+H(X_s|Y_c)=H(X_s|Y_c)$,
where the first equality follows from the functional dependence of
messages $X_s$ on code words $X_c$, the second follows from
Eqn~\ref{totentropy}, and the third follows from the functional
dependence of $X_c$ on $X_s$.  Thus we can simplify the expression
above:
\be
p\left(\mbox{code word in $D_\delta$}\right)=2^{-r(I(X_s:Y_c) - R
-\delta)}
\end{equation}
and we conclude that the probability of an error goes to zero
exponentially (for large $r$) as long as $I(X_s:Y_c)-\delta > R$.
If we in fact employ the code that achieves channel capacity and
choose $\delta$ to be arbitrarily small, then the 
condition for vanishing probability of error becomes
\be
C\geq R
\end{equation}
as desired.

Note that we have shown that the average probability of error can be
made arbitrarily small:
\be
\frac{1}{2^{rR}}\sum_i^{2^rR} p(\mbox{error when code word $x^{(i)}$ is
sent}) < \epsilon
\end{equation}
Let the number of code words for which the probability of error is
greater than $2\epsilon$ be denoted $N_{2\epsilon}$; then
\be
\frac{1}{2^{rR}}N_{2\epsilon}(2\epsilon) < \epsilon
\end{equation}
so that $N_{2\epsilon}<2^{rR-1}$.  If we throw away these
$N_{2\epsilon}$ code words and their messages then all code words have
probability of error less than $2\epsilon$.  There
will now be $2^{rR-1}$ messages communicated, and the effective rate
will be
\begin{eqnarray}
\mbox{Rate} &=& \frac{\log(\mbox{number of messages})}{\mbox{number of
symbols sent}} \nonumber\\
&=& \frac{rR-1}{r} \rightarrow R \label{jellybabes}
\end{eqnarray}
for large $r$.

For the converse, we begin by noting that the channel transition
probability for a string of $r$ symbols factorises (by the memoryless
channel assumption): $p(y_1\ldots y_r|x_1\ldots x_r)=p(y_1|x_1)\ldots
p(y_r|x_r)$.  It is then easy to show that 
\be
H(Y_c^r|X_s^r) = \sum_{i=1}^r H(Y_{c,i}|X_{s,i}).
\end{equation}
Also, since $H(X,Y)\leq H(X)+H(Y)$, we have $H(Y_c^r)\leq \sum_i
H(Y_{c,i})$, so that
\begin{eqnarray*}
I(Y_c^r:X_s^r) &=& H(Y_c^r) - H(Y_c^r|X_s^r) \\
&\leq & \sum_{i=1}^r H(Y_{c,i}|X_{s,i}) - H(Y_{c,i}) \\
&=& \sum I(Y_{c,i}:X_{s,i}) \leq rC.
\end{eqnarray*}
But mutual information is symmetric, $I(X:Y)=I(Y:X)$, and
using the fact that $H(X_s^r)=rR$ we find
\be
I(X_s^r:Y_c^r) = H(X_s^r)-H(X_s^r|Y_c^r)=rR-H(X_s^r|Y_c^r) \leq rC.
\end{equation}
The quantity $\frac{1}{r}H(X_s^r|Y_c^r)$ measures our average
uncertainty about the input after receiving the channel output.  If
our error probability goes to zero as $r$ increases, then this
quantity must become arbitrarily small, whence $R\leq C$.

\section{Quantum Mechanics}

Quantum mechanics is a theory that caught its inventors by surprise.
The main reason for this is that the theory is heavily empirical and
pragmatic in flavour --- the formalism was forced onto us by
experimental evidence --- and so contradicted the principle-based
``natural philosophy'' tradition which had reaped such success in
physics.  In the absence of over-arching principles we are left with a
formalism, fantastically successful, which is mute on several
important subjects.

What is a quantum system?  In the pragmatic spirit of the theory, a
quantum system is one which cannot be described by classical
mechanics.  In general quantum effects become important when small
energy differences become important, but in the absence of {\em a
priori} principles we can give no strict definition. For example, NMR
quantum computing can be described in entirely classical terms
\cite{Schack:1999} and yet is advertised as the first
demonstration of fully {\em quantum} computing; and Kitaev
\cite{Preskill:1999b} has conjectured that some types of quantum
systems can be efficiently simulated by classical systems.  The
quantum-classical distinction is
not crucial to this thesis, where we are dealing with part of the
formal apparatus of the theory; indeed, there is some hope that from
this apparatus can be coaxed some principles to rule the quantum world 
\cite{Fuchs:1999}.

For our purposes, the following axioms serve to define quantum
mechanics:
\begin{enumerate}
\item {\bf States.}  A state of a quantum system is represented by a
bounded linear operator $\rho$ (called a density operator) on a
Hilbert space ${\cal H}$ satisfying the following conditions:
\begin{itemize}
\item $\rho$ is Hermitian.
\item The sum of the eigenvalues is one, $\Tr{}\rho=1$.
\item Every eigenvalue of $\rho$ is nonnegative, which we denote by
$\rho \geq 0$.
\end{itemize}
Vectors of the Hilbert space will be denoted by $|\psi\ran$ (Dirac's
notation), and the inner product of two vectors is denoted $\lan\psi|
\phi\ran$.  In all cases considered in this thesis, the underlying
Hilbert space will be of finite dimension\footnote{For the case of an
infinite-dimensional Hilbert space, additional technical requirements
regarding the completeness of the space and the range of the operator
$\rho$ must be imposed.}. 
\item {\bf Measurement.}  Repeatable, or von Neumann, measurements are
represented by complete sets of projectors\footnote{A measurement is
also frequently associated with a Hermitian operator $E$ on the
Hilbert space; correspondence with the current formalism is achieved
if we associate with $E$ the set of projectors onto its eigenspaces.}
$\Pi_i$ onto orthogonal
subspaces of ${\cal H}$.  By complete we mean that $\sum_i \Pi_i = \unit$
(the identity operator), but the projectors are not required to be
one-dimensional.  Then the probability of outcome $i$ is
\be
p(i) = \Tr{}\Pi_i\rho.
\end{equation}
After the measurement the new state is given by
$\rho_i'=\Pi_i\rho\Pi_i/ \Tr{}\Pi_i\rho$ if we know that  outcome $i$ was
produced, and otherwise by $\rho'=\sum_i \Pi_i\rho\Pi_i$ if we merely
know a measurement was made.
\item {\bf Evolution.} When the system concerned is sufficiently
isolated from the environment and no measurements are being made,
evolution of the density operator is {\em unitary}:
\be
\rho\longrightarrow U\rho U^\dagger,
\end{equation}
where $U$ is unitary so that $U^{-1}=U^\dagger$.  The dynamics of the
system are governed by the Schr\"odinger equation.  In this thesis we
will not be concerned about the specific operator $U$ relevant for a
certain situation since we will be more concerned with evolutions that 
are in principle possible.  However, we observe that for a given
system the unitary operator is given by $U=\exp(iHt)$, where $H$ is the 
Hamiltonian of the system and $t$ is the time.
\end{enumerate}
These are the basic tools of the mathematical formalism of quantum
mechanics.  In the following sections we will consider systems which
are not fully isolated from their environment and in which
measurements occur occasionally.  Our aim is to complete our toolkit
by discovering what evolutions are in principle possible for a real
quantum system within the framework of the axioms above.

\subsection{States and Subsystems}

A density operator $\rho$ which is a one-dimensional projector is
called a {\em pure state}.  In this case $\rho = |\psi\ran\lan\psi|$
for some vector $|\psi\ran$ in the Hilbert space ${\cal H}$, and we
frequently call $|\psi\ran$ the state of the system.  Pure states
occupy a special place in quantum theory because they describe states
of {\em maximal knowledge} \cite{Peres:1993} --- no further experiments
on such a pure state will allow us to predict more of the system's
behaviour.  There is in fact a strong sense in which mixed states (to
be discussed below) involve less knowledge than pure states: Wootters
\cite{Wootters:1990} showed that if the average value of a dynamical
variable of a system is known, the uncertainty in this observable
decreases if we have the additional knowledge that the state is pure.

The density operators form a {\em convex} set, which means that for
any $(0\leq\lambda \leq 1)$, a
convex combination $\lambda\rho_1 +(1-\lambda)\rho_2$ of two density
operators $\rho_1,\rho_2$ is again a density 
operator.  The pure states are also special in this context: they form 
the extreme points of this convex set, which means that there is no
non-trivial way of writing a pure state $|\psi\ran\lan\psi|$ as a
convex combination of other density operators.  Also, since a general
density operator is Hermitian and positive, there exists a set $\{|
\psi_i\ran\}$ of vectors and real numbers $\lambda_i$ such that
\be
\rho = \sum_{i=1}^D \lambda_i |\psi_i\ran\lan\psi_i|
\end{equation}
where $D$ is the dimension of the underlying Hilbert space; this
result is the spectral theorem for Hemitian operators.  The
eigenvalues of $\rho$ are then the $\lambda_i$, and they sum to one.
For this reason mixed states are frequently regarded as classical
probability mixtures of pure states.  Indeed, if we consider the
ensemble in which the pure state $|\psi_i\ran$ occurs with probability 
$\lambda_i$, and we make a measurement represented by the projectors
$\{\Pi_\nu\}$, then the probability of the outcome $\mu$ is
\be
p(\Pi_\mu) = \sum_{i=1}^D \lambda_i \Tr{}|\psi_i\ran\lan\psi_i|
\Pi_\mu = \Tr{} \left(\sum_{i=1}^D \lambda_i
|\psi_i\ran\lan\psi_i| \Pi_\mu\right) = \Tr{}\rho\Pi_\mu
\end{equation}
where we have used the linearity of the trace to take the $\lambda$s
inside.

Thus if a pure state results from {\em maximal} knowledge of a physical
system, then a mixed state arises when our available knowledge of the
system doesn't uniquely identify a pure state --- we have less than
maximal knowledge.  Mixed states arise in two closely related ways:
\begin{itemize}
\item We do not have as much knowledge as we could in principle,
perhaps due to imperfections in our preparation apparatus, and
must therefore characterise the state using a probability distribution
over all pure states compatible with the knowledge available to us.
\item The system $A$ is part of a larger system $AB$ which is in a
pure state.  In this case our knowledge of the whole system is
maximal, but correlations between subsystems don't permit us to make
definite statements about subsystem $A$.
\end{itemize}
From an experimenter's viewpoint these situations are
indistinguishable.  Consider the simplest canonical example of a
quantum system, a two-level system which we will refer to as a {\em
qubit}.  We can arbitrarily label the pure states of the system
$|0\ran$ and $|1\ran$, which could correspond to the two distinct
states of the spin degree of freedom (`up' and `down') in a spin-1/2
particle. The experimenter may produce these particles in an
ionisation chamber, and the density matrix describing these randomly
produced particles would be $\frac{1}{2}|0\ran\lan0|+\frac{1}{2}|1\ran 
\lan1| = \frac{1}{2}\unit$.  If he were extremely skilled, however,
the experimenter could observe the interactions which produced each
electron and write down an enormous pure state vector for the system.
When he traced out the many degrees of freedom not associated with the 
electron to which his apparatus is insensitive, he would again be left
with the state $\frac{1}{2}\unit$. 

The ``tracing out of degrees of freedom'' is achieved in the formalism 
by performing a {\em partial trace} over the ignored system.  Suppose
we consider two subsytems $A$ and $B$ and let $|1_a\ran$ (where 
$a=1,\ldots,D_1$) and $|2_\alpha\ran$ (where $\alpha=1,\ldots,D_2$) be
orthonormal bases over the two systems' Hilbert spaces respectively.
Then if $\rho^{AB}$ is a joint state of the system, the {\em reduced
density matrix} $\rho^A$ of system $A$ is
\be
\rho^A = \sum_{\alpha=1}^{D_2} \lan 2_\alpha|\rho^{AB}|2_\alpha\ran
\equiv \Tr{B} \rho
\end{equation}
with a similar expression holding for subsystem $B$.  We frequently use
the matrix elements $\rho^{AB}_{a\alpha,b\beta} =
\lan1_a| \lan2_\alpha|\rho^{AB}|2_\beta\ran|1_b\ran$ of $\rho^{AB}$ to 
represent the density operator of a system; in this notation, the
operation of partial trace looks like
\be
\rho^A_{ab} = \sum_{\nu=1}^{D_2} \rho^{AB}_{a\nu,b\nu}.
\end{equation}

As a converse to this, one can also consider the {\em purifications} of 
a mixed state.  Suppose our state $\rho$ is an operator on the Hilbert 
space $H_A$ with spectral decomposition $\rho=\sum_i\lambda_i|
\psi_i\ran\lan\psi_i|$.  Then consider the following vector from the space
$H_A\otimes H_B$ ($\dim H_A\leq\dim H_B$):
\be
|\Psi^{AB}\ran = \sum_i \sqrt{\lambda_i}|B_i\ran\otimes|\psi_i\ran,
\end{equation}
with the $|B_i\ran$ any orthogonal set in $H_B$.  Then $\rho =
\Tr{B}|\Psi^{AB}\ran\lan \Psi^{AB}|$, so $\rho$ can be considered to
be one part of a bipartite pure state.  There are of course an
infinite number of alternative purifications of a given mixed state.

\begin{example}{Two spin-1/2 systems}
The simplest bipartite system is a system of two qubits.  The sets $\{
|0_A\ran, |1_A\ran\}$ and $\{|0_B\ran, |1_B\ran\}$ are bases for the
two individual qubit's spaces.  If we denote a vector $|i_A\ran\otimes 
|j_B\ran$ by $|ij\ran$, then a basis for the combined system is
$\{|00\ran, |01\ran, |10\ran, |11\ran\}$.  Consider the pure state of the
combined system
\be
|\psi^-\ran = \frac{1}{\sqrt{2}}\left(|10\ran - |01\ran\right);
\end{equation}
which has density matrix 
\be
\rho^{AB}=|\psi^-\ran\lan\psi^-| = \frac{1}{2}\bigg(|10\ran\lan10| +
|01\ran\lan01| - |10\ran\lan01| - |01\ran\lan10|\bigg);
\end{equation}
tracing out system $B$ removes the second index and retains only those 
terms where the second index of the bra and ket are the same.  So the
reduced density matrix of system $A$ is $\rho^A =
\frac{1}{2}(|1\ran\lan1| + |0\ran\lan0|) = \frac{1}{2}\unit$.
\end{example}

The singlet state $|\psi^-\ran$ is a state of maximum uncertainty in
each subsystem.  This is because of the high degree of entanglement
between its two subsystems --- and this state, along with the triplet
states $|\psi^+\ran$ and $|\phi^\pm\ran$, is a canonical example of
all the marvel and mystery behind quantum mechanics.  We will return
to investigate some properties of this state in later chapters. 

\subsection{Generalised Measurements}

Von Neumann measurements turn out to be unnecessarily restrictive: the 
number of distinct outcomes is limited by the dimensionality of the
system considered, and we can't answer simple questions that have
useful classical formulations \cite[p. 280]{Peres:1993}.  Our remedy
for this is to adopt a new measurement technique.  Suppose we wish to
make a measurement on a quantum state $\rho^s$ with Hilbert space
$H_s$.  We will carry out the following steps:
\begin{itemize}
\item Attach an {\em ancilla} system in a known state $\rho^a$.  The
state of the combined system will be the product state $\rho^s\otimes
\rho^a$.  Note that our ancilla Hilbert space could have as many
dimensions as we require. 
\item Evolve the combined system unitarily, $\rho^s\otimes \rho^a
\longrightarrow U\rho^s\otimes\rho^a U^\dagger$.  The resulting state is
likely going to be entangled.
\item Make a von Neumann measurement on just the ancilla represented
by the projectors $\unit_s\otimes\Pi_\alpha$, where $\Pi_\alpha$ acts
on the ancilla.  The probability of outcome $\alpha$ will then be 
\be
p(\alpha)=\Tr{}\left(\unit\otimes\Pi_\alpha U\rho^s
\otimes\rho^a U^\dagger\right).
\end{equation}
\end{itemize}
Of course we are not interested in the (hypothetical) ancilla system
we have introduced, and so we can simplify our formalism by tracing it 
out earlier.  If we write
\be\label{london}
E_\alpha = \Tr{\mbox{\scriptsize ancilla}} \left(\unit\otimes\Pi_\alpha U\unit
\otimes\rho^a U^\dagger\right)
\end{equation}
then $E_\alpha$ is an operator over the system Hilbert space $H_s$,
and the probability of outcome $\alpha$ is given by the much simpler
formula $p(\alpha)=\Tr{} (E_\alpha\rho^s)$, where the trace is over
just the system degrees of freedom.  Note that the final two steps can 
be amalgamated: unitary evolution followed by measurement is exactly
the same as a measurement in a different basis.  However in this case
we will have to allow a von Neumann measurement on the combined
system, not just the ancilla.

The set of operators $\{E_\alpha\}$ is called a POVM, for Positive
Operator-Valued Measure, and represents the most general type of
measurement possible on a quantum system.  Note that, in contrast with 
a von Neumann measurement, these operators are not orthogonal and at
first glance there doesn't appear to be any simple way to represent
the state just after a measurement has been made.  We will return to
this in the next section.

The set $\{E_\alpha\}$ has the following characteristics:
\begin{enumerate}
\item $\sum_\alpha E_\alpha = \unit$.
\item Each $E_\alpha$ is a Hermitian operator.
\item All the eigenvalues of the $E_\alpha$ are non-negative
i.e. $E_\alpha \geq 0$.
\end{enumerate}
These are in some sense the minimum specifications required to extract
probabilities from a density operator.  The first requirement ensures
that the probability of obtaining some outcome is one; the second
ensures that the probabilities are real numbers, and the third
guarantees that these real numbers are non-negative.  It is therefore
pleasing that any set of operators satisfying these 3 conditions can
be realised using the technique described at the start of this
section: this is the content of {\em Kraus' Theorem}
\cite{Peres:1993}.  So if we
{\em define} a POVM by the above three requirements, we are guaranteed 
that this measurement can be realised by a repeatable measurement on a 
larger system.  And the above characterisation is a lot easier to work 
with!

\subsection{Evolution}

What quantum state are we left with after we have performed a POVM on
a given quantum state?   We will first look at this question for the
case where the system starts in a pure state,
$\rho^s=|\psi^s\ran\lan\psi^s|$, and the ancilla is in a mixed
state\footnote{We could, without loss of generality, assume the
ancilla starts in a pure state.} 
$\rho^a=\sum_i \mu_i|\phi^a_i\ran\lan\phi^a_i|$.
Then according to the measurement axiom presented previously, the final
state of the combined system will be 
\be
\rho^{sa}_\alpha = \frac{1}{\Tr{}(E_\alpha \rho^s)}
\sum_i\mu_i UP_\alpha|\psi^s\ran\otimes| \phi^a_i\ran \lan 
\psi^s| \otimes \lan\phi^a_i| P_\alpha U^\dagger
\end{equation}
where $P_\alpha=U^\dagger(\unit\otimes\Pi_\alpha)U$ represents a
projection onto some combined basis of the system and ancilla.  Now
we define a set of operators on the system,
\be
M_{(i,k)\alpha} = \sqrt{\mu_i}\:
(\unit\otimes\lan\xi_k|)P_\alpha(\unit\otimes |\phi^a_i\ran),
\end{equation}
where $|\xi_k\ran$ is an arbitrary orthonormal basis for the ancilla Hilbert
space and $E_\alpha$ is the corresponding POVM element. Then the state
of the system after measurement can be expressed 
in terms of these operators:
\begin{eqnarray}
\rho^s_\alpha &=& \frac{1}{\Tr{}(E_\alpha\rho^s)}\Tr{\mbox{\scriptsize 
ancilla}}
\bigg(\sum_i \mu_i P_\alpha|\psi^s\ran\otimes|\phi^a_i\ran \lan
\psi^s| \otimes \lan\phi^a_i|
P_\alpha\bigg) \nonumber\\
&=& \frac{1}{\Tr{}(E_\alpha\rho^s)} \sum_{\bf b} M_{{\bf
b}\alpha}|\psi^s\ran\lan\psi^s|M_{{\bf b}\alpha}^\dagger 
\end{eqnarray}
where we have amalgamated the indices $(i,k)$ into one index ${\bf
b}$.  This operation is linear so that for any mixed state $\rho^s$,
the state after measurement is
\be\label{chinatown}
\rho^s_\alpha = \frac{1}{\Tr{}(E_\alpha\rho^s)} \sum_{\bf b} M_{{\bf
b}\alpha}\rho^s M_{{\bf b}\alpha}^\dagger 
\end{equation}
{\em if} we know that the outcome is $\alpha$, and 
\be\label{eastwing}
\rho^s  = \sum_\alpha p(\alpha) \rho^s_\alpha = \sum_{{\bf b}\alpha}
M_{{\bf b}\alpha}|\psi^s\ran\lan\psi^s|M_{{\bf b}\alpha}^\dagger
\end{equation}
if we only know that a measurement has been made.  Note that we can
choose different orthonormal bases $|\xi_k\ran$ for each value of
$\alpha$ to make the representation as simple as possible --- and that 
in general, the state resulting from a POVM depends on exactly how the 
POVM was implemented.  The operators $M_{{\bf b}\alpha}$ are not free, 
however; they must satisfy
\begin{eqnarray}
\sum_{\bf b}M^\dagger_{{\bf b}\alpha}M_{{\bf b}\alpha} &=& \sum_{i,k}
\mu_i (\unit\otimes\lan\xi_k|)P_\alpha(\unit\otimes
|\phi^a_i\ran)(\unit\otimes \lan\phi^a_i|) P_\alpha (\unit\otimes|\xi_k\ran)
\nonumber\\
&=& \Tr{\mbox{\scriptsize ancilla}} \bigg[ P_\alpha(\unit\otimes\rho^a)P_\alpha
\bigg] \nonumber \\
&=& \Tr{\mbox{\scriptsize ancilla}} \bigg[ \unit\otimes\Pi_\alpha U\unit\otimes
\rho^a U^\dagger \bigg] = E_\alpha
\end{eqnarray}
(from Eqn~\ref{london}).  
We will call a generalised measurement {\em
efficient} if for each value of $\alpha$, there is only one value of
${\bf b}$ in the sum in Eqn~\ref{chinatown}; such a measurement is
called efficient because pure states are mapped to pure states, so if
we have maximal knowledge of the system before the measurement this is 
not ruined by our actions.

Suppose now we have an arbitrary set of operators $N_\mu$ satisfying
(as do the $M_{{\bf b}\alpha}$ above) $\sum_\mu N_\mu^\dagger N_\mu =
\unit$.  Then the mapping defined by 
\be
\rho \longrightarrow \$(\rho) = \sum_\mu N_\mu\rho N_\mu^\dagger
\end{equation}
is called an {\em operator-sum} and has the following convenient
properties:
\begin{enumerate}
\item If $\Tr{}\rho = 1$ then $\Tr{}\$(\rho)=1$ ($\$$ is
trace-preserving).
\item $\$$ is linear on the space of bounded linear operators on a
given Hilbert space.
\item If $\rho$ is Hermitian then so is $\$(\rho)$.
\item If $\rho\geq 0$ then $\$(\rho)\geq 0$ (we say $\$$ is positive).
\end{enumerate}
If we add one more condition to this list, then this list defines an
object called a {\em superoperator}.  This extra condition is
\begin{enumerate}\addtocounter{enumi}{4}
\item Let $\unit_n$ be the identity on the $n$-dimensional Hilbert space.  We
require that the mapping $T_n = \$\otimes\unit_n$ be positive for all
$n$.  This is called {\em complete positivity}.
\end{enumerate}
Physically, this means that if we include for consideration any extra
system $X$ --- perhaps some part of the environment --- so that the
{\em combined} system is possibly entangled, but the system $X$ evolves 
trivially (it doesn't evolve), the
resulting state should still be a valid density operator.  It turns out that 
an operator-sum does satisfy this requirement and so any operator-sum
is also a superoperator.

Superoperators occupy a special place in the hearts of quantum
mechanicians precisely because they represent the most general possible
mappings of valid density operators to valid density operators, and
thus the most general allowed evolution of physical states.  It is
thus particularly pleasing that we have the following theorems (proved
in \cite{Schumacher:1996}): 
\begin{enumerate}
\item Every superoperator has an operator-sum representation.
\item Every superoperator can be physically realised as a unitary
evolution of a larger system.
\end{enumerate}
The first theorem is a technical one, giving us a concrete
mathematical representation for the ``general evolution'' of an
operator.  The second theorem has physical content, and tells us that
what we have plucked out of the air to be ``the most general possible
evolution'' is consistent with the physical axioms presented previously.


\chapter{Information in Quantum Systems}

At the start of the previous chapter, we identified information as an
abstract quantity, useful in situations of uncertainty, which was in a
particular sense independent of the symbols used to represent it.  In
one example we had the option of using symbols {\bf Y} and {\bf N} or
integers $\{0,1,\ldots,365\}$.  But we in fact have even more freedom
than this: we could communicate an integer by sending a bowl with $n$
nuts in it or as an $n$ volt signal on a wire; we could relay {\bf Y}
or {\bf N} as letters on a piece of paper or by sending Yevgeny
instead of Nigel.

Physicists are accumstomed to finding, and exploiting, such
invariances in nature.  The term ``Energy''
represents what we have to give to water to heat it, or what a
rollercoaster has at the top of a hump, or what an exotic particle
possesses as it is ejected from a nuclear reaction.  Information thus
seems like a prime candidate for the attention of physicists, and the
sort of questions we might ask are ``Is information conserved in
interactions?'', ``What restrictions are there to the amount of
information we can put into a physical system, and how much can we
take out of one?'' or ``Does information have links to any other
useful concepts in physics, like energy?''.  This chapter addresses
some of these questions.

The physical theory which we will use to investigate them is of 
course quantum mechanics.  But --- and this is another reason for
studying quantum information theory --- such considerations are also
giving us a new perspective on quantum theory, which may in time lead
to a set of principles from which this theory can be derived.

One of the major new perspectives presented by quantum information
theory is the idea of analysing quantum theory from within.  Often
such analyses take the form of algorithms or cyclic processes, or in
some circumstances even games \cite{Eisert:1999}; in short, the
situations considered are {\em finite} and {\em completely
specified}.  The aims of such investigations are generally:
\begin{itemize}
\item To discover situations (games, communication problems,
computations) in which a system behaving quantum mechanically yields 
qualitatively different results from {\em any} similar system described
classically.
\item To investigate extremal cases of such ``quantum violation'' and
perhaps deduce fundamental limits on information storage, transmission 
or retrieval.  Such extremal principles could also be useful in
uniquely characterising quantum mechanics.
\item To identify and quantify the quantum resources (such as
entanglement or superposition) required to achieve such qualitative
differences, and investigate the general properties of these resources.
\end{itemize}
In many of these investigations, the co-operating parties Alice and
Bob (and sometimes their conniving acquaintance Eve) are frequently
evoked and this thesis will be no exception.  In this chapter we
consider a message to be sent from Alice to Bob\footnote{The
contrasting of preparation, missing and accessible information
presented here is based on \cite{Caves:1996}.}, encoded in the
quantum state of a system.  For the moment we will be assuming that
the quantum state passes unchanged (and in particular is not subject
to degradation or environmental interaction) between them.

\section{Physical Implementation of Communication}

\begin{figure}
\begin{center}
\epsfig{file=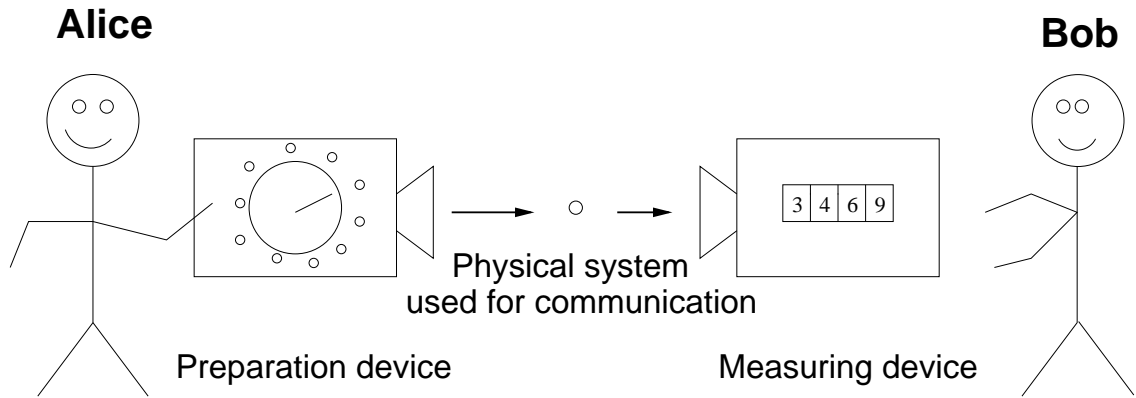, width=15cm}
\end{center}
\caption{Alice and Bob using a physical system to convey information}
\label{alicebob}
\end{figure}

In classical information theory, the {\em bit} is a primitive notion.
It is an abstract unit of information corresponding to two
alternatives, which are conveniently represented as 0 and 1.  Some of
the characteristics of information encoded in bits are that it can be
freely observed and copied (without changing it), and that after
observing it one has 
exactly the right amount of information to prepare a new instance of
the same information.  Also, in order to specify one among $N$
alternatives we require $\log N$ bits.

The quantum version of a two-state system has somewhat different
properties.  Typical examples of two state systems are spin-1/2
systems, where the `spin up' state may be written $\spinupvec$ and
`spin down' $\spindownvec$, or the polarisation states of photons,
which may be written $|\!\updownarrow\ran$ for vertical polarisation and 
$|\!\leftrightarrow\ran$ for horizontal.  For convenience we will label
these states $|0\ran$ and $|1\ran$, using the freedom of
representation discussed previously. A new feature arises
immediately in that the quantum two-state system --- or
qubit\footnote{This word was coined by Schumacher
\cite{Schumacher:1995}.} --- can exist in the superposition state
$|0\ran +|1\ran$ which, if measured in the conventional basis
$\{|0\ran, |1\ran\}$ will yield a completely random outcome.

We will analyse the communication setup represented in
Figure~\ref{alicebob}.  Alice prepares a state of the physical system
A and sends the system to Bob who is free to choose the measurement he
makes on A.  Bob is aware of the possible preparations available to
Alice, but of course is uncertain of which one in particular was
used.  We will identify three different types of information in
physical systems (preparation, missing and accessible information)
and briefly contrast the quantum information content with the
information of an equivalent classically described system.

\subsection{Preparation Information}

Let us suppose for a moment that Alice and Bob use only pure states of 
a quantum system to communicate (mixed states will be discussed
later).  To be more specific, suppose they are using a two-level
system and the $k^{\rm th}$ message state is represented by 
\be
|\psi_k\ran = a_k|0\ran + b_k|1\ran
\end{equation}
where $a_k$ and $b_k$ are complex amplitudes.  We can choose the
overall phase so that $a_k\in \R$ in which case $|\psi_k\ran$ is
specified by two real numbers ($a_k$ and the phase of $b_k$).  How
much information in a real number?  An infinite amount,
unfortunately.  We overcome this problem, as frequently done in
statistical physics, by coarse-graining --- in this case by giving the 
real number to a fixed number of decimal places.  In practice, Alice's
preparation equipment will in any case have some finite resolution so
{\em she} can only distinguish between a finite number ${\cal N}$ of
signals. So to prepare a signal state of this qubit Alice must specify
$\log {\cal N}$ bits of information --- and if she wants Bob to have
complete knowledge of which among the signal states was prepared, she
must send him exactly $\log {\cal N}$ classical bits.

If Alice and Bob are using an $N$-level quantum system 
the state of the system is represented by a ray in a projective
Hilbert space\footnote{$N$-dimensional projective Hilbert space is the
set of equivalence classes of elements of some $N$-dimensional Hilbert
space, where two vectors are regarded as equivalent if they are
complex multiples of each other.} ${\cal H}$.  A natural metric on
${\cal H}$, given by Wootters \cite{Wootters:1981}, is
\be\label{bryan}
d(\psi_0,\psi_1) = \cos^{-1}|\lan\psi_0|\psi_1\ran|
\end{equation}
where the $|\psi_i\ran$ are normalised representatives of their
equivalence classes.  With this metric, the compact space ${\cal H}$
can be partitioned into a finite number ${\cal N}$ of cells each with
volume less than a fixed resolution $d{\cal V}$, and these cells can
be indexed by integers $j=1,2,\ldots,{\cal N}$.  By making $d{\cal V}$
small enough,
any ray in cell $j$ can be arbitrarily well approximated by a fixed
pure state $|\psi_j\ran$ which is inside cell $j$.  Thus our signal
states are the finite set $\{|\psi_1\ran,\ldots,|\psi_{\cal N}\ran\}$, 
and they occur with some pre-agreed probabilities $p_1,\ldots, p_{\cal
N}$.

At this point we can compare the quantum realisation of signal states
with an equivalent classical situation.  In the classical case, the
system will be described by canonical co-ordinates $P_1,\ldots, P_F,
Q_1,\ldots, Q_F$ in a phase space.  We can select some compact subset
$W$ to represent practically realisable states, and partition $W$ into 
a finite number ${\cal N}$ of cells of phase space volume smaller than
$d{\cal V}_c$.  A typical state of the system corresponds to a point
in the phase space, but on this level of description we associate any
state in cell $j$ with some fixed point in that cell.

By ``preparation information'' we mean the amount of information, once 
we know the details of the coarse-graining discussed above, required
to unambiguously specify one of the cells.  From the discussion of
information theory presented in Chapter 1, we know that the classical
information required, on average, to describe a preparation is
\be
H(p) = - \sum p(j)\log p(j)
\end{equation}
which is bounded above by $\log {\cal N}$.  Thus by making our
resolution volume ($d{\cal V}$ or $d{\cal V}_c$) small enough, we can
make the preparation information of the system as large as we want.

How small can we make it?

\paragraph{Von Neumann Entropy and Missing Information}

In classical statistical mechanics, there is a state function which is
identified as ``missing information''.  Given some incomplete
information about the state of a complicated system (perhaps we know
its temperature $T$, pressure $P$ and volume $V$), the {\em
thermodynamic entropy} is
defined\footnote{There are many ways of approaching entropy;
see \cite{Lieb:2000} and \cite{Wehrl:1978}.}, up to an additive
constant, to be 
\be\label{apples}
S(T,V,P) = k \log W(T,V,P)
\end{equation}
where $W$ is the statistical weight of states which have the prescribed
values of $T$,$V$ and $P$; this is the Boltzmann definition of
entropy.  The idea is very similar to that expressed 
in Chapter~1: the function $S(T,V,P)$ is, loosely speaking, the number
of extra bits of information required --- once we know $T$,$V$ and $P$
--- in order to determine the {\em exact} state of the system.
According to the Bayesian view, these macroscopic variables are
background information which allow us to assign a prior probability
distribution; our communication setup is completely equivalent to
this except for the fact that we prescribe the probability
distribution ourselves.  So it is reasonable to calculate a function
similar to Eqn~\ref{apples} for our probability distribution and call
it ``missing information''.

Consider an ensemble of $\nu \gg 1$ classical systems as described
above.  Since $\nu$ is large, the number $\nu_j$ of systems in this
ensemble which occupy cell $j$ is approximately $p(j)\nu$.  The
statistical weight is then the number of different ensembles of $\nu$
systems which have the same number of systems in each cell
\cite{Mandl:1971}:
\be
\frac{\nu !}{\nu_1 ! \ldots \nu_{\cal N}!}.
\end{equation}
The information missing towards a ``microstate'' description of the
classical state is the average information required to describe the
ensemble: 
\begin{eqnarray}
S(p) &=& \frac{1}{\nu} k \log \frac{\nu !}{\nu_1 ! \ldots \nu_{\cal
N}!} \nonumber\\ 
&=& - k \sum p(j)\log p(j)
\end{eqnarray}
where we have used Stirling's approximation.  Thus, up to a factor
which expresses entropy in thermodynamic units, the information
missing towards a microstate description of a classical system is
equal to the preparation information.

The {\em von Neumann entropy} of a quantum state $\rho$ is defined to
be
\be
S(\rho) = -\Tr{} (\rho\log\rho).
\end{equation}
While this definition shares many properties with classical
thermodynamic entropy, the field of quantum information theory exists
largely because of the differences between these functions.  However,
there are still strong reasons for associating von Neumann entropy
with missing information.  Firstly, if the ensemble used for
communication is the eigenensemble of $\rho$ (i.e. the eigenvectors
$|\phi_i\ran$ of $\rho$ appearing with probabilities equal to the
eigenvalues $\lambda_i$) then an orthogonal measurement in the
eigenbasis of $\rho$ will tell us exactly which signal was sent.
Secondly, if $S(\rho)>0$ then any measurement (orthogonal or a POVM)
whose outcome probabilities $q_k = \Tr{}(\rho E_k)$ yield {\em less}
information than $S(\rho)$ cannot leave the system in a pure quantum
state \cite{Caves:1996}.  Intuitively, we start off missing $S(\rho)$ bits
of information to a maximal description of the system, and discovering
less information than this is insufficient to place the system in a
pure state.

Some properties of the von Neumann entropy are \cite{Preskill:1998},
\cite{Wehrl:1978}:
\begin{enumerate}
\item Pure states are the unique zeros of the entropy.  The unique
maxima are those states proportional to the unit matrix, and the
maximum value is $\log D$ where $D$ is the dimension of the Hilbert
space.
\item If $U$ is a unitary transformation, then $S(\rho)=S(U\rho
U^\dagger)$.
\item $S$ is {\em concave} i.e. if $\lambda_1,\ldots,\lambda_n$ are
positive numbers whose sum is 1, and $\rho_1,\ldots,\rho_n$ are
density operators, then
\be
S(\sum_{i=1}^n \lambda_i\rho_i) \geq \sum_{i=1}^n \lambda_i S(\rho_i).
\end{equation}
Intuitively, this means that our missing knowledge must increase if we 
throw in additional uncertainties, represented by the $\lambda_i$'s.
\item $S$ is {\em additive}, so that if $\rho_A$ is a state of system
$A$ and $\rho_B$ a state of system $B$, then $S(\rho_A\otimes\rho_B) =
S(\rho_A) + S(\rho_B)$.  If $\rho_{AB}$ is some (possibly entangled)
state of the systems, and $\rho_A$ and $\rho_B$ are the reduced
density matrices of the two subsystems, then 
\be
S(\rho_{AB}) \leq S(\rho_A) + S(\rho_B).
\end{equation}
This property is known as {\em subadditivity}, and means that there
can be more predictability in the whole than in the sum of the parts.
\item Von Neumann entropy is {\em strongly subadditive}, which means
that for a tripartite system in the state $\rho_{ABC}$,
\be
S(\rho_{ABC}) + S(\rho_B) \leq S(\rho_{AB}) + S(\rho_{BC}).
\end{equation}
This technical property, which is difficult to prove (see
\cite{Wehrl:1978}), reduces to subadditivity in the case where system
$B$ is one-dimensional.
\end{enumerate}

So what is the relationship between preparation information and
missing information in the quantum mechanical version of the
communication system?  Well, given the background information
discussed above (i.e.\ knowledge of the coarse-grained cells and their 
{\it a priori} probabilities)
but no preparation information (we don't know which coarse-grained
cell was chosen), the density operator
we assign to the system\footnote{Note that this density operator
encapsulates the best possible predictions we can make of any
measurement of the system --- this is why we employ it.} is $\rho =
\sum p(j)|\psi_j\ran\lan\psi_j|$;
so the missing information is $S(\rho)$.  Now it can be shown 
\cite{Wehrl:1978} that
\be\label{chips}
S(\rho) = -\sum_{k=1}^D \lambda_k\log\lambda_k \leq -\sum_{j=1}^{\cal
N} p(j)\log p(j) = I_P.
\end{equation}
In this expression, the $\lambda_k$ are eigenvalues of $\rho$, $D$ is
the dimension of $\rho$, and $I_P$ is the preparation information for
the ensemble.  Equality holds if and only if the ensemble is the
eigenensemble of $\rho$, that is, if all the signal states are
orthogonal.

The conclusion is that preparation information $I_P$ and missing
information $S$ are equal in classical physics, but in quantum physics 
$S\leq I_P$, and the preparation information can be made arbitrarily
large. One question still remains: How much information can be
extracted from the quantum state that Alice has prepared?

\subsection{Accessible Information}

Alice and Bob share the ``background'' information concerning the
partitioning of the projective Hilbert space and the probability
distribution of signals $p(j)$. Bob, knowing this information and no
more, will ascribe a mixed state $\rho = \sum_i p(i)|\psi_i\ran\lan
\psi_i|$to the system he receives.  His duty in the communication
scenario is to determine as well as he can which pure state he
received.  Of course, if he has very precise measuring equipment which 
enables him to perform von Neumann (orthogonal) measurements, he can
say immediately after the measurement that the system is in a pure
state --- if the measurement outcome was `3', he knows the system is
now exactly in state $|3\ran$.  But this information is not
terribly useful for communication.

Unfortunately for Bob, exact recognition of unknown quantum states is
not possible.  This is related to the fact that there is no universal
quantum cloning machine \cite{Wootters:1982}; that
is, there doesn't exist a superoperator 
$\$$ which acts on an arbitrary state $|\psi\ran$ and an ancilla
in standard state $|s\ran$ as
\be
\$:|\psi\ran|s\ran \longrightarrow |\psi\ran|\psi\ran.
\end{equation}
Cloning violates either the unitarity or the linearity of quantum
mechanics (see Section~\ref{states}), and could be used for
instantaneous signalling \cite{Dieks:1982} or discrimination between
non-orthogonal quantum states.

The only option open to Bob, once he receives the quantum system from
Alice, is to perform a POVM on the system.  Suppose Bob chooses to
perform the POVM $\{E_b\}$; the probability of outcome $b$ is $\Tr{}
(\rho E_b)$.  His task is to infer which state was sent, and 
so he uses Bayes' Theorem (Eqn~\ref{bayesthm}):
\be
p({\rm P}_j|{\rm O}_b) = \frac{p({\rm O}_b|{\rm P}_j)p({\rm P}_j)}
{p({\rm O}_b)}
\end{equation}
where ${\rm O}_b$ represents the event ``The outcome is $b$'' and
${\rm P}_j$ the event ``The preparation was in cell $j$''.  We can
calculate all the quantities on the right side of this expression:
$p({\rm O}_b)=\Tr{} \rho E_b$, $p({\rm O}_b|{\rm P}_j)=\Tr{} |\psi_j\ran
\lan\psi_j| E_b = \lan\psi_j|E_b|\psi_j\ran$ and $p({\rm P}_j)=p(j)$.
Thus we can calculate the post-measurement entropy
\be
H({\rm P}|{\rm O}) = -\sum_b p({\rm O}_b) \sum_j p({\rm P}_j|{\rm
O}_b) \log p({\rm P}_j|{\rm O}_b).
\end{equation}
We can now define the information gain due to the measurement
$\{E_b\}$ as
\be
I(\{E_b\}) = I_P - H({\rm P}|{\rm O})
\end{equation}
and the {\em accessible information} \cite{Schumacher:1990} is defined 
as
\be\label{syrup}
J = \max_{\{E_b\}} I(\{E_b\}).
\end{equation}

The accessible information is the crucial characteristic of a physical 
communication system.  In classical physics, states in different phase 
space cells are perfectly distinguishable as long as Bob's measuring
apparatus has fine enough resolution; hence the accessible information 
is in principle equal to the preparation information and the missing
information.  What can be said in the quantum situation?

The accessible information is unfortunately very difficult to
calculate, as is the measurement which realises the maximum in
Eqn~\ref{syrup}.  Davies \cite{Davies:1978} has shown that the optimal 
measurement consists of unnormalised projectors $E_1, \ldots, E_N$,
where the number of such projectors is bounded by $D\leq N\leq D^2$,
where $D$ is the Hilbert space dimension\footnote{Davies also showed
how to calculate or constrain the measurement if the ensemble exhibits
symmetry.}.  Several upper and lower bounds have been derived
\cite{Fuchs:1995}, some of which yield measurements that attain the
bound. The most well-known bound is the {\em Holevo Bound}, first
proved by Holevo in 1973 \cite{Holevo:1973}.  The bound is
\be
J\leq S(\rho);
\end{equation}
accessible information is always less than or equal to missing
information.  This is in fact the tightest bound which depends only on
the density operator $\rho$ (and not on the specific ensemble), since
the eigenensemble, measured in the eigenbasis, realises this bound,
$J= S(\rho)$.  The bound is not tight, and in many situations there is
a significant difference between $J$ and $S(\rho)$ \cite{Peres:1991},
\cite{Fuchs:1994}. 
The Holevo Bound will be proved below --- in more generality, when we
discuss mixed state messages --- and a physical interpretation will be 
discussed later in Section~\ref{notagain}.

\subsection{Generalisation to Mixed States}\label{egypt}

We have found that the accessible information $J$ and the preparation
information $I_P$ for an ensemble of pure states obey
\be
0 \leq J\leq S(\rho)\leq I_P.
\end{equation}
What can be said in the situation when the states comprising the
ensemble are mixed states?

Consider the ensemble of states $\rho_i$ occurring with probabilities
$p(i)$, and suppose that $\rho_i = \sum_k
\lambda_k^{(i)}|\phi_k^{(i)}\ran \lan\phi_k^{(i)}|$ is the spectral
decomposition of each signal state.  We could then substitute the {\em 
pure state} ensemble $|\phi_k^{(i)}\ran$ occurring with probabilities
$p(i)\lambda_k^{(i)}$; then
\begin{eqnarray}
S(\rho) &=& S\left(\sum_i p(i)\rho_i\right) = S\left(\sum_{i,k}
p(i)\lambda_k^{(i)}
|\phi_k^{(i)}\ran \lan\phi_k^{(i)}|\right) \nonumber \\
&\leq & -\sum_{i,k} p(i)\lambda_k^{(i)} \log p(i)\lambda_k^{(i)}
\label{tomatopaste}\\
&=& -\sum_i p(i)\log p(i) - \sum_i p(i)\sum_k \lambda_k^{(i)}
\log\lambda_k^{(i)} = I_P + \sum_i p(i)S(\rho_i)
\end{eqnarray}
where Eqn~\ref{tomatopaste} follows from the pure state result,
Eqn~\ref{chips}.  We thus conclude that the preparation information is 
bounded below by
\be
I_P \geq S(\rho) - \sum_i p(i)S(\rho_i) \equiv \chi({\cal E})
\end{equation}
where ${\cal E} = \{\rho_i,p(i)\}$ denotes the ensemble of signal
states.  The function $\chi({\cal E})$ is known by various names,
including {\em Holevo information} and {\em entropy defect}.  It
shares many properties with von Neumann entropy, and reduces to $S$ in 
the case of a pure state ensemble.  We can compare the Holevo
information with the definition of mutual information,
Eqn~\ref{pizza},
\begin{equation*}
I(X:Y) = H(X) - H(X|Y),
\end{equation*}
and we see that Holevo information quantifies the reduction in
``missing information'' on learning which state (among the $\rho_i$)
was received.

\paragraph{The Holevo Bound}
For pure states, the accessible information is bounded above by the
von Neumann entropy; it turns out that, as with preparation
information, the generalisation to mixed states is achieved by
substituting the Holevo information for $S$.
This more general Holevo bound can be proved fairly easily once the
property of strong subadditivity of $S$ has been shown
\cite{Preskill:1998}.  We will assume this property.

Alice is going to prepare a quantum state of system $Q$ drawn from the
ensemble ${\cal E} = \{\rho_i,p(i)\}$.  These states are messy ---
they may be nonorthogonal or mixed --- but in general Alice will
keep a classical record of her preparations, perhaps by writing in her 
notebook.  We will call the notebook quantum system $X$, and assume
that she writes one of a set of pure orthogonal states in her
notebook.  Thus for each  
value of $i$, there is a pure state $|i\ran$ of the notebook $X$;  to
send message $i$, Alice prepares $|i\ran\lan i|\otimes \rho_i$ with
probability $p(i)$, and {\em these} state are orthogonal and perfectly 
distinguishable.

Bob receives the system $Q$ from Alice and performs a POVM $\{E_b\}$
on it.  Bob finds POVMs distasteful, so he decides to rather fill in
all the steps of the measurement.  He appends a system $W$ onto $Q$
and performs an orthogonal measurement on $QW$, represented by the
unitary operators $\{F_b\}$ which are mutually orthogonal.  Lastly, in
order to preserve a record 
of the measurement, Bob has his notebook $Y$ which contains as many
orthogonal states as measurement outcomes $b$.  His measurement will
project out an orthogonal state of $QW$ which he will transcribe into
an orthogonal state of his notebook\footnote{Transcribing is another
way of saying cloning, and quantum mechanics forbids universal
cloning \cite{Wootters:1982}.  However, cloning one of a known set of
orthogonal states is allowed.}.

The initial state of the entire setup is
\be
\rho_{XQWY} = \sum_i p(i)|i_X\ran\lan i_X| \otimes\rho_i \otimes |0_W\ran
\lan 0_W| \otimes |0_Y\ran \lan 0_Y|.
\end{equation}
When Bob receives system $Q$ from Alice, he acts on the combined
system $QWY$ with the unitary operation
\be
U_{QWY}:|\phi\ran_Q\otimes|0_W\ran \otimes|0_Y\ran \longrightarrow 
\sum_b F_b \big(|\phi_Q\ran\otimes|0_W\ran\big)\otimes |b_Y\ran
\end{equation}
for any pure state $|\phi_Q\ran$ in $Q$, where the $|b\ran_Y$ are
mutually orthogonal. The state of the combined system after Bob
performs this transformation is
\be
\rho'_{XQWY} = \sum_{i,b,b'} p(i)|i_X\ran\lan i_X| \otimes\ F_b
\big[\rho_i \otimes |0_W\ran \lan 0_W|\big]F_{b'} \otimes |b_Y\ran
\lan b'_Y|.
\end{equation}

We will be using strong subadditivity in the form
\be
S(\rho'_{XQWY}) + S(\rho'_Y) \leq S(\rho'_{XY}) +
S(\rho'_{QWY}). \label{chicken} 
\end{equation}
We note first that due to unitary invariance $S(\rho'_{XQWY})
= S(\rho_{XQWY})$, and these are equal to $S(\rho_{XQ})$ since the
systems $W$ and $Y$ are in pure states (with zero entropy).  Thus
\begin{eqnarray}
S(\rho'_{XQWY}) &=& S\left(\sum_i p(i)|i_X\ran\lan i_X| \otimes\rho_i
\right) \nonumber\\
&=& -\sum_i \Tr{} \big[p(i)\rho_i \log p(i)\rho_i\big] \label{pudding}\\
&=& - \sum_i p(i)\log p(i) - \sum_i p(i) \rho_i\log \rho_i \nonumber\\
&=& H(X) + \sum_i p(i)S(\rho_i) \label{salmonmousse}
\end{eqnarray}
where \ref{pudding} follows from the fact that $\rho_{XQ}$ is block
diagonal in the index $i$.  To calculate $\rho'_{XY}$, we note that
\begin{eqnarray}
\Tr{}\left(F_b [\rho_i\otimes |0_W\ran \lan 0_W|] F_{b'}\right) &=&
\Tr{}\left(F_{b'}F_b \rho_i\otimes |0_W\ran \lan 0_W|\right) \nonumber\\
&=& \delta_{bb'}\; \Tr{} \left(F_b \rho_i\otimes |0_W\ran \lan 0_W|\right)
= \delta_{bb'}\; p(O_b|P_i)
\end{eqnarray}
where the second equality follows from the orthogonality of the
measurement. Thus we have that
\begin{eqnarray}
\rho'_{XY} = \sum_{i,b} p(i)p(O_b|P_i)\; |i_x\ran\lan i_X|\otimes
|b_y\ran\lan b_Y| \nonumber\\
\Longrightarrow S(\rho'_{XY}) = -\sum_{i,b} p(i,b)\log p(i,b) =
H(X,Y), \label{hotdogs}
\end{eqnarray}
and by taking another partial trace (over $X$) we find
\begin{eqnarray}
\rho'_Y = \Tr{X} \sum_{i,b} p(i,b) |i_x\ran\lan i_X|\otimes
|b_y\ran\lan b_Y| = \sum_b p(b) |b_Y\ran\lan b_Y| \nonumber\\
\Longrightarrow S(\rho'_Y) = -\sum_b p(b)\log p(b) = H(Y). \label{chutney}
\end{eqnarray}
The transformation $\rho_{QWY}\longrightarrow \rho'_{QWY}$ is unitary, 
so
\begin{eqnarray}
S(\rho'_{QWY}) &=& S(\rho_{QWY}) = S(\rho_{Q}) \nonumber\\
&=& S(\rho) \label{fruitcake}
\end{eqnarray}
where the second equality follows from the purity of the initial
states of $W$ and $Y$, and we have used the previous notation $\rho =
\sum_i p(i)\rho_i$.  Combining Eqns \ref{chicken} through
\ref{fruitcake}, we find 
\be
H(X) + \sum_i p(i)S(\rho_i) + H(Y) \leq H(X,Y) + S(\rho);
\end{equation}
recalling the definition of mutual information, we end up with the
Holevo bound:
\be
I(X:Y) = H(X)+H(Y)-H(X,Y) \leq S(\rho)-\sum_i p(i)S(\rho_i) =
\chi({\cal E}).
\end{equation}

\subsection{Quantum Channel Capacity}

The original, practical problem which motivated this chapter was: How
does the quantum nature of the information carrier affect
communication between Alice and Bob?  We have found that, in contrast
with classical states, information in quantum states is slippery and
often inaccessible.  So why would we want to employ non-orthogonal
states, or even mixed states, in a communication system?

The practical answer is that sometimes we cannot avoid it.   If we send 
photons down an optical fibre, then in order to achieve a high
transmission rate we will need to overlap the photon packets slightly, 
which means we are using nonorthogonal states.   In fact, Fuchs
\cite{Fuchs:1997} has shown that for some noisy channels the rate of
transmission of classical information is maximised by using {\em
non}orthogonal states! And if we hope to
maximise transmission rate, we are going to have to have a deeper 
understanding of how errors are introduced into the photon packets,
which requires us to deal with the mixed states emerging from the
optical fibre --- even if the input states were pure and orthogonal
and so essentially ``classical''.

The question we now turn to is: What is the maximum information
communicated with states drawn from the ensemble $\{\rho_i, p_i\}$?
The answer to this question was given by Hausladen {\it et al}
\cite{Hausladen:1996} for ensembles of pure states and independently
by Holevo \cite{Holevo:1998} and Schumacher and Westmoreland
\cite{Schumacher:1997} for mixed states.  The Holevo Bound can be {\em
attained} asymptotically, if we allow collective measurements by the
receiver.  We will present the main concepts from the proof of the
pure state result, without exercising complete rigour.

The essential idea is that of a {\em typical subspace}.  A sequence of 
$n$ signals from the source is represented by a vector
$|\phi_{i_1}\ran \ldots |\phi_{i_n}\ran$ in the product Hilbert space
$H^n$, and the ensemble of such signals is represented by the density
operator
\begin{eqnarray}
\rho^{(n)} &=& \sum_{i_1,\ldots,i_n}p_{i_1}\ldots p_{i_n}|\phi_{i_1}\ran
\ldots |\phi_{i_n}\ran \lan\phi_{i_n}|\ldots\lan\phi_{i_1}|    \\
&=& \rho\otimes\rho\otimes\ldots\otimes\rho
\end{eqnarray}
where $\rho$ is the single system density matrix.
Then for given $\epsilon,\delta >0$, and for $n$ large enough, 
there exists a typical subspace $\Lambda$ of $H^n$ such that
\cite{Preskill:1998}
\begin{enumerate}
\item Both $\Lambda$ and $\Lambda^\bot$ are spanned by eigenstates of
$\rho^{(n)}$.
\item Almost all of the weight of the ensemble lies in $\Lambda$, in
the sense that
\be
\Tr{} \Pi_\Lambda\rho^{(n)} > 1-\epsilon \mbox{ and } \Tr{}
\Pi_{\Lambda^\bot}\rho^{(n)} < \epsilon 
\end{equation}
(where $\Pi_A$ is used to denote the projection onto the subspace $A$).
\item The eigenvalues $\lambda_l$ of $\rho^{(n)}$ within $\Lambda$
satisfy
\be
2^{-n[S(\rho)+\delta]} < \lambda_l < 2^{-n[S(\rho)-\delta]}.
\end{equation}
\item The number of dimensions of $\Lambda$ is bounded between
\be
(1-\epsilon)2^{n[S(\rho)-\delta]}\leq \dim\Lambda \leq
2^{n[S(\rho)+\delta]}.
\end{equation}
\end{enumerate}
To see how this typical subspace is constructed, we suppose that the
signals sent are indeed eigenstates of $\rho$.  Then the signals are
orthogonal and essentially classical, governed by the probability
distribution $\mu_i$ given by the eigenvalues; for the properties
listed above it makes no difference if the signal states are
indeed eigenstates or some other (nonorthogonal) states.  The
eigenvalues of $\rho^{(n)}$ will then be $\mu_{\bf i}=\mu_{i_1}\ldots
\mu_{i_n}$ and by the weak law
of large numbers (mentioned in Section~\ref{largenumbers}) the set of
these eigenvalues satisfies
\be
P\left(\frac{1}{n}\left| \log\mu_{\bf i}-S(\rho^{(n)})\right| > \delta 
\right) < \epsilon.
\end{equation}
Let $A$ be the set of eigenvalues satifisfying $\frac{1}{n}\left|
\log\mu_{\bf i}-S(\rho^{(n)})\right| \leq \delta$, and define
$\Lambda$ to be the eigenvectors corresponding to these eigenvalues;
then a moment's thought reveals $\Lambda$ to have the properties
listed above.

\begin{figure}
\begin{center}
\epsfig{file=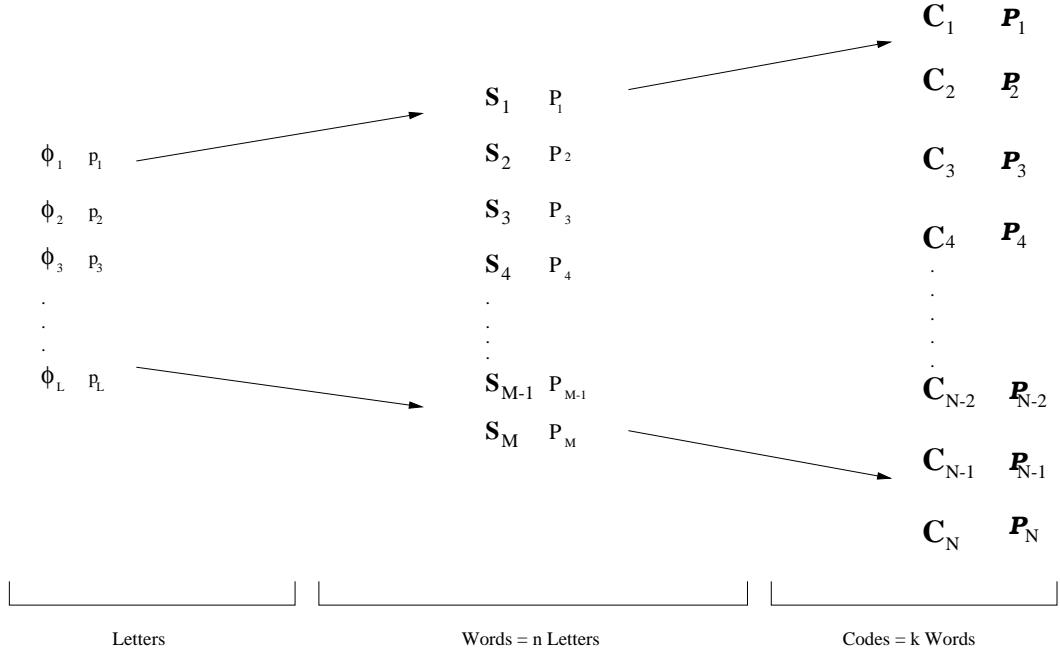, width=14cm}
\caption{A schematic representation of random codes and their
probabilities}\label{randomcode} 
\end{center}
\end{figure}

The technique of the proof is very similar to that for Shannon's Noisy 
Coding theorem: we use the idea of random coding to symmetrise the
calculation of error probability.  The technique of random coding is
illustrated in Figure~\ref{randomcode}.
We begin with an ensemble of letter states ($\phi_1,\ldots \phi_L$ in
the Figure) and generate the ensemble of all possible $n$-letter
words --- the probability of a particular word just being the product
of its letter probabilities.  In the same way we construct codes each
containing $k$ words, with $k$ left unspecified for the moment.

The important feature of this heirarchy is that, just as there was a
typical subspace of words, so too is there a typical subspace of codes 
which has the following properties: (a) the overall letter frequencies
of each code are close to those of the letter ensemble, (b) the
words in each typical code are almost all from the typical subspace of 
words and (c) the set of atypical codes has negligible probability for 
large enough $n$ and $k$.  This means that in calculating error
probability averaged over {\em all} codes, we can calculate the error only
for these typical codes and include an arbitrarily small
$\epsilon$-error for the contribution from non-typical codes.

To calculate this error, note that the expected overlap of any two code words
$|u\ran = |\phi_{x_1}\ran\ldots |\phi_{x_n}\ran$ and $|v\ran =
|\phi_{y_1}\ran\ldots |\phi_{y_n}\ran$ is
\begin{eqnarray}
E\; |\lan u|v\ran|^2 &=&\sum_{x_1,\ldots,x_n}\sum_{y_1,\ldots,y_n}
p_{x_1}p_{y_1}\ldots 
p_{x_n}p_{y_n} |\lan\phi_{x_1}|\phi_{y_1}\ran|^2 \ldots
|\lan\phi_{x_n}|\phi_{y_n}\ran|^2 \nonumber\\
&=& \Tr{} (\rho^{(n)})^2.
\end{eqnarray}
By the above typicality argument, we need only consider the codes
which consist of typical codewords when calculating averages over
codes.  So the expected overlap between code words in a typical code
is\footnote{Heuristically if we choose two unit vectors at random from 
a vector space of dimension $D$, their overlap will be $1/D$ on average.}
\begin{eqnarray}
E_\Lambda \; |\lan u|v\ran|^2 &=& \Tr{} \Lambda(\rho^{(n)})^2 \\
&<& 2^{n[S(\rho)+\delta]}\left(2^{-n[S(\rho)-\delta]}\right)^2 \\
&=& 2^{-n[S(\rho)-3\delta]}
\end{eqnarray}
where we have used the bounds on the dimension of the typical subspace 
and the eigenvalues contained in it.
So if we choose $k=nS(\rho)-\delta'$ then for any fixed codeword
$|S_j\ran$ the overlap of $|S_j\ran$ with all other codewords,
averaged over {\em all} codes, is
\be \label{booboo}
E\; \sum_{i\neq j} |\lan S_i|S_j\ran|^2 \leq 2^{n[S(\rho)-\delta']}
2^{-n [S(\rho)-3\delta]} +\epsilon = 2^{-n[\delta' -3\delta]} +\epsilon;
\end{equation}
the $\epsilon$ is included for the contribution of atypical codes.
For fixed $\delta'$ we can make $\epsilon$ and $\delta$ as small as we 
choose by making $n$ large; so the expected overlap can be made
arbitrarily small.

We now invoke some arguments encountered previously in connection with 
Shannon's Noisy Coding theorem.  Since this result (Eqn~\ref{booboo})
holds on average, 
there is at least one code which has small average overlap between
code words; and by throwing away at most half the code words (as
discussed at Eqn~\ref{jellybabes}) we can ensure that {\em every}
codeword has less than $\epsilon$ overlap with all other codewords.
And of course, since almost all the codes are typical, we can choose
the code so that the letter frequencies of the code are close to those
of the original ensemble.

The resulting code will consist of $nS(\rho)$ codewords which are
almost equally likely; thus $S(\rho)$ bits are communicated per
quantum system received.  It is interesting to note how this
communication is achieved: as the number of signals becomes large, it
becomes possible to choose words which are almost orthogonal and hence 
highly distinguishable.  However, decoding these message states will
require sophisticated joint measurements by Bob --- and the problem of 
finding the optimal measurement is very difficult.  Hausladen {\it et
al} \cite{Hausladen:1996} employ a specific POVM in their proof which, 
although not optimal, is easier to work with.  They find a rigorous
bound on the average error probability similar to that motivated here.

We now have the result that, given an ensemble of message states
${\cal E} = \{\rho_i,p_i\}$, information can be communicated through a 
channel at the rate
\be
R = S(\sum p_i\rho_i) - \sum p_iS(\rho_i) = \chi({\cal E})
\end{equation}
and no higher.  We can define the {\em quantum channel capacity}
(relative to a fixed alphabet) to be
\be
C_Q = \max_{p_i} \chi({\cal E})
\end{equation}
similar to the definition of channel capacity in Eqn~\ref{joubert}.
This is then the maximum rate at which information can be communicated 
using the given physical resources.

\section{Distinguishability}\label{states}

In considering the information contained in a quantum state, we are
dealing closely with issues of distinguishability.  For example,
suppose we use nonorthogonal states in a communication channel, and
suppose there was a POVM which could distinguish {\em with certainty}
between these states; then the information capacity of the channel
could be calculated using the classical theory!  In this connection
several questions arise: Can we quantitatively measure the
``distinguishability'' of a set of states?  Are there physical
processes which increase distinguishability?  Can we find the optimal
measurement for distinguishing between states?  And how can we use
imperfect distinguishability to our advantage?

Before looking at these issues, however, there is an important caveat
to this relationship between 
accessible information and distinguishability.  Suppose Alice and Bob
communicate using two pure state signals, $|\psi_0\ran$ and
$|\psi_1\ran$ with probabilities $p_0$ and $p_1$.  It can be shown
that the von Neumann entropy of this ensemble --- and hence the
ability of these states to convey information --- is a monotonically
decreasing function of the overlap $|\lan\psi_0|\psi_1\ran|$.  This
makes sense: if we make the signal states less distinguishable they
should convey less information.  Intuitively, this should be a
universal property: for any ensemble in any number of dimensions, if
we make all the members of the ensemble {\em more parallel} we should
decrease the von Neumann entropy.  But Jozsa and Schlienz
\cite{Jozsa:2000}, in the course of investigating compression of
quantum information (see the next chapter), have found that this is
generically not true in more than two dimensions.  Specifically, for
almost any ensemble ${\cal E} = \{p_i, |\psi_i\ran\}$ in three or more
dimensions, there is an ensemble ${\cal E}' = \{p_i, |\psi_i'\ran\}$
such that
\begin{enumerate}
\item All the pairwise overlaps of states in ${\cal E}'$ are not smaller
than those in ${\cal E}$, i.e. for all $i$ and $j$
\be
|\lan\psi_i|\psi_j\ran| \leq |\lan\psi_i'|\psi_j'\ran|;
\end{equation}
\item $S({\cal E}') > S({\cal E})$.
\end{enumerate}
The relationship between distinguishability and information capacity
is therefore not entirely straightforward; and in particular, the
distinguishability of a set of states is a global property of the set, 
and can't be simply reduced to pairwise distinguishability of the states.

Before even discussing measures of distinguishability, we discuss one
possible way of increasing distinguishability.  Suppose Alice has sent 
me a photon which I know is in a pure state of {\em either} horizontal 
or diagonal polarisation, each with probability 1/2.  I decide to make
$2N$ copies of the state and pass $N$ through a horizontally oriented
calcite crystal\footnote{A calcite crystal has different refractive
indices along two of its symmetry axes.  So when photons are incident
on the crystal, they are refracted different amounts depending on their
polarisation --- hence the two beams emerging from the crystal have orthogonal
polarisations.} and $N$ through a diagonally oriented crystal.  I can
deduce the polarisation state of Alice's original photon by observing
which orientation of the crystal has all $N$ photons emerging in one
beam.  But something is wrong: Alice will thus have communicated one
bit of information to me, despite the fact that the Holevo bound
dictates that the most information she can transmit is $S=0.60$.

The problem lies in the assumption of copying.  Wootters and Zurek
\cite{Wootters:1982} considered such a ``universal cloning machine'',
that is, a machine which acts on a system in state $|\psi\ran$, an 
ancilla in a standard state $|0\ran$ and an environment in state
$|E\ran$ as
\be
|\psi\ran\otimes|0\ran\otimes|E\ran \longrightarrow
|\psi\ran\otimes|\psi\ran\otimes|E'\ran
\end{equation}
(where $|E'\ran$ is any other state of the environment).  They showed
that if the machine is expected to clone just {\em two} specific
nonorthogonal states $|\psi_1\ran$ and $|\psi_2\ran$ faithfully then
it violates the unitarity of quantum mechanics, since
\begin{eqnarray}
\lan\psi_1|\psi_2\ran &=& \bigg(\lan\psi_1|\otimes\lan 0|\otimes\lan
E|\bigg) \;
\bigg(|\psi_2\ran\otimes|0\ran\otimes|E\ran\bigg) \\
&=& \bigg(\lan\psi_1|\otimes\lan \psi_1|\otimes\lan E'|\bigg) \;
\bigg(|\psi_2\ran\otimes|\psi_2\ran\otimes|E'\ran\bigg) =
\lan\psi_1|\psi_2\ran^2
\end{eqnarray}
which can only be satisfied if $\lan\psi_1|\psi_2\ran = 0$ or 1 (the
states are orthogonal or identical).  The
first inequality here follows from the normalisation of the extra
states, and the second from the unitarity of the process.
And if we expect the machine to copy more than two 
nonorthogonal states then it must violate the superposition
principle, since if $|\psi_1\ran$ and $|\psi_2\ran$ can both be
copied then their superposition is ``copied'' as
\begin{eqnarray}
\big(|\psi_1\ran+|\psi_2\ran\big)\otimes|0\ran\otimes|E\ran&\longrightarrow&
\big(|\psi_1\ran\otimes|\psi_1\ran+|\psi_2\ran\otimes|\psi_2\ran\big)
\otimes|E'\ran \\
&\neq & \big(|\psi_1\ran+|\psi_2\ran\big) \otimes
\big(|\psi_1\ran+|\psi_2\ran\big) \otimes |E'\ran:
\end{eqnarray}
superpositions can't be copied by a unitary process.

It has also been shown that cloning would allow
superluminal signalling \cite{Dieks:1982}.  A generalisation of this
No-Cloning theorem is the No-Broadcasting theorem \cite{Barnum:1996},
which states that there is no physical process which achieves
\be
\rho_A\otimes\Sigma \longrightarrow \rho_{AB}
\end{equation}
where $\Sigma$ is some standard state and the entangled state
$\rho_{AB}$ satisfies $\Tr{A} \rho_{AB} = \rho_A$ and $\Tr{B}
\rho_{AB} = \rho_A$.
Despite the impossibility of cloning, it has become a fruitful area of 
research.  Bounds on the probability of success of an attempt to clone 
have been studied \cite{Gisin:1998}, and cloning has been used in many
studies of quantum information and computation\footnote{Copious
references to this can be found on the LANL preprint server,
http://xxx.lanl.gov/.}.

\subsection{Wootters' Problem}

Wootters \cite{Wootters:1980} considered the problem of distinguishing 
signals in quantum mechanics.  The scenario he considered (which is
also graphically recounted in \cite{Wheeler:1982}) was the use of
two-state probabilistic information carriers for transmitting
information.  Suppose we know that these information carriers ---
which we can call ``infons'' --- are represented by unit vectors in a
{\em real} two dimensional space: 
for example, they may be photons which are guaranteed to be in a state of 
plane polarisation.  The state $|\psi\ran$ of one of these infons is thus
completely characterised by one real parameter,
\be
|\psi\ran = \cos\theta |0\ran + \sin\theta |1\ran
\end{equation}
where $|0\ran,|1\ran$ is some orthonormal basis for the two dimensional space.
Exactly like in quantum mechanics, $|0\ran$ and $|1\ran$ correspond to 
some maximal test \cite{Peres:1993} on the infon --- in the case of a
plane polarised photon these tests may be for horizontal or vertical
polarisation, using a calcite crystal.  

Where Wootters departs from conventional quantum mechanics, however,
is in the probabilities of each outcome as a function of the state
parameter $\theta$.  If the infon was indeed a photon, then the
probability of obtaining outcome corresponding to $|0\ran$ (say,
horizontal polarisation) would be
\be
p(\theta) = \cos^2\theta.
\end{equation}
Wootters leaves undecided the exact probability function, but instead
asks the question: If Alice sends $N$ infons to Bob, all with exactly
the same value of $\theta$, how does the number of ``distinguishable'' 
values of $\theta$ vary with the probability function $p(\theta)$?
And what optimal function $p^*(\theta)$ gives the most number of
``distinguishable'' states?

The concept of distinguishability used here was one of practical
interest.  Suppose Alice would like to send one of $l$ signals to Bob
by preparing the $N$ infons with one of the parameter values
$\theta_1,\theta_2,\ldots \theta_l$.  If Alice chooses parameter value 
$\theta_k$, then the probability that Bob measures $n$ of them to be
in state $|0\ran$ (and hence $N-n$ in state $|1\ran$) is given by the
binomial distribution,
\be
p(n|\theta_k) = \frac{N!}{n!(N-n)!}[p(\theta_k)]^n[1-p(\theta_k)]^{N-n},
\end{equation}
where the function $p(\theta)$ is as yet unspecified.  Of course, if
the signals are to be distinguished reliably, they must be a {\em
standard deviation} $\sigma$ apart (or some multiple of $\sigma$,
depending on how reliable we require the communication to be).
Communication using infons is thus subject to two competing
requirements: that there be as many signal states as possible (and
thus that they be very close together) and that they be at least a
standard deviation apart.  Clearly the exact form of the probability
function $p(\theta)$ will have a great impact on how many
distinguishable states are available; if $p(\theta)$ is almost uniform 
over $0\leq\theta\leq 1$ then almost all values of $\theta$ will give
the same outcome statistics, and Bob will have no idea what value of
$\theta$ Alice used!

It turns out, unsurprisingly, that {\em which} function maximises the number
of distinguishable states\footnote{The function extremised by Wootters
was in fact the mutual information between $\theta$ and the number $n$
of $|0\ran$ outcomes.  This is closely related to ``number of
distinguishable parameter values''.} depends on $N$.  But the limit of 
these optimal functions $p_N^*(\theta)$ for large $N$ turns out to be
one of the functions
\be
\cos^2\frac{m}{2}(\theta-\theta_0)
\end{equation}
for some positive integer $m$.  The actual probability law for photon
polarisation measurements is of this form, so the universe in fact
{\em does} operate in a way that maximises the amount of information
we (asympotically) obtain from photons!

Can more of quantum mechanics be ``derived'' from such a principle of
extremisation of information, similar to Fermat's principle of least
time?  And, as in the case of light ``finding the quickest route''
between two points as a consequence of its wave nature, is there an
underlying reason why photons maximise the amount of information we
can obtain from them?  In a sense, information is an ideal candidate
for capturing the essence of quantum mechanics because it is a natural 
characterisation of uncertainty --- and quantum mechanics is an
intrinsically uncertain theory.  Unfortunately, even Wootters'
interesting result is not entirely convincing: his extremisation
principle is valid only in a real Hilbert space; the actual quantum
mechanical law does {\em not} maximise information in a two
dimensional complex Hilbert space.

In passing, we note that this work provided motivation for the
``natural metric'' on projective Hilbert space mentioned previously
(Eqn~\ref{bryan}). Suppose we know that a system is in one of two
states, but we don't know which, and let $N$ be the minimum number of
copies of the state we require in order to be able to distinguish
reliably between the two states (reliability is again defined by a
fixed number of standard deviations).  Then Wootters
\cite{Wootters:1980}, \cite{Wootters:1981} defines {\em
statistical distance} to be $1/\sqrt{N}$, and goes on to show that
statistical distance is proportional to the ``actual distance'' given
in Eqn~\ref{bryan}.  This endows the actual distance with a practical
interpretation.

\subsection{Measures of Distinguishability}

The problem of distinguishing between probability distributions has
been well studied, and several measures of distinguishability are
discussed by Fuchs \cite{Fuchs:1995}.  Some of these will be briefly
described below.  These notions of statistical distinguishability can
be easily transplanted to quantum mechanics to refer to quantum
states: all we do is consider the distinguishability of the
probability distributions $p_0(b)=\Tr{}\rho_0E_b$ and $p_1(b)=
\Tr{}\rho_1E_b$ for some POVM $\{E_b\}$, and then extremise the
resulting function over all POVMs.  It is also useful from a practical 
point of view to ask which POVM maximises a given distinguishability
measure.

The distinguishability problem for probability distributions is formulated
as follows.  Our {\em a priori} knowledge is that a process is
governed either by distribution $p_0(b)$ (with probability $\pi_0$) or
by distribution $p_1(b)$ (with probability $\pi_1$).  From our
position of ignorance, we ascribe the probability distribution $p(b) = 
\pi_0p_0(b)+\pi_1p_1(b)$ to the process.  A distinguishability measure 
will be a functional of the functions $p_0$ and $p_1$, and possibly
(but not necessarily) also of $\pi_0$ and $\pi_1$ --- and in order to
be useful this measure should have some instrumental interpretation,
as mutual information could be interpreted as number of bits of
information conveyed reliably.

Fuchs considers five such measures of distinguishability of {\em two}
states\footnote{But recall, as mentioned earlier, that
distinguishability of states in more than two dimensions is not a
simple function of pairwise distinguishability of the constituent
states; little is known of the more general problem.}:
\begin{enumerate}
\item {\bf Probability of error.}  The simplest way of distinguishing
the two distributions is to sample the process once, and guess which
distribution occurred.  Let the possible outcomes of the sampling be
$\{1,2,\ldots, n\}$ and suppose $\delta:\{1,\ldots,n\}\longrightarrow\{0,1\}$
called a {\em decision function}, represents such a guess (that is,
$\delta(3)=1$ means that whenever event 3 occurs we guess that
distribution $p_1$ was the one sampled).  The
probability of error is defined to be
\be
P_e = \max_\delta \pi_0P(\delta=1|0) + \pi_1P(\delta=0|1)
\end{equation}
where $P(\delta=i|j)$ is the probability that the guess is $p_i$ when
the true distribution is $p_j$.  This measure has the advantage that
it is very easy to compute.

\item {\bf Chernoff bound.}  Probability of error is limited by the
fact that it is determined by exactly one sampling, which is not
always the best way of distinguishing.  It would make more sense to
sample the process $N$ times; in effect we will be sampling the
$N$-fold distributions ($p_o^N$ or $p_1^N$) {\em once}, and we can
apply almost the same reasoning as above to calculate the $N$-fold
error probability.  It turns out that for $N\longrightarrow \infty$,
\be\label{layla}
P_e \longrightarrow \lambda^N \mbox{ where } \lambda =
\min_{0\leq\alpha\leq1} \sum_{b=1}^np_0(b)^\alpha p_1(b)^{1-\alpha}.
\end{equation}
We call $\lambda$ the {\em Chernoff bound}, since it is also an upper
bound on the probability of error for any $N$.  While perhaps having
more applicability, the Chernoff bound is very difficult to calculate.

\item {\bf Statistical overlap.}  Since $\lambda$ is hard to
calculate we could consider bounds of the form $\lambda_\alpha$,
similar to Eqn~\ref{layla} which are not maximised over $\alpha$.  Of
these, the most useful is the statistical overlap
\be
{\cal F}(p_0,p_1) = \sum_{b=1}^n \sqrt{p_0(b)}\sqrt{p_1(b)}
\end{equation}
which is also, conveniently, symmetric.  This function has previously
been studied in connection with the Riemannian metric on the
probability simplex; so while it is not practically useful it has
compelling mathematical application.

\item {\bf Kullback-Leibler information.}  The Kullback-Leibler
information was mentioned previously in connection with mutual
information (Eqn~\ref{crow}):
\be
D(p_0||p_1) = \sum_{b=1}^n p_0(x) \log \frac{p_0(x)}{p_1(x)}.
\end{equation}
The interpretation of this measure is slightly more abstract, and may
be called ``Keeping the Expert Honest'' \cite{Fuchs:1995}.  Suppose we 
have an expert weather forecaster who can perfectly predict
probability distributions for tomorrow's weather --- the only problem
is, he enjoys watching people get wet and so is prone to distorting
the truth.  We wish to solve the problem by paying him according to
his forecast, and we choose a particular payment function which is
maximised, on average, if and only if he tells the truth i.e. if the
forecasted probabilities usually coincide with those that occur.  The
Kullback-Leibler information measures his loss, relative to this
expected maximum payment, if he tries to pass off the distribution
$p_1$ for the weather when in fact $p_0$ occurs.

\item {\bf Mutual information.}  We have already encountered mutual
information in a slightly different context, in Chapter 1.  Here we
consider the mutual information between the 
outcome and the mystery distribution index, 0 or 1:
\begin{eqnarray}
J &=& H(p) - \pi_0H(p_0) - \pi_1H(p_1) \\
&=& \pi_0D(p_0||p) + \pi_1D(p_1||p).
\end{eqnarray}
From the last form of the mutual information, we see that $J$ is the
expected loss of the expert when the weather is guaranteed to be
either $p_0$ or $p_1$, and he tries to make us believe it's always
their average, $p$.
\end{enumerate}

In the quantum case, we wish to minimise the first three measures
(since in these cases a measure of 0 corresponds to maximum
distinguishability) and maximise the final two over all possible
POVMs.  There is a very thorough discussion of these extremisations in
\cite{Fuchs:1995}.  The only measures which have a closed form in
terms of quantum states are the probability of error and the
statistical overlap.  The Chernoff bound becomes ambiguous in the
quantum case, since collective measurements become possible; not only
this, but the optimal single-system measurement depends expressly on
the number of samplings allowed, so this measure loses some of its
meaning.

For curiosity we note that the statistical overlap of two quantum
states has a simple closed form expression.  The statistical overlap,
maximised over all POVMs, of $\rho_0$ and $\rho_1$ is
\be\label{noddie}
{\cal F}(\rho_0, \rho_1) = \Tr{} \sqrt{\rho_1^{1/2} \rho_0
\rho_1^{1/2}} \equiv \sqrt{F(\rho_0, \rho_1)},
\end{equation}
where the quantity defined on the right is called the {\em fidelity}
between the two states. The quantum statistical overlap has the
following  useful significance.  Let $|\psi_0\ran$ and $|\psi_1\ran$
be any two purifications (see Chapter 1) of the same dimensions of states 
$\rho_0,\rho_1$.  Then $F(\rho_0,\rho_1)$ is an upper bound to the
overlaps $|\lan\psi_0|\psi_1\ran|$ for all purifications, and moreover
this bound is achievable \cite{Uhlmann:1976}, \cite{Fuchs:1995}.

The final two measures are very difficult to work with, and the best
that can be done in the quantum situation is to find upper and lower
bounds on the distinguishability, of which one useful bound is the
Holevo bound.  Another measure worth mentioning, which is related to 
the probability of error in the quantum case \cite{Fuchs:1996b} 
and has the useful property of being a metric, is the {\em distortion}
\cite{Horodecki:1998}.  The distortion between $\rho_0$ and $\rho_1$
is defined to be
\be
d(\rho_0, \rho_1) = ||\rho_0 - \rho_1||
\end{equation}
where $||\cdot||$ is some appropriate operator norm.  We may choose
the trace norm, $||A|| = \Tr{} |A|$, where $|A| = \sqrt{A^\dagger A}$.

One important requirement for any feature of distinguishability is its 
monotonicity under evolution of the quantum system.  If under the action
of the same superoperator two states become {\em more}
distinguishable, then the distinguishability measure may not have a
useful interpretation --- since the second law of thermodynamics
implies, roughly, that states lose information as time increases.  In
this regard, one may also investigate the change in distinguishability 
upon adding an ancilla, or on tracing out a system, or under unitary
evolution.  We may also ask that a distinguishability measure be
monotonic in another measure --- or if this is not the case, we should 
ask where they disagree.

\subsection{Unambiguous State Discrimination}

All discrimination is not necessarily lost in quantum mechanics,
however.  Peres and Terno \cite{Terno:1999}, \cite{Peres:1998} have
investigated how, given a set of linearly independent states, one
might be able to distinguish between them.  They have found that there
is a method to unambiguously distinguish them, but that this method
has a finite probability of failure unless the states are mutually
orthogonal.  If the method always succeeded, we would have a technique
for cloning nonorthogonal states (find out which state it is, and make 
replicas of it).

Given the set $\{|\psi_1\ran,\ldots, |\psi_n\ran\}$ of linearly
independent states, consider the $(n-1)$-dimensional\footnote{We
assume the vector space is exactly $n$-dimensional.} subspace $V_1 =
\mbox{span}\{|\psi_2\ran,\ldots, |\psi_n\ran\}$.  Let $E_1$ be a
non-normalised projection onto the 1-dimensional subspace $V_1^\bot$.
Then $\Tr{}E_1|\psi_j\ran\lan\psi_j|=0$ for all $j\neq 1$: this
operator unambiguously selects the state $|\psi_1\ran$.  Continuing
this way, we can generate a set of non-normalised projectors $E_1,
\ldots, E_n$ which select their corresponding states.  Surely if we now
define $E_0 = \unit-\sum E_j$, then these operators will form a POVM
--- one which can discriminate unambiguously between all the given states!

Unfortunately, these operators do not necessarily form a POVM.  It is
quite possible that the operator $E_1+\ldots+E_n$ has an eigenvalue
larger than 1 --- which means $E_0$ will not be a positive operator.
But even if it {\em is} a POVM, we have another catch:
the operators $E_i$ are not normalised, so $\Tr{}E_j|
\psi_j\ran\lan\psi_j|\neq 1$: we are not guaranteed that the POVM will 
recognise the state at all. Hence the necessity of the catch-all operator
$E_0$ which yields almost no information about the identity of the
state.  In passing we note that the actual normalisations of the
(unnormalised) projectors $E_i$ will be
given by the requirements that $E_0$ be a positive operator, and
possibly that the probability of the $E_0$ outcome be minimised; more
detail about this is to be found in \cite{Peres:1998}

\section{Quantum Key Distribution}\label{sneeze}

The discrepancy between preparation information and accessible
information is the basis for {\em quantum key distribution} (QKD).  The term 
{\em quantum cryptography} is also used in this connection, but in
fact the quantum aspect is merely a useful protocol within the larger field
of cryptography.  Using the quantum mechanical properties of
information carriers, Alice and Bob can generate a random sequence of
classical bits --- 1's and 0's --- which they both know perfectly and
which they can {\em guarantee} is not known by any third party.

If Alice and Bob share a secret key in this way, they can transmit
information completely securely over a public (insecure) channel.
They do this by using the Vernam Cipher or One-Time Pad, which is the 
only guaranteed unbreakable code known.  If Alice wishes to send the
secret message $101110$ to Bob and they share a secret key
$001011$, then by bitwise-adding the message and the key she arrives
at the encrypted message $100101$ which she sends to Bob.  By
bitwise-adding the secret key to the message, Bob uncovers the
original message.  It can be shown that, as long as Alice and Bob use
the secret key only once, an eavesdropper Eve can obtain no
information about the message --- and even if the key is {\em partially}
known to Eve, there are protocols which Alice and Bob can use to
communicate completely securely \cite{Bennett:1995}.  The problem, as
experienced by nameless spies throughout the Cold War, is how to share 
a secret key with someone when it is difficult or impossible to find a 
trusted courier to carry it.

\begin{figure}
\begin{center}
\epsfig{file=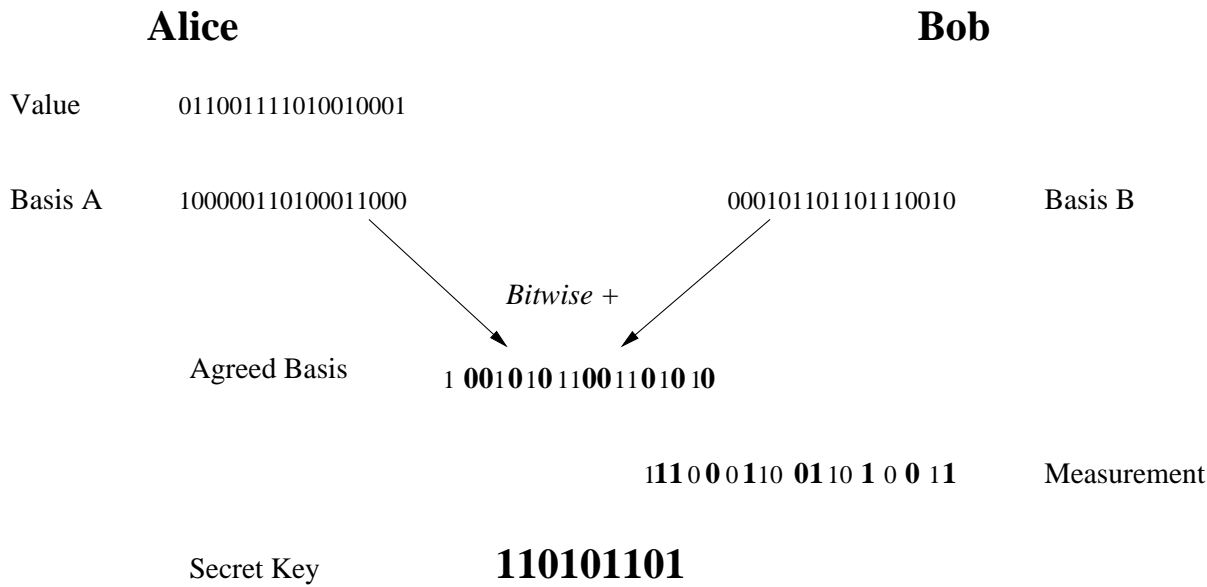, width=\textwidth}
\caption{The private random bit strings required for quantum key
distribution.}\label{buttercup} 
\end{center} 
\end{figure}

QKD solves this problem using properties of
quantum mechanics.  The first step in this protocol\footnote{This
protocol is known as BB84, after its inventors Bennett and Brassard
and its year of publication.} requires Alice to 
write down two random, independent strings of $n$ bits, as shown in
Figure~\ref{buttercup}: a {\em value} string and a {\em basis} string.
She then prepares $n$ qubits according to the bits in her two strings, 
as shown in the table below; in this table, $\{|0\ran,|1\ran\}$ is some
orthonormal basis and $\{|\psi_0\ran,|\psi_1\ran\}$ is any other
basis.\newline
\begin{center}
\begin{tabular}{c|c||c}
Basis A bit & Value bit & State prepared \\ \hline
0 & 0 & $|0\ran$ \\
0 & 1 & $|1\ran$ \\
1 & 0 & $|\psi_0\ran$ \\
1 & 1 & $|\psi_1\ran$ \\
\end{tabular}
\end{center}
Alice sends this system to Bob, who must make a measurement on it to
find out about Alice's preparation.  This step also requires Bob to
have written down a random, independent $n$-bit string, which is also
a {\em basis} string.  If Bob's $i^{\rm th}$ bit is 0, he measures the 
$i^{\rm th}$ string he receives in the $\{|0\ran,|1\ran\}$
basis; if this bit is 1, he measures in the
$\{|\psi_0\ran,|\psi_1\ran\}$ basis.  If the result is $|0\ran$ or
$|\psi_0\ran$, he writes down a 0 as his measurement result, otherwise 
he writes down a 1.

Now Alice and Bob both publicly announce their basis
strings\footnote{It is crucial that this announcement only happen {\em 
after} Bob has made his measurement.}.  By
performing a bitwise-addition of the two basis strings, Alice and Bob
(and indeed anyone paying attention) can find out when Alice's
preparation and Bob's measurement were in the same basis; this will
occur on average half the time.  When this happens Alice knows with
100\% certainty what result Bob got from his measurement, so Alice and
Bob will share one bit of a secret key.  In Figure~\ref{buttercup},
whenever a 0 occurs in the Agreed Basis string the measurement
outcomes agree, so Alice and Bob end up with a 9 bit secret key (in
general, an $n/2$-bit key).

In what sense is the key ``secret''?  Can Eve not surreptitiously
observe the proceedings, intercept the states and make measurements on 
them?  In fact Eve can do so, but the problem is that any measurement
which obtains a non-negligible amount of information will disturb the
message state {\em in a detectable way}.  By applying certain random
hashing functions on the bits \cite{Bennett:1996} and sharing the
values of the hashing functions, Alice and Bob can discover if their
mutual information is less than it should be --- a sign of
eavesdropping --- in which case they discard the key.

Considering that the original proposal for a QKD
protocol was formulated by Bennett and Brassard in 1984 
\cite{Bennett:1984}, it took a long time for a general proof of the
security of the protocol to be given.  In 1998, Lo and Chau
\cite{Lo:1999} showed that if all the parties, including Eve,  are
equipped with quantum computers, then no measurement made by Eve which 
yields any information whatever can remain undetected by Alice and
Bob.  Almost simultaneously Mayers \cite{Mayers:1998} independently
showed that unconditional security was attainable without quantum
computers.  A comprehensive discussion of the nuts and bolts of QKD is
given in \cite{Lo:1998}. 

The implementation of this protocol using photon polarisation as the
qubit requires efficient single-photon detectors, long coherence times
in very small photon packets and highly accurate polarisation
control.  The experimental efforts in this field have been very
successful, both through open air and using 23 km of commercial
optical fibre.  A survey of results can be found in \cite{Zbinden:1998}.

\subsection{The Inference-Disturbance Tradeoff}

Information gathering causes disturbance to quantum systems.  This is
the principle which powers QKD, and the idea has
been with us since the early days of quantum theory:
\begin{quote}
\ldots every act of observation is an interference, of undeterminable
extent, with the instruments of observation as well as with the system 
observed, and interrupts the causal connection between the phenomena
preceding and succeeding it. \cite[p. 132]{Pauli:1995}
\end{quote}
This necessary disturbance is such a part of quantum physics that
undergraduates are told about it in a first course on the theory.
Unfortunately, this statement is frequently followed by an explanation 
that this is a consequence of the Heisenberg uncertainty principle --- 
a misconception that has been with us since 1927 when Heisenberg
\cite{Heisenberg:1927} first explained his principle with a
semiclassical model\footnote{In the same article quoted above, Pauli
wrote that the observer can choose from ``two mutually exclusive
experimental arrangements,'' indicating that he also believed the
disturbance was related to conjugate variables in the theory.}.

The meaning of Heisenberg's principle is far from the idea of
disturbance considered here, and has to do with the inability to
ascribe classical states of motion to a quantum system.  Until the
observer interacts with the system the classical variables $x$ and $p$ 
are not defined for the system.  Specifically, 
no matter how accurately or cunningly you prepare $N$ copies of a
state, measurements of position on half of them and momentum on the
other half will yield measurement statistics obeying $\Delta x\cdot
\Delta p \geq h$.  Trying to consider the ``disturbance'' to a
variable that does not have an objective meaning before, during or
after the measurement is clearly an exercise in futility.

The perspective of statistical inference, however, allows us to
investigate the disturbance objectively.  When Alice prepares a system 
in a state $\rho$, that means that she has certain predictive power
about any measurement we might make on the system (if $\rho$ is a
pure state there is a measurement which will yield a completely
certain result).  This is what allows her to synchronise her
predictions with Bob's measurements; in fact, as long as there is nonzero 
mutual information between her preparation and Bob's measurement, they 
will be able to distill a shared key.  We can measure the {\it
disturbance} to the state by Alice's loss of predictive power, in a
statistical sense, perhaps by measuring the distinguishability of the
prepared state from the disturbed state.  Similarly we can measure the 
inferential power that Eve gains from her interference with some
statistical measure; perhaps we consider the mutual information
between her measurement and the state preparation.  In the
eavesdropping context all Alice might want is some knowledge of the
correlation between Alice's preparation and Bob's measurements.

Fuchs and Jacobs \cite{Fuchs:2000} highlight two radically different features
of this model from the ``disturbance'' discussed by the founding
fathers of quantum theory.  Firstly, in order to make this a
well-posed problem of inference, all the observers must have
well-defined prior information.  The disturbance is then grounded with 
respect to Alice's prior information, and the inference is grounded with
respect to Eve's.  In this way we avoid reference to disturbance of
``the state'' --- which is after all defined in terms of our knowledge 
of the outcomes of the same experiments which ``disturb'' it --- and we
consider the predictive power of various observers --- since statistical 
prediction of observers' results is what quantum theory is all about.

Secondly one must consider at least two nonorthogonal states; the
question ``How much is the coherent state $|\psi\ran$ disturbed by
measuring its position?'' is ill-posed.  This is because if we already 
know the state, we can measure it and simply remanufacture the state
$|\psi\ran$ afterwards.  Likewise, if we know that a system is in one
of a given set of orthogonal states, we can perform a {\em nondemolition
measurement} \cite{Unruh:1978} and cause no disturbance to the state at 
all.  The disturbance we are considering here is thus disturbance to
the {\em set} of possible states.

The situation of QKD can be described as follows.  Alice and Eve have
differing knowledge of a system's preparation, and therefore assign
different states to it. In this situation, Alice has more predictive
power than Eve.  Alice passes the system to Eve, who is not happy that 
Alice knows more than her.  So Eve attempts to interact with the
system in such a way that she can bring her predictions more into
alignment with Alice's, {\em without} influencing Alice's predictive
power (so that Eve will not be detected).  But, unfortunately for Eve, 
quantum mechanics is such that any such attempt to surreptitiously
align her predictions with Alice's is doomed to failure.  This seems
to be a very strong statement about quantum theory.

Few quantitative studies have been carried out to quantify this
trade-off \cite{Fuchs:1996}, \cite{Fuchs:1997b}.  A promising direction
for this work is to consider measures of entanglement (to be discussed 
in the next chapter), and consider how entanglement may be used in
improving inferential power of observers in quantum mechanics
\cite{Schumacher:1998}, \cite{DiVincenzo:1998}.

\section{Maxwell's Demon and Landauer's Principle}

In a sense, the considerations of information in physics date back to
the $19^{\rm th}$ century --- before quantum mechanics or information
theory.  Maxwell found an apparent paradox between the laws of physics 
and the ability to gather information, a paradox which was resolved
more than a century later with the discovery of a connection between
physics and the gathering of information.

Maxwell considered a sentient being (or an appropriately programmed
machine) later named a ``Demon'' whose aim was to violate a law of
physics.  Szilard \cite{Szilard:1929} refined the conceptual model
proposed by Maxwell into what is now known as {\em Szilard's engine}.
This is a box with movable pistons at either end, and a removable
parition in the middle.  The walls of the box are maintained at
constant temperature $T$, and the single (classical) particle inside
the box remains at this temperature through collisions with the
walls.  A cycle of the engine begins with the Demon inserting the
partition into the box and observing which side the particle is on.
He then moves the piston in the empty side of the box up to the
partition, removes the partition, and allows the particle to push the
piston back to its starting position isothermally.  The engine
supplies $k_BT\ln 2$ energy per cycle, apparently violating the Second
Law of Thermodynamics (Kelvin's form \cite{Mandl:1971}).  Szilard
deduced that, if this law is not to be violated, the entropy of the
Demon must increase, and conjectured that this would be a result of
the (assumed irreversible) measurement process.

To rescue the Second Law --- as opposed to assuming its validity and
supposing that measurement introduces entropy
 --- we clearly need to analyse the Demon's actions to discover 
where the corresponding {\em increase} in entropy occurs.  Many
efforts were made in this direction, mostly involving analyses of the
measurement process \cite{Brillouin:1956}.  An important step in
resolving this paradox was Landauer's \cite{Landauer:1961} 1961
analysis of thermodynamic irreversibility of computing, which led to
{\em Landauer's Principle}: erasure of a bit of information in an
environment at temperature $T$ leads to a dissipation of energy no
smaller than $k_BT\ln 2$.

Bennett \cite{Bennett:1982} exorcised the Demon in 1982. He noted that 
measurement does not necessarily involve an overall increase in
entropy, since the measurement performed by the Demon can in principle 
be performed
reversibly.  The entropy of the Demon, considered alone, does
increase, but the overall entropy is reduced through the correlation
between the position of the particle and the Demon's knowledge.  The
important realisation is that the thermodynamic accounting is
corrected by returning the demon to his ``standard memory state'' at
the end of the cycle.  At this stage the Demon erases one bit of
knowledge and hence loses energy at least $k_BT\ln 2$; so the Szilard
engine cannot produce useful work. 
\begin{figure}
\begin{center}
\epsfig{file=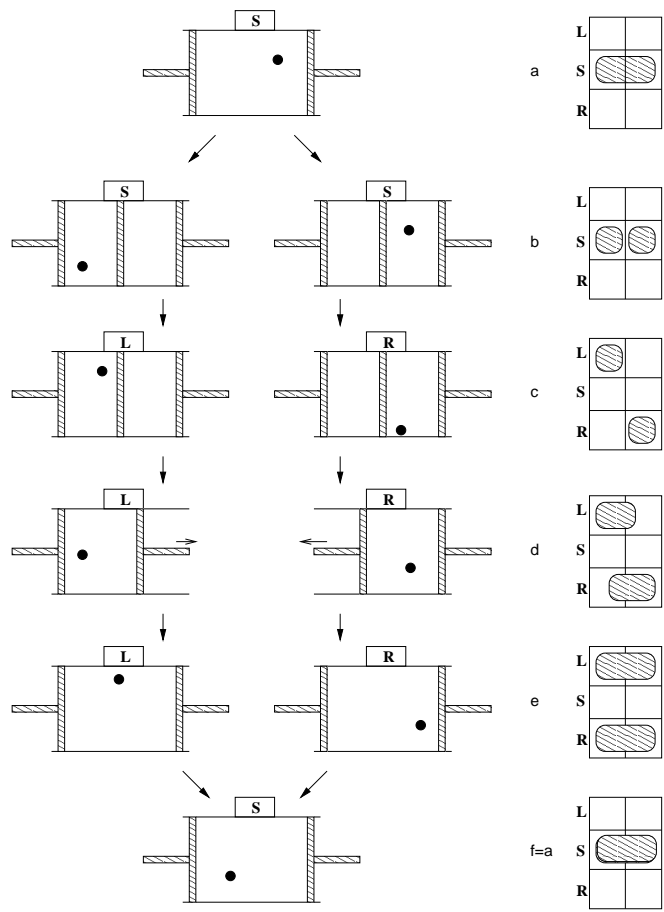, height=12cm}
\caption{One cycle of a Szilard engine operation (after
\cite{Bennett:1982}).  On the right is a phase diagram with the Demon's 
co-ordinates on the vertical axis and the box's on the left.}\label{tulip} 
\end{center} 
\end{figure}

Figure~\ref{tulip} is reproduced from Bennett's paper, and follows the 
phase space changes through the cycle.  In (a), the Demon is in a
standard state and the particle is anywhere in the box; the entropy of 
the system is proportional to the phase space volume occupied.  In (b) 
the partition is inserted and in (c) the Demon makes his measurement:
his state becomes correlated with the state of the particle.  Note
that the overall entropy has not changed.  Isothermal expansion takes
place in (e), and the entropy of the particle+Demon increases.  After
expansion the Demon remembers some information, but this is not
correlated to the particle's position.  In returning the Demon to his
standard state in (f) we dissipate energy into the environment,
increasing its entropy.

\subsection{Erasing Information in Quantum Systems}\label{notagain}

Szilard's analysis has been criticised by Jauch and Baron
\cite{Jauch:1972} for introducing an operation which cannot be treated
by equilibrium statistical mechanics: at the moment when the piston
is inserted the gas violates the law of Gay-Lussac \cite{Jauch:1972}.
These authors even go so far as to denounce any relation between
thermodynamic entropy and informational entropy, much to the peril of
the author of the current chapter.

However, a more careful quantum analysis of Szilard's engine supports
Szilard's idea of a connection between these concepts.  Zurek
\cite{Zurek:1984} has performed this analysis.  He considers a
particle in an infinite square well potential of width $L$, and the
piston is replaced by a slowly inserted barrier of width $\delta\ll
L$ and height $U\gg kT$.  The main result is that, in an appropriate
limit, the system can at all times be described by its partition
function $Z$: the thermodynamic approximation is valid.  Zurek then
analyses a ``classical'' demon to illustrate that Szilard's conclusion 
is correct if the Second Law is valid, and a ``quantum'' demon which
explicitly has to reset itself and so demonstrates that Bennett's
conclusion regarding the destination of the excess entropy is also
correct.

The moral of the story is that erasure of information is a process
which costs free energy --- or equivalently, which causes an increase
in overall entropy.  This is useful for us, because if we can find a
way to {\em efficiently} erase information (i.e. to saturate the bound 
in Landauer's principle) then we can give a physical interpretation of 
the Holevo bound\footnote{This is a slightly handwaving
interpretation, and certainly not rigorous.}.

\begin{figure}
\begin{center}
\epsfig{file=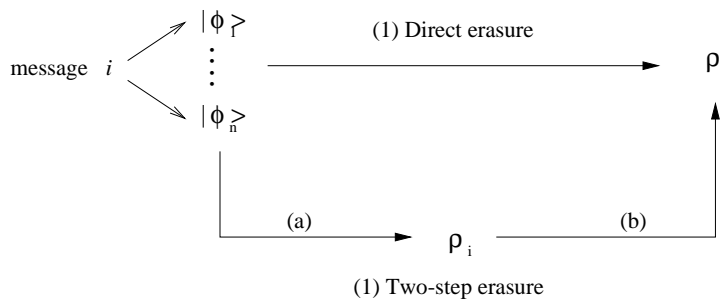, height=4cm}
\caption{Two ways of erasing information.}\label{nooooo!} 
\end{center} 
\end{figure}

Plenio \cite{Plenio:1999}, exploiting an erasure protocol described by
Vedral \cite{Vedral:1999}, gives such an optimal erasure protocol
based on placing the demon (or the measurement apparatus) in contact
with a heat bath with an appropriate density operator.  Plenio then
describes two ways to erase information (illustrated in
Figure~\ref{nooooo!}).
Alice and Bob are going to communicate by sending mixed states from
the ensemble $\{\rho_i; p_i\}$ (so the overall state is $\rho=\sum
p_i\rho_i$.  However, Alice is feeling whimsical:
she decides to send {\em pure} states $|\phi^i_k\ran$ drawn from a
probability distribution $r^i_k$.  These pure states are chosen so
that $\rho_i = \sum_k r^i_k |\phi^i_k\ran\lan \phi^i_k|$.
Once Bob receives the system, he can erase the information\footnote{He
must erase the information and return the state to Alice if this is to
be a cyclic process} in one of two ways: (1) by directly erasing all
the pure states or (2) by first erasing the irrelevant information
regarding which pure state he received (a) and then erasing the
resulting known mixed state.

By erasing in one step, Bob incurs an entropy cost $S_1 = S(\rho)$
since he has erased efficiently.  In part (b) of the  second step, he
incurs a cost $S_{2b}^i = S(\rho_i)$ when erasing a particular message
$\rho_i$, after determining that this was indeed the message Alice
sent.  However, on average this step will cost Bob $S_{2b} = \sum_i
p_iS(\rho_i)$.  The best Bob can therefore do on step 2(a) is the
difference:
\be
S_{2a} = S_1 - S_{2b}.
\end{equation}
But Landauer's principle tells us that the best measurement Bob makes
to determine the value of $i$ that Alice sent can yield no more
information that the entropy of erasure:
\be
I \leq S_{2a} = S(\rho) - \sum_i p_i S(\rho_i).
\end{equation}

We thus see that the Holevo bound, proved in Section~\ref{egypt},
which was based on an abstract relation satisfied by the von Neumann
entropy, has an interpretation in terms of a very deep principle in
information physics, namely Landauer's principle.  This connection
provides further support for the latter principle and possibly also
some guidance for the question, still unresolved, of the capacity of a 
channel to transmit intact quantum states.
\chapter{Entanglement and Quantum Information}

In the previous chapter we considered how classical information is
manifested in quantum systems.  One of the major results was the
discrepancy between the amount of information used to prepare a
quantum system, and that which can usefully be obtained from it.
Clearly the concept of ``classical'' information is missing something, 
just as ``classical'' orbits became awkward in quantum mechanics.

A ``classical'' information carrier is usually a two-state system, or
a system with two distinct phase space regions.  This is regarded as a 
resource for conveying the abstract notion of a bit.  By direct
analogy, a two-level quantum system can be regarded as a resource for
carrying {\em one two-level quantum state's} worth of information.  We 
call the abstract quantity embodied by this quantum resource a
{quantum bit}, or {\em qubit}.

In this chapter we will consider the properties of quantum
information.  The Schumacher Coding Theorem \cite{Schumacher:1995}
will give us a 
``system-independent'' way of measuring information.  We will discuss 
the relation between quantum information and quantum entanglement, and 
how the latter can be measured --- and more importantly, what it can
be used for.  We will also look at whether quantum information is
useful in a practical sense: can it be protected from errors?

\section{Schumacher's Noiseless Coding Theorem}

A first question which arises is: How does Alice send quantum
information to Bob?  Classically, she would compress the information
according to Shannon's Noiseless Coding Theorem, encode this into the
state of her classical physical systems, and convey these to Bob ---
we assume there is no noise in the channel they are using ---  who
would perform an inverse operation to extract Alice's message.  In the 
quantum case, we immediately run into a problem.  As soon as Alice
attempts to determine (through measurement) her quantum state, she has lost
exactly that quantum information she wished to convey to Bob --- we
discovered this in the previous chapter.  If Alice happens to have the 
preparation information in classical form, she can send this to Bob;
but this information can be arbitrarily large.
Also, there are situations in which Alice doesn't know what state
she's got (perhaps the state is the product of a quantum computation
which now requires Bob's input \cite{Preskill:1999}).

An obvious solution is to simply convey the physical system to Bob.
If for example the system is a two-level quantum system, she will
incur no more effort in doing this than in conveying one classical bit 
of information, which also requires a two-state system.  But once
again we encounter a redundancy: suppose the system has four
orthogonal states but we happen to know that all the ``signal''
quantum states Alice needs to send are contained in a two-dimensional
subspace.  Then sending the entire system is not necessary, since Alice
could {\em transpose} the state onto a two-dimensional system without
losing any of this ``quantum information''.  So the question is: Can
we generalise this result in some way?  How big does the system
conveyed to Bob {\em have} to be?  Does it depend on whether the
quantum states Alice wants to send are pure or mixed?

The answer for pure states is that the system can be compressed to a
Hilbert space of dimension $2^{S(\rho)}$, where once again $S(\rho)$ is 
the entropy of the signals.  And 
for mixed states?  One would be tempted to guess that the answer is
similar to that for pure states, with the Holevo information
substituted for the entropy, but this is not known.
This is certainly a lower bound, but the question becomes
rather more complicated for mixed states, as will be discussed later.
But for now, before discussing Schumacher's result, we will look at
how quantum information is encoded, and what measure we will use for
judging the accuracy of the decoded state.

\begin{figure}
\begin{center}
\epsfig{file=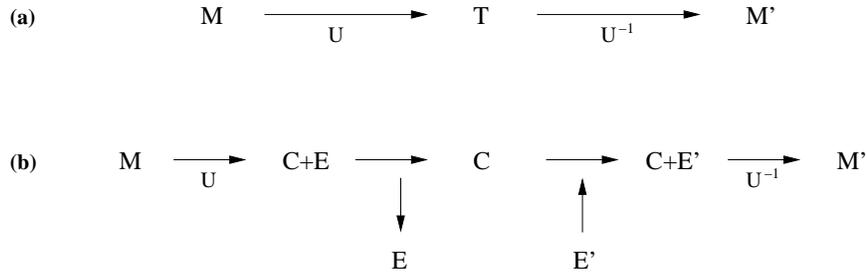, width=12cm}
\caption{The transmission of quantum information.  In (a) no
information is discarded, but in (b) the communication is
approximate.}\label{hunnybunny}
\end{center} 
\end{figure}

Suppose the state Alices wants to send is in the Hilbert space $H_M$, which 
we call the message system; this could be a subspace of some larger
system space as long as we {\em know} that the signals are in $H_M$.
She wishes to transfer this quantum 
information to a transmission system represented by $H_T$.  {\em Copying}
the information is not allowed, but transposition is; that is, the
operation
\be
|a_M, 0_T\ran \longrightarrow |0_M, a_T\ran
\end{equation}
can be performed, where $|0_M\ran$ and $|0_T\ran$ are fixed standard
states.  This operation can be made unitary if $\dim H_T\geq \dim
H_M$, simply by specifying the (ordered) basis in $H_T$ to which some
(ordered) basis in $H_M$ gets mapped.  This is the exact analogue of
classical coding.  Alice then passes the system $T$ to Bob, who
performs an inverse swap operation onto his decoding system $H_{M'}$
--- and no quantum information has been discarded.
The communication process is represented in Figure~\ref{hunnybunny}(a).

If we wish to {\em compress} the information so that it occupies a
smaller Hilbert space, we will have to consider an approximate
transposition, as represented in Figure~\ref{hunnybunny}(b).  In this
case our transmission system is decomposed into a code system $H_C$ of 
dimension $d$, and an ancilla $H_E$.  Only the code system is conveyed to
Bob, and the ancilla is discarded.  On the receiving end, Bob attaches 
an ancilla $H_{E'}$ in a standard state and performs a decoding
operation, which could be the inverse $U^{-1}$ of the coding
operation\footnote{We can without loss of generality assume that the
coding is unitary, but the decoding performed by Bob could in general
be a superoperator.  We will not consider this complication here.}.
In this case the exact form of the encoding operation $U$ becomes very 
important; as much of the ``weight'' of Alice's signal states as
possible must end up in the quotient space $H_C$ that Bob receives,
and indeed the dimension of this space must be large enough to receive 
almost all of the ``quantum messages'' Alice would like to send.

Suppose Alice sends signal states $|a_M\ran$, each with probability
$p(a)$.  Then we can trace the path of the signal as follows:
\begin{eqnarray}
|a_M\ran\lan a_M| & \longrightarrow & U|a_M\ran\lan a_M|U^{-1} = \Pi_a
\nonumber\\
& \longrightarrow & \Tr{E} \Pi_a \nonumber\\
& \longrightarrow & \Tr{E} \Pi_a \otimes |0_{E'}\ran\lan0_{E'}| = W_a \nonumber\\
& \longrightarrow & U^{-1} W_a U = \rho_a.
\end{eqnarray}
So the final state that Bob receives will be a mixed state.

Clearly Bob will not get all the quantum information Alice started
with.  But how do we measure what got to Bob?  Any measure of
distinguishability will work (where now Alice and Bob's aim is clearly
to minimise distinguishability).  For mathematical reasons, the
fidelity is usually used, and in this case it has a very convenient
interpretation.  Suppose that Bob knew precisely 
which pure state $|a_M\ran$ Alice started with --- for example, they are
calibrating their equipment.  Then there is a maximal test
\cite{Peres:1993} for which the outcome can be predicted with
certainty.  Suppose the (mixed) state received by Bob after decoding
is $\rho$; if Bob performs this maximal test on the decoded state then
the probability that he will think that the state is {\em exactly the
one sent} is $\Tr{} \rho\,|a_M\ran\lan a_M|$.
This is therefore a practical measure of the accuracy of the received state.
The average fidelity of the transmission is then defined as
\be
\bar{F} = \sum_a p(a) \Tr{} |a_M\ran\lan a_M| \rho_a.
\end{equation}
The fidelity is between 0 and 1, and equal to one if and only if all
signals are transmitted perfectly, as in Figure~\ref{hunnybunny}(a).
Note that the fidelity can also be computed in terms of the code
system states (rather than the message and decoding systems):
\begin{eqnarray}
\bar{F} &=& \sum_a p(a) \Tr{} |a_M\ran\lan a_M| \rho_a \nonumber\\
&=& \sum_a p(a) \Tr{} U^{-1}\Pi_a U\; U^{-1}W_a U \nonumber\\
&=& \sum_a p(a) \Tr{} \Pi_aW_a.
\end{eqnarray}

Schumacher then goes on to prove two lemmas:
\begin{lemma}
Suppose that $\dim H_C = d+1$ and denote the ensemble of states on $M$ by 
$\rho_M = \sum p(a)|a_M\ran\lan a_M|$.  Suppose there exists a
projection $\Lambda$ onto a $d$-dimensional subspace of $H_M$ with
$\Tr{} \rho_M\Lambda > 1-\eta$.  Then there exists a transposition
scheme with fidelity $\bar{F}>1-2\eta$.
\end{lemma}
\begin{lemma}
Suppose $\dim H_C = d$ and suppose that for {\em any} projection
$\Lambda$ onto a $d$-dimensional subspace of $H_M$, $\Tr{}
\rho_M\Lambda < \eta$ for some fixed $\eta$.  Then the fidelity $\bar{F}<\eta$.
\end{lemma}
Both of these results are highly plausible.  The first says that if
almost all of the weight of the ensemble of message states is
contained in some subspace no larger than the code system, then
accurate transposition will be possible.  This is proved
constructively by demonstrating a transposition scheme with this
fidelity.  The second lemma deals with the situation in which any
subspace of the same size as code system {\em doesn't} contain enough
of the message ensemble, in which case the fidelity is bounded from
above.

No mention is made in these lemmas of the ensemble of states
comprising $\rho_M$; they could be arbitrary states, or they could be
the eigenstates of $\rho_M$; but according to the lemma the
transposition fidelity can always be bounded as appropriate.  Focus is 
thus shifted away from the ensemble, and all we have to deal with is
the density operator $\rho_M$.  Schumacher's theorem then says:

\begin{thm}{Quantum Noiseless Coding Theorem \cite{Schumacher:1995}}
Let $M$ be a quantum signal source with a signal ensemble described by 
the density operator $\rho_M$ and let $\delta,\epsilon>0$.

(i) Suppose that $S(\rho_M)+\delta$ qubits are available per signal.
Then for sufficiently large $n$, groups of $n$ signals can be
transposed via the available qubits with fidelity $\bar{F}>1-\epsilon$.

(ii) Suppose that $S(\rho_M)-\delta$ qubits are available for coding
each signal.  Then for sufficiently large $n$, if groups of $n$
signals are transposed via the available qubits, then the fidelity
$\bar{F}<\epsilon$.
\end{thm}

To prove this, we begin by noting that the maximum value of $\Tr{}
\rho_M\Lambda$ is realised 
when $\Lambda$ is a projection onto the subspace spanned by
eigenstates of $\rho_M$ with the largest eigenvalues.  Now, as in the
case of the classical theorem, we consider {\em block coding} i.e. we
focus on the density operator $\rho^N =
\rho\otimes\ldots\otimes\rho$.  The eigenvalues of this operator are
products of the single system eigenvalues.  As in the discussion of
the law of large numbers following Eqn~\ref{palmer}, we can find a set 
of approximately $2^{n[s(\rho_M)+\delta]}$ such product eigenvalues
with total sum $> 1-\epsilon/2$.  Then by Lemma 1, there is a
transposition scheme with fidelity $\bar{F}>1-\epsilon$.

Conversely, if we can only add together $2^{n[S(\rho_M)-\delta]}$ of
the largest eigenvalues, this sum can be made arbitrarily small (see
for example \cite{Schumacher:1995}, \cite{Cover:1991}).  Hence any
projection onto a subspace of this size will yield arbitrarily low fidelity.

Schumacher's result is valid for unitary encoding and decoding
operations.  In full generality, one should allow Alice and Bob to
perform any operation (represented by a superoperator) on their
systems and check whether further compression can be achieved than
discussed above.  This was considered by Barnum {\it et al}
\cite{Barnum:1996}, who showed that Schumacher coding gives the
maximum possible compression.  In particular, they showed that even if 
Alice has knowledge of which state she is sending, and applies
different encodings depending on this knowledge, she can do no better
than if she coded without knowledge of the signals.

\subsection{Compression of Mixed States}

What is the optimal compression for ensembles of {\em mixed} states?
That is, what size Hilbert space do we need to communicate signals
from the ensemble $\{\rho_a; p_a\}$ accurately?  Once
again, we will measure accuracy using the fidelity, although in this
case we must use the form for mixed states given in Eqn~\ref{noddie}.

The protocol for pure state compression suggested by Jozsa and
Schumacher in \cite{Jozsa:1994} can also be applied to mixed state
ensembles, and compresses an ensemble down to the value of its von
Neumann entropy.  However, it is very simple to show that this is not
optimal, simply by considering an ensemble of mixed states with
disjoint support.  In this case our coding technique could be simply
to measure the system --- we can distinguish perfectly between signal
states since they are disjoint --- and prepare it in a pure state,
such that the ensemble of pure states has lower entropy

For the problem of mixed state compression, the signal ensemble is
${\cal E}^0=\{\rho_a; p_a\}$, where  
the $\rho_a$ are now mixed states.  In order to code this, Alice
accumulates a block of $n$ signals $\rho_{a_1}\otimes\ldots \otimes
\rho_{a_n} = \sigma_{\bf a}$, where ${\bf a}$ is a multiple index.
The block ensemble is ${\cal E}_n = \{\sigma_{\bf a}; q_{\bf a}\}$, where
$q_{\bf a}$ is a product of the appropriate $p$ probabilities, and it
has a density matrix $\rho_n = \sum_{\bf a} q_{\bf a}\rho_{\bf a}$.
Alice will perform some operation on these states to code them.

We now have to distinguish two types of coding: if Alice {\em knows}
which states she is sending to Bob, then she can code each state
differently and we call this {\em arbitrary} or {\em nonblind} coding;
if she doesn't 
know then she treats all states equally, which is called {\em blind}
coding.  Alice's allowed operations, in the case of blind coding, are
completely positive maps (represented by superoperators).  For
arbitrary coding, she can use {\em any} map; she could even, if she
wanted, replace each code state $\rho_{\bf a}$ by a pure state
$|\psi_{\bf a}\ran$.  Bob, on the other hand, is always constrained to 
using a completely positive map, since he can have no knowledge of the
state he receives.

A {\em protocol} is then defined as a sequence $(\Lambda_n, \$_n)$ of
Alice's encoding operations $\Lambda_n$ and Bob's decoding
superoperators $\$_n$ which yield arbitrarily good fidelity as
block size increases i.e.
\be
\bar{F}_n = \sum_{\bf a} F\big(\rho_{\bf a}, \$_n\left[
\Lambda_n(\rho_{\bf a})\right]\big) 
\end{equation}
satisfies $\bar{F}_n\rightarrow 1$ as $n\rightarrow\infty$.  The {\em
rate} of a protocol ${\cal P}$ is 
\be
R_{\cal P} = \lim_{n\rightarrow\infty} \frac{1}{n} \log \dim \tilde{\rho},
\end{equation}
where $\dim \tilde{\rho}$ is the dimension of the support of $\tilde{\rho}$.

Horodecki then distinguishes between the {\em passive information}
$I_p = \inf_{\cal P} R_{\cal P}$ where the infimum is taken over all
blind protocols (where $\Lambda_n$ is a superoperator), and the {\em
effective information} $I_e$ which is 
defined similarly but with the infimum taken over arbitrary
protocols (where $\Lambda_n$ is an arbitrary map).  Clearly $I_e\leq
I_p$ since all blind protocols are also 
arbitrary.  Horodecki \cite{Horodecki:1998} shows that the effective
information is bounded below by the Holevo information, $\chi({\cal
E})\leq I_e$, and notes that the bound can be achieved if the
ensemble $\tilde{\cal E}$ of code system states has asymptotically
vanishing mean entropy (the code states become almost pure).  We may also
consider the ``information defect'' $I_d = I_p - I_e$, which
characterises the stinginess of the ensemble: a nonzero $I_d$
indicates that more resources are needed to convey the quantum
information than the actual value of the information.  It is not known 
if the information defect is ever nonzero.

A tantalising clue is given in Schumacher's original paper where  he showed 
that {\em entanglement} could be compressed using his protocol.  Suppose
Alice and Charlie share an EPR pair between them; for definiteness,
suppose each of them holds an electron and the {\em joint} state of
the electrons is
\be
|\psi^-\ran = \frac{1}{\sqrt{2}}\left( \spinupvec_A \spindownvec
_C - \spindownvec_A \spinupvec_C \right)
\end{equation}
($\spinupvec$ and $\spindownvec$ refer to spin polarisation
states of the electrons).
It is well-known that such a state exhibits strong correlations (see
e.g. Bell \cite{Bell:1964}).  The question is, do such correlations
persist between Bob and Charlie if Alice conveys her system to Bob
using the methods above?  Schumacher showed that they do.  We are in
effect conveying the full pure state $|\psi^-\ran$ to Bob, but only
compressing one half of it --- Charlie's half undergoes the identity
transformation.  This can be done with arbitrarily good fidelity, so
the correlations between Bob and Charlie will be arbitrarily close to
the original EPR correlations.

Horodecki \cite{Horodecki:1999} then showed that in full generality,
the problem of arbitrary (nonblind) compression of an ensemble of mixed states
could be reduced to minimising the entropy of any {\em extension}
ensemble.  An extension of density operator $\rho$ is a density
operator $\sigma$ over a larger Hilbert space with the property that
$\rho = \Tr{\mbox{\scriptsize ancilla}} \sigma$, and the extension of
an ensemble is another 
ensemble with the same probabilities whose elements are extensions of
the original. In particular, it seems fruitful to investigate the
purifications of a given ensemble --- since any purification is also
an extension.  We would then be able to use the provably optimal
techniques of pure state coding to solve the problem.  This is
extremely handy reduction, enabling us to ignore the unwieldy
representation of protocols and focus on states and their purifications.

However, minimisation of von Neumann entropy $S$ over all extensions
is also a very difficult problem.  A first, and counter-intuitive, result was
that of Jozsa and Sclienz \cite{Jozsa:2000} mentioned in Sec.~\ref{states}.
Suppose we start with a particular purification of ${\cal E}$ and ask how
we can distort the states, leaving the probabilities fixed, to
decrease $S$.  A reasonable first guess, 
based on our knowledge of two-state ensembles, is that any distortion
which increases pairwise overlap will decrease von Neumann entropy.
As disussed previously, Jozsa and Schlienz showed that the von Neumann 
entropy of almost any pure state ensemble in more than two dimensions
can be increased by making all the states {\em more} parallel.  The
compressibility --- and the distinguishability --- are global properties of
the ensemble which cannot be reduced to accumulated local properties
of members of the ensemble.

There are still many unanswered questions in this area, and many
tantalising clues. Several of these clues, together with some partial
results have been collected together by Barnum, Caves, Fuchs, Jozsa
and Schumacher \cite{Barnum:2000}.  One of their most striking insights 
is that the fidelity between two strings of states $\rho_{\bf a}$ and
$\sigma_{\bf a}$ depends on whether we are content to compute the
fidelity of each state as it arrives (LOCAL-FID) or whether we compute 
the fidelity of blocks of states (GLOBAL-FID).  The latter is a
stringer condition that (LOCAL-FID) since it not only requires the
individual states to be very similar but also their entanglements
should not be altered.  For pure states the two types of fidelity are
the same, but for mixed states Barnum {\it et al} exhibit an example
with high (LOCAL-FID) but zero (GLOBAL-FID).

An important bound proved by Barnum {\it et al} \cite{Barnum:2000} is
that the Holevo information is a lower bound for compression under the 
criterion (GLOBAL-FID), although there are strong reasons to believe
that under this fidelity criterion this bound is not generally
attainable, but it has been shown that for a weaker fidelity this
bound can always be achieved \cite{Fuchs:2000b}.  At the moment there
appear to be too many trees for us to see the wood in mixed state
compression.

\section{Entanglement}

Entanglement is one of the most intruiging aspects of quantum theory.
First labeled {\em verschr\"ankung} by Schr\"odinger
\cite{Schrodinger:1935}, entanglement became one of the features most
uncomfortable to Einstein.  It was the paper by Einstein, Podolsky and 
Rosen \cite{Einstein:1935} (describing the famous EPR paradox) which expressed
some of this dissatisfaction with the ``action at a distance'' of
quantum mechanics.  The crucial blow to Einstein's requirement of
``local reality'' came from John Bell in 1964 \cite{Bell:1964}, and a
version of Bell's theorem will be mentioned below.

First, however, we look at what entanglement is; and we begin by
looking at the canonical example of entangled qubits.  Suppose we are
considering two electrons $A$ and $B$; then a basis for the electron
spin states is
\begin{eqnarray*}
|\phi^+\ran = \frac{1}{\sqrt{2}}\left(|\!\UP\UP\ran + |\!\DN\DN\ran\right)\\
|\psi^+\ran = \frac{1}{\sqrt{2}}\left(|\!\UP\DN\ran + |\!\DN\UP\ran\right)\\
|\phi^-\ran = \frac{1}{\sqrt{2}}\left(|\!\UP\UP\ran - |\!\DN\DN\ran\right)\\
|\psi^-\ran = \frac{1}{\sqrt{2}}\left(|\!\UP\DN\ran - |\!\DN\UP\ran\right)
\end{eqnarray*}
where, for example, we denote the state 
$\spinupvec_A\otimes\spindownvec_B$ by $|\!\UP\DN\ran$. 
These states form the {\em Bell basis} (or alternatively they are {\em 
EPR pairs}), and they have numerous interesting properties that will
be briefly discussed here.

A bipartite system in a Bell state has a peculiar lack of identity.
If we
consider just one half of the system and trace out the other system's
variables, we find 
\begin{eqnarray}
\Tr{A}|\Upsilon\ran\lan \Upsilon| &=& {}_A\lan\UP\!|\Upsilon\ran\lan
\Upsilon \spinupvec_A + {}_A\lan\DN\!|\Upsilon\ran\lan
\Upsilon\spindownvec_A \\ 
&=&  \frac{1}{2}\unit = \Tr{B}|\Upsilon\ran\lan \Upsilon|
\end{eqnarray}
(where $|\Upsilon\ran$ is one of the Bell states): so
any orthogonal measurement on a single particle will yield a completely
random outcome.  Of course neither half of the system, considered
alone, is in a pure state --- so despite the fact that we have maximal 
information about both systems, there is no test we can perform on
either one which will yield a definite outcome.
Each of the basis states represents (optimally) two bits of
information, which we may conveniently call the {\em amplitude} bit
(the $|\psi\ran$ states both have amplitude 0, the $|\phi\ran$ states
1) and the {\em phase} bit (the $+$ states have phase 0, the $-$
states have phase 1).  The amplitude bit is the eigenvalue of the
\label{bellow}
observable\footnote{The operators $\sigma_i$ are the Pauli matrices,
namely $\sigma_x = \left( \begin{array}{cc} 0&1\\1&0 \end{array}
\right),\sigma_y = \left( \begin{array}{cc} 0&-i\\i&0 \end{array}
\right),\sigma_z = \left( \begin{array}{cc} 1&0\\0&-1 \end{array}
\right)$.}
$\sigma_x^A\sigma_x^B$ and the phase bit of the observable
$\sigma_z^A \sigma_z^B$ --- and these are commuting operators.  The
problem in determining their identity comes when the systems $A$ and
$B$ are spatially separated, because the local operators $\sigma_x^A$
and $\sigma_x^B$ which could be used to discover the amplitude bit
{\em do not} commute with the phase operator $\sigma_z^A \sigma_z^B$.
Determining the amplitude bit through local measurements will disturb
the phase bit; so the information is practically inaccessible.  If the 
particles are brought together, we can measure the amplitude bit {\em
without} determining the values of $\sigma_x$ individually, and hence
fully determine the identity of the state.

So: what makes these states entangled?  And what other states are
entangled?  The answer to these questions is rather an answer in the
negative: a pure state is {\em not} entangled (i.e. it is separated)
if it can be written as a product state over its constituent
subsystems (we will mention mixed state entanglement later).  An alert
student may protest that by changing bases we could change a state
from entangled to separated -- so we should be sure of our facts
first.

Suppose $|\Psi\ran$ is a pure state over two subsystems $A$ and
$B$ (that is, it is a vector from $H_A\otimes H_B$).  Then let
$\{|i\ran\}$ be the basis for $H_A$ in which the reduced density
matrix for subsystem $A$ is diagonal:
\be\label{whoopdedoo}
\rho_A = \Tr{B}|\Psi\ran\lan\Psi| = \sum_i \lambda_i |i\ran \lan i|.
\end{equation}
Of course if $\{|\alpha\ran\}$ is any basis for $H_B$, then we can
write
\be
|\Psi\ran = \sum_{i, \alpha} a_{i,\alpha}|i\ran\otimes |\alpha\ran
= \sum_i |i\ran\otimes |\tilde{i}\ran
\end{equation}
where we have defined $|\tilde{i}\ran = \sum_\alpha a_{i,\alpha}
|\alpha\ran$.  The $|\tilde{i}\ran$ do not necessarily form an
orthonormal basis, but we can calculate $A$'s reduced density matrix
in terms of them:
\begin{eqnarray}
|\Psi\ran\lan\Psi| &=& \sum_{i,j} |i\ran_A|\tilde{i}\ran_B\; {}_A\lan j|{}_B
\lan\tilde{j}| \\
\Rightarrow \rho_A &=& \sum_k {}_B\lan k|\Psi\ran\lan\Psi|k \ran_B \\
&=& \sum_{i,j} {}_B\lan k|\tilde{i}\ran_B \; {}_B\lan\tilde{j}|k\ran_B 
\; |i\ran_A{}_A\lan j| \\
&=& \sum_{i,j} \lan\tilde{j}|\tilde{i}\ran \; |i\ran\lan j|
\end{eqnarray}
where $\{|k\ran_B\}$ is any orthonormal basis for $H_B$.  Comparing
this with Eqn~\ref{whoopdedoo}, we see that
\be
\lan\tilde{j}|\tilde{i}\ran = \lambda_i \delta_{ij}:
\end{equation}
the $|\tilde{i}\ran$ are orthogonal!  Defining $\sqrt{\lambda_i}
|i'\ran = |\tilde{i}\ran$ (an orthonormal basis) we find that
\be\label{schmidt}
|\Psi\ran = \sum_i \sqrt{\lambda_i} |i\ran\otimes|i'\ran.
\end{equation}
This is called the {\em Schmidt decomposition} of the
bipartite state\footnote{{\em Bipartite} means that the state is a
joint state over two systems.}
$|\Psi\ran$.  Notice that this decomposition is unique, and tells us
also that the reduced density matrices $\rho_A$ and $\rho_B$ have the
same non-zero eigenvalue spectrum --- if $H_A$ and $H_B$ have
different dimensions, then the remaining eigenvalues are zero.  The
Schmidt decomposition is related to a standard result in linear
algebra, known as the singular value decomposition.

We now have a very simple method of defining whether a pure state is
entangled.  Define the {\em Schmidt number} $N_S$ to be the number of
non-zero terms in the expansion Eqn~\ref{schmidt}.  Then the state is
entangled if $N_S>1$.  The special features of such entangled states
are related to Bell's theorem.

\paragraph{Bell's Theorem}
Einstein and others thought that, although quantum physics appeared to 
be a highly accurate theory, there was something missing.  One
proposed way of remedying this problem was to introduce hidden
variables; variables lying at a deeper level than the Hilbert space
structure of quantum mechanics, and to which quantum theory would
be a statistical approximation.  The idea is then that quantum
mechanics describes
a sort of averaging over these unknown variables, in much the same way 
as thermodynamics is obtained from statistical mechanics.  In fact,
the variables could even in principle be inaccessible to experiment:
the important feature is that these variables would form a complete
description of reality.  Of course, to be palatable we should require
these variables to be constrained by locality, to make the resulting
theory ``local realist''.  In the words of Einstein
\cite[p. 85]{Einstein:1949}:
\begin{quote}
But on one supposition we should, in my opinion, absolutely hold fast: 
the real factual situation of the system $S_2$ is independent of what
is done with the system $S_1$, which is spatially separated from the
former.
\end{quote}
Einstein was convinced that the world was local, deterministic and
real.

John Bell \cite{Bell:1964} struck a fatal blow to this view.  Bell
considered a very simple inequality which any local deterministic
theory must obey --- an inequality that was a miracle of brevity and
could be explained to a high school student, and was indeed known to
the $19^{\rm th}$ century logician Boole \cite{Pitowsky:1994} --- and
he showed that 
quantum mechanics violates this inequality.  We are thus forced to one 
of several possible conclusions: (i) quantum mechanics gives an
incorrect prediction, and a ``Bell-type'' experiment will demonstrate
this; (ii) quantum mechanics is correct, and any deterministic hidden
variable theory replicating the predictions of quantum mechanics must
be non-local.  With the overwhelming experimental evidence supporting
quantum mechanics, it would be a foolhardy punter who put his money on 
option (i), and indeed several Bell-type experiments performed to date
have agreed with the quantum predictions \cite{Aspect:1984}.  Bell's
theorem refers to the conclusion that any deterministic hidden
variable theory that reproduces quantum mechanical predictions must be
nonlocal.

An excellent discussion of Bell's theorem and its implications is
found in \cite[Ch. 6]{Peres:1993}.  Several papers have since appeared
on ``Bell's theorem without inequalities'' (for a readable discussion, 
see \cite{Mermin:1994}), which eliminate the statistical element of
the argument and make Bell's theorem purely logical.  In this case a
single outcome of a particular quantum mechanical experiment is enough 
to render a hidden variable description impossible.

The importance of entangled states, as defined above, is that every
entangled pure state violates a Bell inequality \cite{Gisin:1991}.
However, the definition of entangled states given above is not very
helpful when we need to {\em compare} the degree of entanglement of
bipartite states.  However, we 
have all we need to find a quantitative measure of entanglement: a
probability distribution, given by the eigenvalues $\lambda_i$.  We
define the entanglement of a bipartite state $|\Psi\ran$ to be
\begin{eqnarray}
E(|\Psi\ran) &=& H(\lambda_i) \label{finish}\\
&=& S(\rho_A) = S(\rho_B)
\end{eqnarray}
where $H$ is the familiar Shannon entropy.  Happily, the entanglement
of a state with $N_S=1$ is zero.  The maximally entangled states are
those of the form $\sum_i |i\ran|i'\ran/\sqrt{N}$ which are equal
superpositions of orthogonal product states\footnote{Any one of
the Bell states can be written in this way by redefining the
single-particle basis; in this basis none of the remaining Bell states 
has this form.}.  The Bell state basis
consists of maximally entangled states, and these are the only
maximally entangled states of two qubits.  In analogy with the
primitive notion of a qubit being a measure of quantum entanglement,
we define an {\em ebit} to be the amount of entanglement shared
between systems in a Bell state.

Bennett {\it et al} \cite{Bennett:1996} point out that this
entanglement measure has the pleasing properties that (1) the
entanglement of independent systems is additive and (2) the amount of
entanglement is left unchanged under local unitary transformations,
that is, those that can be expressed as products: $U = U_A\otimes
U_B$.  The entanglement cannot be increased by more general
operations either (see below), and has a convenient interpretation
that one system with entanglement $E=E(|\Psi\ran)$ is completely
equivalent (in a sense to be made clear later) to $E$ maximally
entangled qubits. 

\paragraph{Mixed state entanglement}
The case for ``a measure of entanglement'' of a bipartite mixed state
$\rho_{AB}$ is not quite as clear cut.  Indeed, just about the only
useful definition is that of a {\em separable} state, which is one
which can be written as
\be\label{mixedness}
\rho_{AB} = \sum_i p_i \;\rho_A^i \otimes \rho_B^i.
\end{equation}

We will return to measures of entanglement for mixed states later,
once we have developed more practical notions of the uses of
entanglement and an idea of what we are in fact hoping to quantify.
For now, we turn our attention to these issues.

\subsection{Quantum Key Distribution (again)}

We have already seen a protocol for QKD in Section~\ref{sneeze}.  Here 
we describe a variant based on Bell states, due to Ekert
\cite{Ekert:1991} which illustrates the interchangeability of quantum
information with entanglement.

\begin{figure}
\begin{center}
\epsfig{file=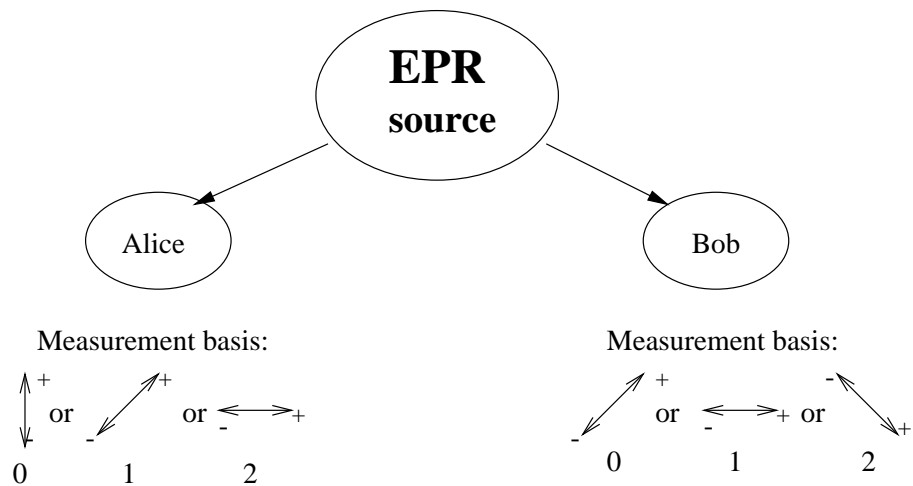, width=12cm}
\caption{Illustration of quantum key distribution using EPR
pairs.}\label{choochoo} 
\end{center} 
\end{figure}

The protocol is illustrated in Figure~\ref{choochoo}.
We assume Alice and Bob start off sharing a large number $n$ of Bell
state pairs, which we will assume are the singlet state $|\psi^-\ran$.
These pairs could be generated by a central ``quantum software'' distributor 
and delivered to Alice and Bob, or Alice could manufacture them and
transmit one half of each pair to Bob.  Once again, Alice and Bob have 
written down strings of $n$ digits, but this time the digits could be
0, 1 or 2 --- and once again, 0, 1 and 2 correspond to different
nonorthogonal measurement bases (these bases are illustrated for spin-1/2
measurements in Figure~\ref{choochoo}).  The outcome of any
measurement is either ``$+$'' or ``$-$''.

We denote by $P_{-+}(i,j)$ the probability that Alice gets a ``$-$'' when
measuring in the $i$ basis at the same time as Bob finds ``$+$'' in the
$j$ basis.  The {\em correlation coefficient} between two measurement
bases is 
\be
E(i,j) = P_{++}(i,j) + P_{--}(i,j) - P_{+-}(i,j) - P_{-+}(i,j).
\end{equation}
If we do the calculations we find for our choice of bases
\be
E(i,j) = -{\bf a}_i \cdot {\bf b}_j,
\end{equation}
where ${\bf a}_i$ and ${\bf b}_j$ are the direction vectors of the ``$+$'' 
outcomes of Alice's $i^{\rm th}$ and Bob's $j^{\rm th}$ basis
respectively.  (For example, ${\bf a}_0 = (0,1)$ and ${\bf b}_0 =
(\frac{1}{\sqrt{2}},\frac{1}{\sqrt{2}})$.)   So we define the
quantity
\be
S = E(0,0) - E(0,2) + E(2,0) + E(2,2)
\end{equation}
and calculate that for the bases illustrated in Figure~\ref{choochoo}, 
$S=-2\sqrt{2}$. The quantity 
$S$ is a generalisation, due to Clauser, Horne, Shimony and Holt
\cite{Clauser:1969}, of the correlation function
originally used by Bell\footnote{Clauser {\it et al} showed that a local
deterministic theory must have $|S|\leq 2$. Thus this particular
measurement outcome, if it occurs, cannot be replicated by a local
deterministic theory.}.

Once Alice and Bob have made all their measurements on the EPR pairs,
they announce their measurement bases (i.e. their random strings of 0, 
1 and 2).  The {\em only cases} in which their measurements will agree
are (i) Alice measured in basis 1 and Bob in basis 0; (ii) Alice
measured in basis 2 and Bob basis 1.  They will thus know that when
these measurements were made they got anti-correlated results, and
they can use this to generate a secret key.  And how can they tell if
someone was eavesdropping?  This is where the quantity $S$ comes in.
Notice that $S$ is defined only for cases where the measurement bases
were different.  Bob can publicly announce all his measurement results 
for when both their bases were either 0 or 2, and Alice can use this
to calculate the correlation functions and hence $S$.  Since $S$ is
bounded by $-2\sqrt{2}\leq S \leq 2\sqrt{2}$, any intereference by an
eavesdropper will reduce the (anti)correlation between Alice and Bob's 
measurements and hence be statistically detectable (although this is
quite difficult to prove; see \cite{Inamori:2000}).  In this case
Alice will abort the key distribution.

A useful aspect of this protocol is that the ``element of reality''
from which the secret key is constructed does not come into existence
until Alice and Bob make their measurements.  This contrasts with the
BB84 scheme of Section~\ref{sneeze}, in which Alice is required to
impose a known state onto the quantum system she sends to Bob.
Practically, this means that the BB84 protocol is vulnerable to
someone breaking into Alice's safe and (without being detected by
Alice) stealing the preparation basis information --- in which case
the eavesdropper simply needs to wait for Alice and Bob's measurement
basis announcement and can calculate the secret key.  In Ekert's
protocol Alice and Bob can hold onto their EPR pair until they need a 
key, then use the protocol, send the message and destroy the key all
within a short time.

\subsection{Quantum Superdense Coding}

Quantum information can also be used to send classical information
more efficiently.  This is known as quantum superdense
coding, and is due to Bennett and Wiesner \cite{Bennett:1992}.

Suppose Alice wanted to send two bits of classical information to
Bob.  One way she could do this is by sending him two electrons with
information encoded in their polarisation states --- and as we've seen 
already, two bits is the maximum amount of classical information she
can send in this way.  One method of doing this would be, for example, to 
send one of the states $|\psi^+\ran$, $|\psi^-\ran$, $|\phi^+\ran$ or
$|\phi^-\ran$ each with probability $1/4$.

Suppose instead that Alice and Bob share the Bell state
$|\phi^+\ran$.  Define the following local actions which Alice can
perform on her side of the pair: $U_0=\unit\otimes\unit$,
$U_1=\sigma_x\otimes\unit$, $U_2=\sigma_y\otimes\unit$ and
$U_3=\sigma_z\otimes\unit$.  The effects of the operations on the
shared EPR pair can be calculated:
\begin{eqnarray*}
& U_0 |\phi^+\ran = |\phi^+\ran &
U_1 |\phi^+\ran = |\psi^+\ran \\
& U_2 |\phi^+\ran = -i|\psi^-\ran &
U_3 |\phi^+\ran = |\phi^-\ran.
\end{eqnarray*}

A method for sending information suggests itself: Alice and Bob take
out their old EPR pairs, which they've had for years now, and Alice
performs one of these operations on her side of the pair.  She then
sends her one electron to Bob, who makes the measurements
$\sigma_x^A\sigma_x^B$ and $\sigma_z^A \sigma_z^B$ on the two
electrons now in his possession.  As mentioned previously
(Section~\ref{bellow}), all the
Bell states are simultaneous eigenstates of these operators
corresponding to {\em different} eigenvalues, so these measurements
completely identify the state.  From this outcome he can infer which
of the four local operations Alice used.  He has received {\em two}
bits of information despite the fact that Alice only sent him {\em
one} two-level system!

One might argue that this still requires the transmission of {\em
both} halves of the EPR pair and so we don't really gain anything ---
we send two two-level systems and get two bits of information.
The important 
point here is that the shared EPR pair is a prior resource which
exists before the classical information needed to be communicated.
Once Alice knew the outcome of the soccer match she wanted to report
to Bob, she only needed to send him one qubit.

What is the maximum ``compression factor'' possible?  And what other
states can be used for superdense coding?  Hausladen {\it et al}
\cite{Hausladen:1996} showed that any pure entangled state can be used
for superdense coding, and that the maximum information that can be
superencoded into a bipartite state $|\Psi_{AB}\ran$ is
$E(|\Psi_{AB}\ran)+\log N$, where $N$ is the dimension of the system
Alice sends to Bob.  Clearly by sending one $N$-state system Alice can 
communicate $\log N$ bits of information; the excess from superdense
coding is exactly equal to the entanglement of the state.

\subsection{Quantum Teleportation}

What is so special about collective measurements, and can we draw a
line between what is possible with separate and with joint measurements?
As mentioned earlier, no local (separated) measurements on a system
in one of the Bell states can identify {\em which} Bell state it is.
Similarly, Peres and Wootters \cite{Peres:1991} showed an explicit
situation in which the accessible information from a set of states
appeared to be higher using a joint measurement than using any set of
separated measurements.  In an 
attempt ``to identify what other resource, besides actually being in
the same place, would enable Alice and Bob to make an optimal
measurement'' \cite[p. 1071]{Bennett:1999} on a bipartite system,
Bennett {\it et al} stumbled across the idea of quantum teleportation.

We suppose, as before, that Alice and Bob share an EPR pair which we
suppose to be the state $|\phi^+\ran_{AB}$.  Alice is also holding a
quantum state $|\mu\ran = a\spinupvec_C + b\spindownvec_C$ of another qubit
which we call system $C$.  Alice might not know $a$ and $b$, and
measuring the state will destroy the quantum information.  However,
she notices something beautiful about the state of the three particles 
$A$, $B$ and $C$ considered together:
\begin{eqnarray}
|\phi^+\ran_{AB}|\mu\ran_C &=& \frac{1}{\sqrt{2}}\big( a|\!\UP_A\UP_B\UP_C\ran
+ a|\!\DN_A\DN_B\UP_C\ran + b|\!\UP_A\UP_B\DN_C\ran +
b|\!\DN_A\DN_B\DN_C\ran\big) \nonumber \\ 
&=& 
\begin{array}[t]{cccc}
\frac{1}{2\sqrt{2}}\bigg\{ && |\phi^+\ran_{AC}\;\big(a\spinupvec_B +
b\spindownvec_B\big) & \\[-2mm]
& + & |\phi^-\ran_{AC}\;\big(a\spinupvec_B - b\spindownvec_B\big) & \\
& + & |\psi^+\ran_{AC}\;\big(b\spinupvec_B + a\spindownvec_B\big) & \\[-2mm]
& + & |\psi^-\ran_{AC}\;\big(b\spinupvec_B - a\spindownvec_B\big) & \bigg\}.
\end{array}\label{whatnow}
\end{eqnarray}
Thus if Alice performs a Bell basis measurement on the system $AC$,
she projects the system into a state represented by one of the terms
in Eqn~\ref{whatnow}.  In particular, Bob's half of the EPR pair will
be projected into something that looks similar to the original state of
$|\mu\ran_C$ --- in fact, by applying one of the operators $\unit,
\sigma_x, \sigma_y, \sigma_z$ Bob can rotate his qubit to be {\em
exactly} $|\mu\ran$!

The protocol runs as follows: Alice measures the system $AC$ in the
Bell basis, and gets one of four possible outcomes, and communicates
this outcome to Bob.  Once Bob knows the result of Alice's
measurement --- and not before this --- he performs the appropriate
operation on his qubit and ends up with the state $|\mu\ran$.

There are many remarkable features to this method of communication.
Firstly, as long as Alice and Bob can reliably store separated EPR
particles and can exchange classical information, they can perform
teleportation.  In particular, there can be all manner of ``noise''
between them (classical information can be made robust to this noise)
and an arbitrarily large distance.  Alice doesn't even need to know
where Bob is, as she would if she wished to send him a qubit directly
--- she merely needs access to a broadcasting channel.  

Secondly, the speed of transmission of the state of a massive particle
is limited only by the speed of classical communication --- which in
turn is limited only by the speed of light.  This is an intriguing
subversion of the idea that massive particles must travel slower than
light speed, although of course, only the {\em state} is transmitted.
We could even imagine a qubit realised as an electron in system $C$
and as a photon in system $B$, in which case we teleport a massive
particle onto a massless particle.  Quantum information is a highly
interchangeable resource!

Thirdly, we have developed a hierarchy of information resources.  The
lowest resource is a classical bit which cannot generally be used to
share quantum information and which cannot be used to
create entanglement (which is a consequence of Bell's theorem).  The highest
resource is a qubit, and an ebit is intermediate.  This classification 
follows from the fact that if Alice and Bob can reliably communicate
qubits, they can create an ebit entanglement (Alice simply creates an
EPR pair and transmits half to Bob); whereas if they share an ebit
they further require classical information to be able to transmit qubits.

Much work has been performed in attempting to implement quantum
teleportation in the laboratory, with varying claims to success.  Some 
of the techniques used are cavity QED \cite{Furusawa:1998}, parametric
down-conversion of laser light \cite{Bouwmeester:1997} and NMR
\cite{Nielsen:1998}.  Experiments are however fraught with
complications, both practical and theoretical; see
\cite{Braunstein:1999} and references therein.

\paragraph{Entanglement swapping}
Quantum teleportation can also be used to swap entanglement
\cite{Bose:1998}.  Suppose Alice and Bob share the EPR pair
$|\phi^+\ran$ and Charlie and Doris share another $|\phi^+\ran$.  Then 
the state they have can be rewritten as
\begin{eqnarray}
|\phi^+\ran_{AB}|\phi^+\ran_{CD} &=& \frac{1}{2}\big( |\!\UP_A\UP_B
\UP_C\UP_D\ran + |\!\UP_A\UP_B \DN_C\DN_D\ran + |\!\DN_A\DN_B \UP_C\UP_D\ran
+ |\!\DN_A\DN_B \DN_C\DN_D\ran\big) \nonumber\\
&=& \frac{1}{2}\big( |\phi^+\ran_{AD}|\phi^+\ran_{BC} +
|\phi^-\ran_{AD}|\phi^-\ran_{BC} \nonumber\\
&& \; + |\psi^+\ran_{AD}|\psi^+\ran_{BC} + |\psi^-\ran_{AD}|\psi^-
\ran_{BC} \big).
\end{eqnarray}
So if Bob and Charlie get together and make a Bell basis measurement
on their qubits, they will get one of the four outcomes with equal
probability.  Once they communicate this classical information to
Alice and Doris, the latter will know that they are holding an EPR
pair --- and Alice can convert it to $|\phi^+\ran$ by a local
rotation.

Of course, all that Bob has done is to teleport his entangled state
onto Doris' system, so entanglement swapping is not much different to
straightforward teleportation.  But entanglement swapping will be
important later when we discuss quantum channels.

\section{Quantum Computing}

The theory of computing dates from work by Turing and others in the
1930's.  However the idea of a computer, as enunciated by Turing,
suffered from the same shortcoming as classical information: there was 
no physical basis for the model.  And there were surprises in store
for those who eventually did investigate the physical nature of
computation.

\begin{figure}
\begin{center}
\epsfig{file=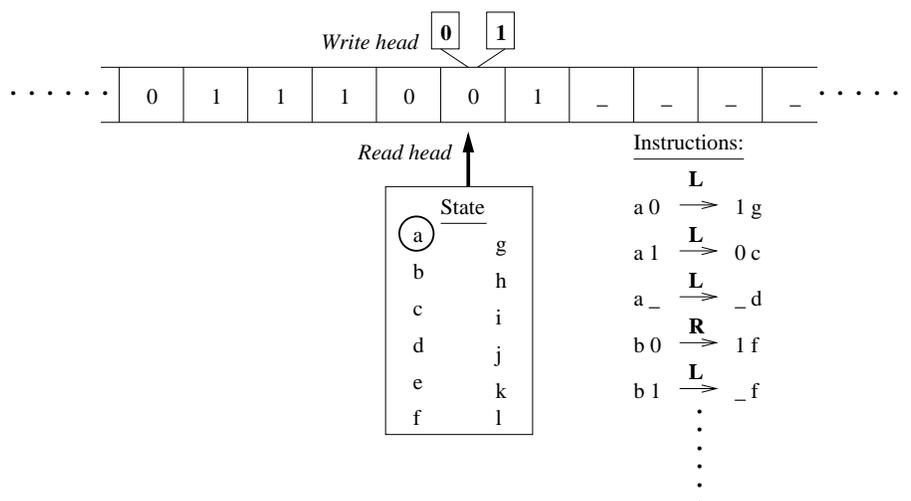, width=12cm}
\caption{A Turing machine.}\label{nobrainer}
\end{center} 
\end{figure}

Turing wanted to formalise the notion of a ``computation'' as it might 
be carried out by a mathematician in such a way that a machine could
carry out all the necessary actions.  His model was a mathematician
sitting in a room with access to an infinite amount of paper, some of
which contains the (finite) description of the problem he is to
solve.  The mathematician is supposed to be in one of a finite number
of definite states, and he changes state only after he reads part of
the problem statement --- and as he changes state he is allowed to
write on the paper.  A {\em Turing machine} (illustrated in
Figure~\ref{nobrainer})
is the machine version of this mathematician.  The Turing machine (TM) has
a finite number of internal states, here labeled $a,b,\ldots,l$; a
``read'' head; and a ``write'' head.  An infinite strip of paper, with
discrete cells on it, runs between the read and write heads; each cell 
can either contain a 0, a 1 or a blank.  Initially the paper is completely
blank except for a finite number of cells which contain the
problem specification.  A cycle of the machine consists of reading the 
contents of the current cell, writing a new value into the cell,
changing state, and moving the tape left or right.  For example,
consider the instruction set shown in the figure.  The machine is
currently in the state $a$ and is reading the value 0; according to
the instructions given, the machine should write a 1, change into state 
$g$ and shift the tape to the left so that the read head sees a 1.

The Turing machine can be used to define the {\em complexity} of
various problems.  If we have a TM which, say, squares any
number it is given, then a reasonable measure of complexity is how
long (how many cycles) it takes for the Turing machine to halt with
the answer.  This of course depends on the size of the input, since
it's a lot harder to square 1564 than 2.  If the
input occupied $n$ cells and the computation took $3n^2$ steps
(asympotically, for large $n$) we refer to the computation as having $n^2$
complexity or more generally {\em polynomial-time} complexity.  Of course, 
the complexity of a problem is identified as the minimum complexity
over all computations which solve that problem.  The two chief divisions
are polynomial-time and exponential-time problems.

Happily, complexity is relatively machine-independent: any other
computational device (a cellular automaton, or perhaps a network of
logic gates like in a desktop PC) can be simulated by a TM
with at most polynomial slowdown.  And in fact there exist {\em
universal} Turing machines that can simulate any other Turing machine
with only a constant increase in complexity --- the universal TM just
requires a ``program'' describing the other TM.  Turing's theory is
very elegant and underpins most of modern computing.

However, we can also start to ask what other resources are required
for computing.  What is the {\em space} complexity of a problem ---
how many cells on the strip of paper does the computation need?
Frequently there is a trade-off between space and time in a particular 
computation.  More practically, we can ask what are the energy
requirements of a computation \cite{Landauer:1961}?  What must the
accuracy of the physical processes underlying computation be?
Questions like these brought complexity theorists to the limits of 
classical reasoning, beyond which the ideas of quantum theory {\em
had} to be employed for accurate answers.

Feynman \cite{Feynman:1982} first noted that simple quantum systems cannot be 
efficiently simulated by Turing machines.  This situation was turned
on its head by Deutsch when he suggested that this 
is not a problem but an opportunity; that the notion of efficient
computing is not entirely captured by classical systems. He
took this idea and proposed first a quantum Turing 
machine \cite{Deutsch:1985} (which can exist in superpositions of
states) and then the quantum computational network \cite{Deutsch:1989}
which is the most prevalent model 
for quantum computing today.  Several problems were discovered which could
be solved faster on quantum computers (such as the Deutsch-Jozsa
problem \cite{Deutsch:1992}), but the ``silver bullet'' application
was an algorithm for polynomial-time factorisation of integers
discovered by Shor \cite{Shor:1995} (for a 
readable account see \cite{Ekert:1996}).  The assumed
exponential complexity of factorisation is the basis for most of
today's public-key encryption systems, so a polynomial algorithm for
cracking these cryptosystems generated a lot of popular interest.

There are many questions which arise about the usefulness of quantum
computation.  For example, how much accuracy is required?  What types
of coherent manipulations of quantum systems are necessary for
completely general quantum computation?  And which problems can be
solved faster on a quantum computer as opposed to a classical
computer?

The first question here requires knowledge of quantum error correction 
techniques, and will be discussed in Section~\ref{damnitall}.

\begin{figure}
\begin{center}
\epsfig{file=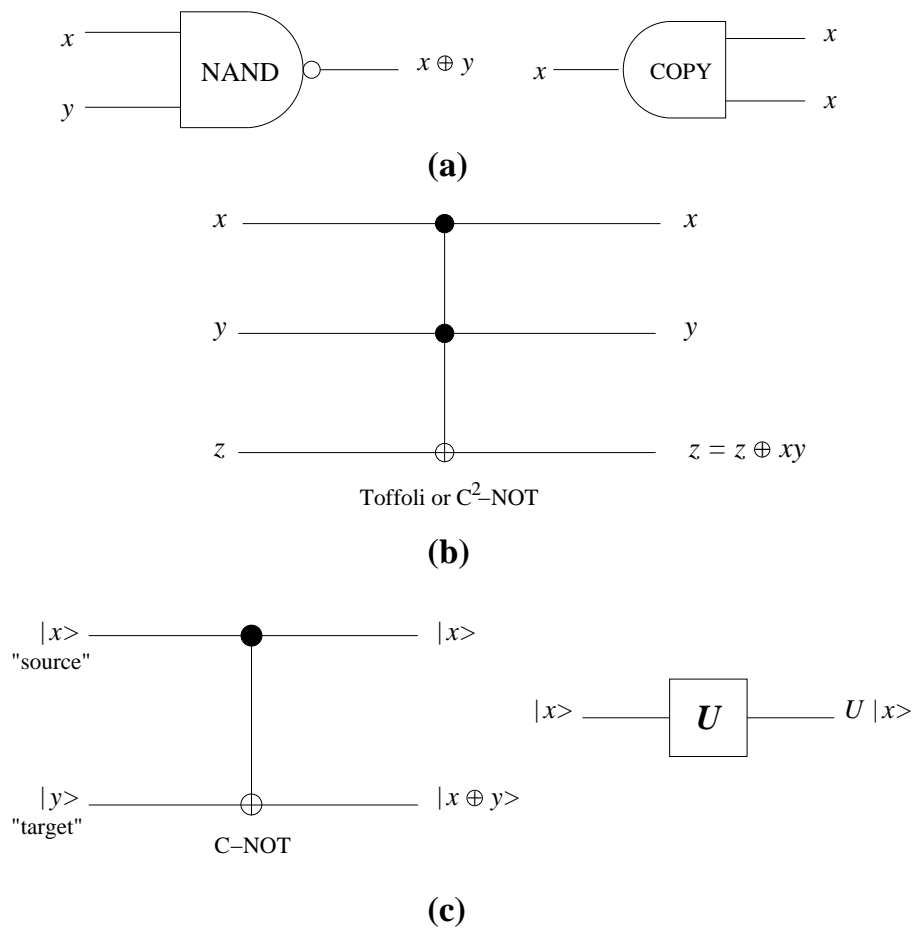, height=5in}
\caption{Gates for universal computation: (a) for irreversible classical
computation (b) for reversible classical computation (c) for quantum
computation.}\label{gates}
\end{center} 
\end{figure}

There is a very simple and intriguing answer to the second question.
To compare, we first consider what operations are required for
universal {\em classical} computation.  If we allow our classical
computation to be logically {\em irreversible} (i.e. we are allowing
information to be discarded) then it suffices\footnote{These gates are 
all universal iff a supply of bits in a standard state ($0$ or
$|0\ran$) are available as well.} to have the operations
COPY and NAND, which are illustrated in Figure~\ref{gates}.
In these diagrams, the bits being operated on ``travel'' from left to
right through the gate, and have an effect described by the outputs;
for instance if the Boolean values $x=1$ and $y=1$ are input into a
NAND gate, the output will be $x\oplus y = 1\oplus 1 = 0$.  Other sets 
of gates are also universal.  For classical {\em reversible} computation,
the three-input Toffoli gate is universal.  This gate is also called a 
``controlled-controlled-NOT'' (C$^2$-NOT), since the ``target''
bit is negated if and only if the two ``source'' bits are both 1;
otherwise the target bit is untouched.

In quantum computing we are no longer dealing with abstract Boolean
values, but with two-level quantum states whose states we denote $|0\ran$ and 
$|1\ran$.  Note that mathematically, any Hilbert space can be
considered as a subspace of the joint space of a set of qubits, so by
considering operations on qubits we are not losing any generality.
The most general quantum evolution is unitary, so the question is:
What type of qubit operations are required to implement an arbitrary
unitary operator on their state space?  Deutsch \cite{Deutsch:1989}
provided an answer: all we need is a ``controlled-controlled-$R$'',
where $R$ is {\em any} qubit rotation by an irrational fraction.  A
more convenient set \cite{Barenco:1995} is that shown in
Figure~\ref{gates}(c), the set comprising the C-NOT and
arbitrary single qubit rotations (such rotations are elements of
$U(2)$ and can be represented by four real parameters).  And
interestingly, 
any single unitary operation acting on two or more qubits is {\em
generically} universal \cite{DiVincenzo:1995}.  While this is a
mathematically simple remark, physically this means that any unitary
operation in $d$ dimensions can be carried out using a single fixed ``local''
rotation --- where by local we mean ``operating in exactly four
dimensions'' (two qubits).

The reader may object that certain gate sets will be more practical
than others; in particular, attempting to compute using just one
two-qubit gate looks like an uphill battle compared with, say, using
the gate set shown in Figure~\ref{gates}(c).  However, the issue of
complexity is implicit in the definition of ``universal'': a gate set
is only universal if it can simulate any other gate set with
polynomial slowdown; indeed, in the analysis of any proposed quantum
algorithm, the complexity of the required gates must be taken into
account.  Such issues are addressed in e.g. \cite{Deutsch:1989},
\cite{Barenco:1995}.

The answer to the final question posed above is difficult, and is
related to the open problem in computing theory of classifying
problems into complexity classes.  Not all classically
exponential-time problems can be reduced to polynomial-time on a
quantum computer --- in fact there appear to be very few of these,
although many problems admit a square-root or polynomial speed-up.
General mathematical frameworks for the quantum factorisation problem
are given by Mosca and Ekert \cite{Mosca:1998} and for the quantum
searching problem by Mosca \cite{Mosca:1998b}. The relation of current
quantum algorithms (in the network model mentioned above) to the
paradigm of multi-particle interferometry is given by Cleve {\it et
al} \cite{Cleve:1998}.

\section{Quantum Channels}

The idea of a quantum channel was implicit in the discussion of
quantum noiseless coding: the channel consisted of Alice sending Bob a 
``coded'' physical system through an effectively noiseless channel.
This is an idealisation; any means that Alice and Bob share for
transmitting information is subject to imperfection and
noise, so we should analyse how this noise can be characterised and
whether transmission of intact quantum states can be achieved in the
presence of noise.  Ideally we would like to arrive at a ``quantum
noisy coding theorem''.

Note that, just as in the classical case, error correction in quantum
channels is not solely important for the obvious task of Alice sending 
information to Bob.  Coding techniques are also used to protect
information which is ``transmitted'' through time --- as for example
the data stored in computer memory, which is constantly refreshed and
checked using parity codes.  If we are to implement useful quantum
computations, we will need similar coding techniques to prevent the
decoherence (loss into the environment) of the quantum information.

It is instructive to look first at how classical communication and
computation are protected against errors.  In general the
``imperfection and noise'' can be well characterised by a stochastic
process acting on the information, whose effect is to flip signal $i$
to signal $j$ with probability $p(j|i)$.  We can remove the errors
through mathematical or physical techniques; by using redundancy or by 
engineering the system to be insensitive to noise.  This insensitivity 
is a result of combining amplification with dissipation
\cite{Steane:1998}: the amplification provides a ``restoring force''
on the system, while dissipation causes oscillations away from
equilibrium to be damped.  This is the primary source of reliability
in a modern classical computer.  Unfortunately, it is precisely these
types of robustness which are not available in quantum systems; we
cannot amplify the state by copying it, and we cannot subject it to
dissipative (non-unitary) evolutions.

What about redundancy?  This plays a more important role in classical
communication than in computing.  Error-correcting codes --- a vast
subject of research in itself \cite{MacWilliams:1978} --- were at
first thought to be inapplicable to quantum systems for the same
reason that amplification is not allowed.  However, once it was
realised that redundancy could be achieved by encoding the state of a
qubit into a much larger space (say that of five qubits) which is
more robust against decoherence, the techniques of classical coding
theory could be used to discover which subspaces should be used.  A
quantum theory of error correction has thus grown with techniques
parallel to the classical theory, as will be discussed in
Section~\ref{damnitall}.

\subsection{Entanglement Purification}

One way to transmit quantum information is to teleport it.  In a
certain sense this is a circular argument, because for teleportation
Alice and Bob require shared entanglement --- a resource which
requires transmission of coherent quantum information in the first
place.  Fortunately, Bennett {\it et al} have had a word with Alice
and Bob and told them how to go about getting almost pure Bell states
out of a noisy channel \cite{Bennett:1996}.

\begin{figure}
\begin{center}
\epsfig{file=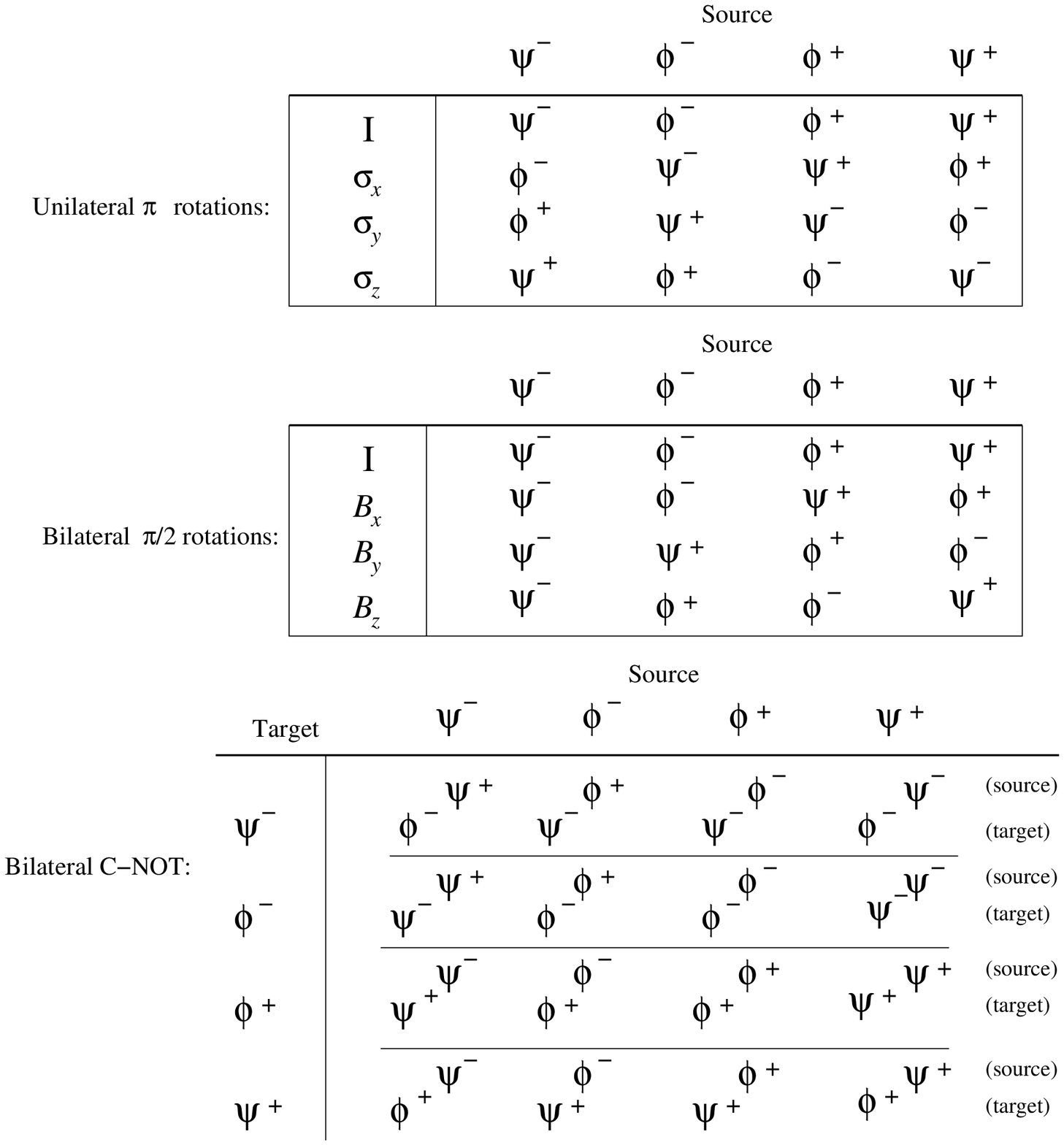, height=12cm}
\caption{The unilateral and bilateral operations used in entanglement
purification (after \cite{Bennett:1996}).}\label{operations}
\end{center} 
\end{figure}

Before we discuss this, we need to know what operations are available
to Alice and Bob.  It turns out that all we will need for this
entanglement purification protocol (EPP) is three types of operations: (i)
unilateral rotations by Alice using the operators $\sigma_x, \sigma_y, 
\sigma_z$; (ii) bilateral rotations, represented by $B_x, B_y, B_z$
which represent Alice and Bob both performing the same rotation on
their qubits; and (iii) a bilateral C-NOT, where Alice and Bob both
perform the operation C-NOT shown in Figure~\ref{gates}(c), where both 
the members of one pair are used as source qubits and both qubits from 
another pair are used as target qubits.  The effects of these
operations are summarised in the table in Figure~\ref{operations}
(complex phases are ignored in this table).
For example, if Alice and Bob hold a pair $AB$ of qubits in the state
$|\phi^+\ran$ and either Alice or Bob (but {\em not} both) perform the 
unitary operation $\sigma_y$ on their half of the pair, they have
performed a unilateral $\pi$ rotation.  According to the table, pair
$AB$ will now be described by the state $|\psi^-\ran$.  If they had
both performed the same $\sigma_y$ operation on their respective
qubits (a bilateral $B_y$ operation) then we see from the tabel that
the state of $AB$ would remain $|\phi^+\ran$.

\begin{figure}
\begin{center}
\mbox{\epsfig{file=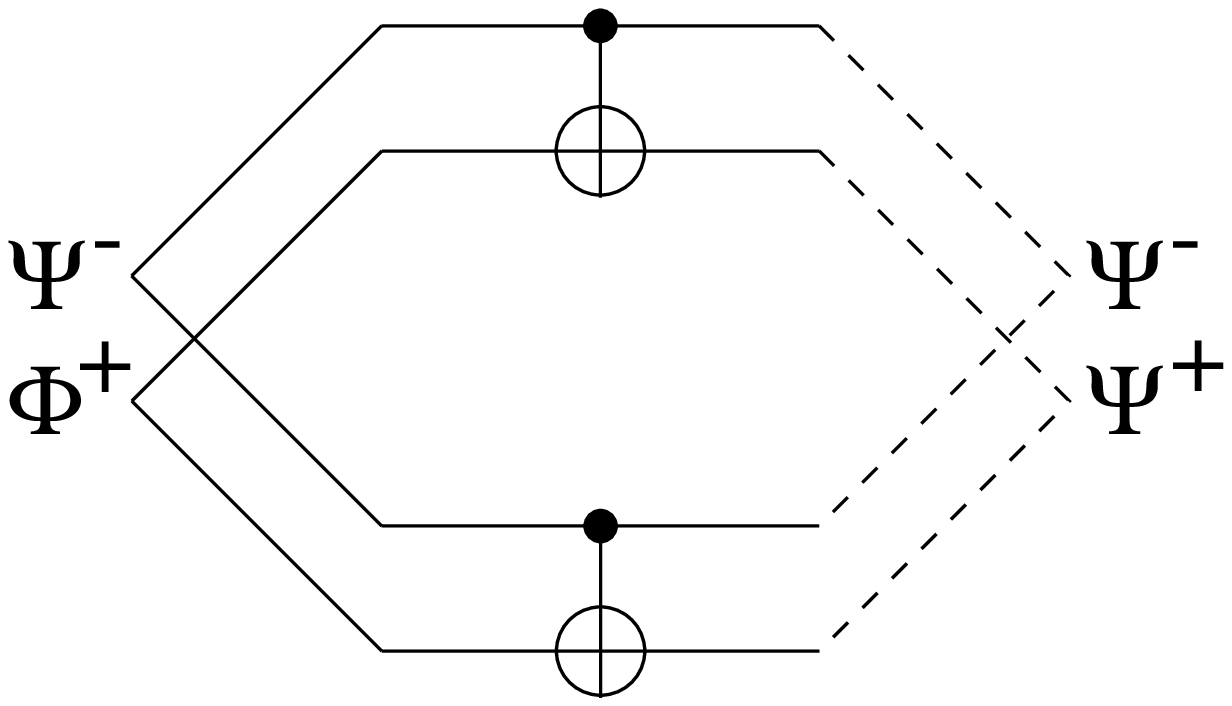, height=3cm}\vspace*{2cm}}
\hspace{2cm}
\mbox{\epsfig{file=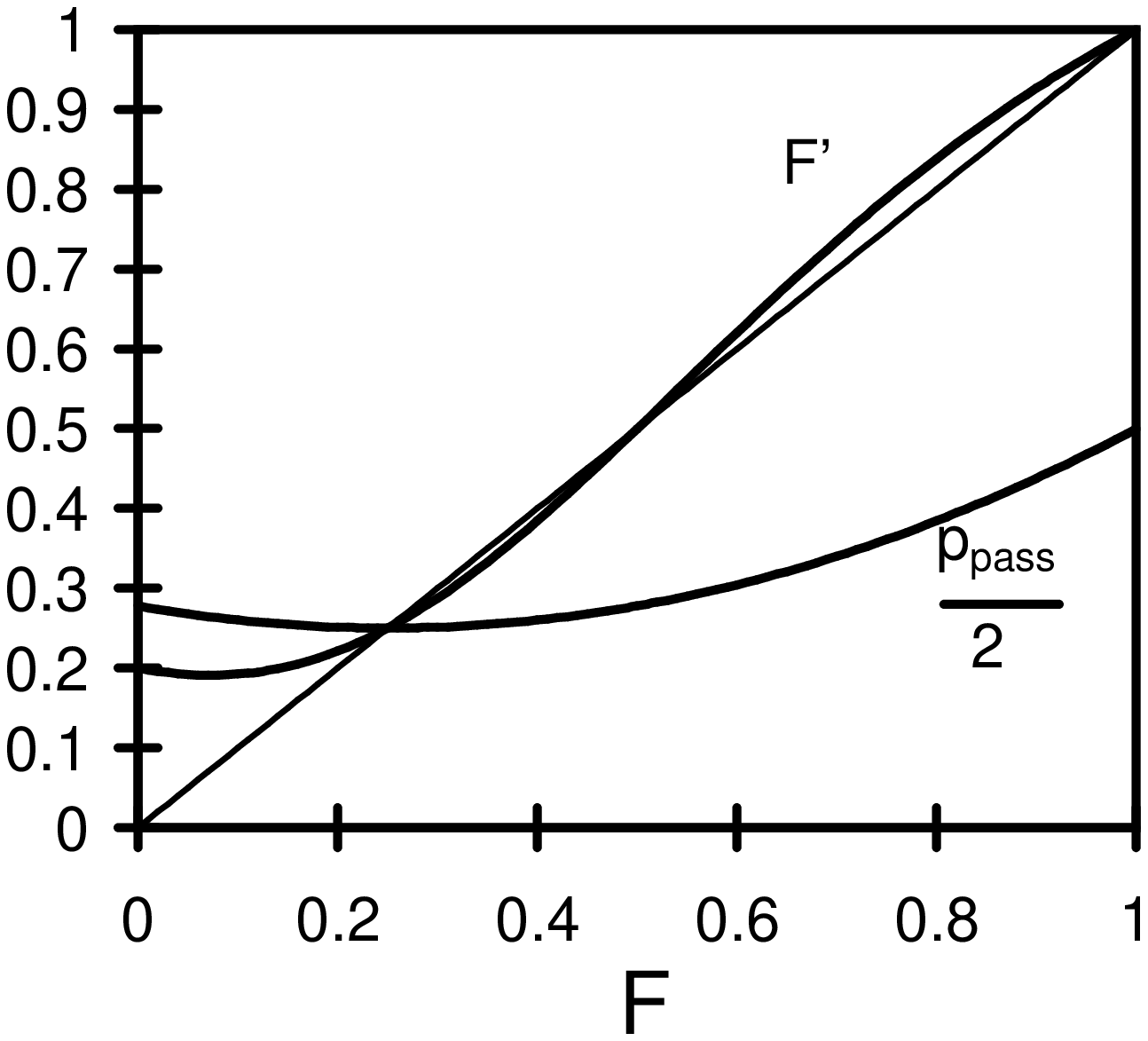, height=5cm}}
\caption{An illustration of the bilateral C-NOT operation, and its
effect on the fidelity of the Werner state (taken from
\cite{Bennett:1996}).}\label{dumbell} 
\end{center} 
\end{figure}

Note how the bilateral C-NOT is implemented: Alice prepares systems
$AB$ and $CD$ in Bell states, and sends $B$ and $D$ to Bob.  Bob uses
$B$ as source and $D$ as target in executing a C-NOT, while Alice uses 
$A$ as source and $C$ as target (this is illustrated in
Figure~\ref{dumbell}).  In general, both pairs $AB$ and $CD$ 
will end up altered, as shown in the table in Figure~\ref{operations}.
For example, if the 
source pair $AB$ is prepared as $|\psi^+\ran$ and the target pair $CD$ 
as $|\phi^+\ran$, then we conclude from the table that after this
operation the pair $AB$ remains in the state $|\psi^+\ran$, but the
pair $CD$ is now described by $|\psi^+\ran$ as well.

The most general possible way of describing the noise is as a
superoperator $\$$ acting on the transmitted states.  We suppose that Alice 
manufactured a large number of pairs of systems in the joint state
$|\psi^-\ran$, and when she sent half of them through the channel to
Bob they were corrupted by the joint superoperator $\unit\otimes \$$
(Alice preserves her half perfectly).  We are only interested in the
state $\rho_{AB}$ which emerges.  This matrix, expressed in
the Bell basis, has three parts which behave differently under
rotation: the $|\psi^-\ran\lan\psi^-|$ behaves as a scalar, 3 terms of 
the form $|\psi^-\ran\lan \nu|$ ($|\nu\ran$ is one of the Bell states) 
which behave as a vector, and the $3\times3$ block which behaves as a
second rank tensor.  Bennett {\it et al} showed that
through a random operation of {\em twirling} any state $\rho_{AB}$ can 
be brought into the so-called {\em Werner form}\footnote{This is not
exactly the Werner form; usually the singlet state $|\psi^-\ran$ is the 
distinguished state.  The state in Eqn~\ref{werner} differs from a
standard Werner state by a simple unilateral rotation.}: 
\be\label{werner}
W_F = F|\phi^+\ran\lan\phi^+| + \frac{1-F}{3}\left(
|\psi^+\ran\lan\psi^+| + |\psi^-\ran\lan\psi^-| + |\phi^-\ran\lan\phi^-|
\right).
\end{equation}
This ``twirling'' is done by performing a type of averaging: for each
EPR pair she generates, Alice chooses randomly one element from a set
of specially chosen combinations of the operations $B_x, B_y, B_z$,
tells Bob which one it is, and they perform the specified operation.
The ``twirls'' are chosen so that the second rank tensor representing
the states, when acted upon by these random twirls, becomes
proportional to the identity and the vector components disappear.  The 
idea is similar to motional averaging over the directional properties of a
fluid, where all vector quantities are zero and tensor properties can
be described by a single parameter.

Alice and Bob can now consider their ensemble of EPR pairs as a
classical mixture of Bell states, with proportion $F$ of the
state $|\phi^+\ran$ and $\frac{1-F}{3}$ of each of the other states.  This is
enormously useful because they can employ classical error
correction techniques to project out and find pure singlet
states. There is a price paid, though: we have introduced additional
uncertainty, and this is reflected by the fact that $S(W_F) >
S(\rho_{AB})$.  It can be shown that the protocol described below
works without the twirl, but this is a more subtle argument.  Note
that $F$ is the fidelity between the Werner state and the state $|\phi^+\ran$:
$F=\lan\phi^+|W_F|\phi^+\ran$, where the fidelity of a transmitted 
state was defined in Eqn~\ref{noddie}.

After this pre-processing stage, Alice and Bob perform the following
steps:
\begin{enumerate}
\item They select the corresponding members of two pairs 
from their ensemble.
\item They perform a bilateral C-NOT from the one pair to the other
(illustrated in Figure~\ref{dumbell}).
\item They make (local) measurements on the {\em target} pair to
determine its amplitude bit.
\item If the amplitude bit is 1 (i.e. if the target pair is in one of
the $\phi$ states) then both pairs are discarded.  If the amplitude
bit is 0, which occurs with probability $p_{\rm pass} =
F^2+\frac{2}{3}F(1-F) + \frac{5}{9}(1-F)^2$, the source pair is in the
state $W_{F'}$ with 
\be
F' = \frac{F^2 + \frac{1}{9}(1-F)^2}{F^2+\frac{2}{3}F(1-F) +
\frac{5}{9}(1-F)^2}
\end{equation}
Proof of this relation is left as an exercise for the reader, using
the information from Figure~\ref{operations} and Eqn~\ref{werner}.
\end{enumerate}
How exactly does this help us?  Well the answer is shown in
Figure~\ref{dumbell}:  for all starting fidelities $F>1/2$, we have
that $F'>F$.  So with probability of success $p_{\rm pass}\leq 1/2$ we
will have succeeded in driving the Werner state closer to a pure Bell
state.

We may want to know how many EPR pairs $m$ can be distilled from $n$
copies of the intial state $\rho_{AB}$ as $n$ gets large; that is, we
define the yield of a protocol $P$ to be
\be
D_P(\rho_{AB}) = \lim_{n\rightarrow\infty} m/n.
\end{equation}
Unfortunately the yield of this protocol is zero.  At each step we are
throwing away at least half of 
our pairs, so this process is very wasteful.  However, this protocol
can be used in combination with other protocols to give a positive
yield.  In particular, Bennett {\it et al} \cite{Bennett:1996}
describe a variation of a classical hashing protocol which gives good
results.

An important feature to note about these protocols is that they
explicitly require two-way communication between Alice and Bob, and so 
we may call them 2-EPP.   What about 1-EPP --- those protocols which
use only communication in one direction?  There is certainly an
important distinction here.  1-EPP protocols can be used to purify
entanglement through time, and thus to teleport information forward in 
time.  This is exactly what one wishes to achieve using quantum
error-correcting codes (QECC).  On the other hand, a 2-EPP can be used 
for communication but not for error-correction.  In fact, we can
define two different {\em rates} of communication using EPP, for a
particular channel desribed by superoperator $\$$:
\begin{eqnarray}
D_1(\$) &=& \sup_{\mbox{\scriptsize P is a 1-EPP}} D_P(\rho_{AB}) \\
D_2(\$) &=& \sup_{\mbox{\scriptsize P is a 2-EPP}} D_P(\rho_{AB}).
\end{eqnarray}
(Note that one EPR pair allows perfect transmission of one qubit,
using teleportation; hence the number of Bell states distilled can be
interpreted as a rate of transmission of quantum information.)
Clearly, since all 1-EPP are also 2-EPP, we have $D_2\geq D_1$.  It can 
be shown that for some mixed states $\rho_{AB}$ --- or equivalently,
for some channels $\$$ --- there is a strict separation between $D_1$
and $D_2$; in fact there are some mixed states with $D_1=0$ and
$D_2>0$ \cite{Bennett:1996}.

We can see now the importance of the entanglement swapping protocol
presented earlier.  For most quantum communication channels, the
fidelity is an exponentially decreasing function of distance $l$.
Ususally this is described by a coherence length $l_0$.  However if we 
divide the length $l$ into $N$ intervals, the fidelity change over each
interval is $\exp(-l/Nl_0)$, which can be made as close to unity as
required.  Using EPP we can improve the fidelity over each interval as 
much as required and then perform the entanglement swapping protocol
approximately $\log_2 N$ times to achieve high fidelity entanglement
between the beginning and end points \cite{Briegel:1999}.  Using this
technique, the fidelity can be made into a polynomially decreasing
function of distance, which shows that is is possible in principle for
us to implement long-distance quantum communication.

Two other EPPs worth noting are the Procrustean method and the Schmidt 
projection method \cite{Bennett:1996b} --- although this more for their
historical interest.  They are both designed to operate on
non-maximally entangled {\em pure} states.

\subsection{Quantum Error Correction}\label{damnitall}

At first sight, quantum errors don't seem amenable to correction ---
at least not in a similar sense to that of classical information.
Firstly, quantum information cannot be copied to make it redundant.
Secondly, quantum errors are analogue (as opposed to digital) errors:
it can make an enormous 
difference if the state $a|0\ran+b|1\ran$ is received as
$(a+\epsilon_1)|0\ran+(b-\epsilon_2)|1\ran$, whereas in classical
error correction we just need to prevent a system in the state `0'
from entering the state `1' and vice versa.

The first problem is countered by using {\em logical} states, which
correspond to the $|0\ran$ and $|1\ran$ states of a two-level system,
but which are in fact states of a much larger quantum system.  In
effect, we can use $n$ qubits with a $2^n$-dimensional Hilbert space,
but only employ two states $|0_L\ran$ and $|1_L\ran$ in
communication.  We aim to design these states so that $a|0_L\ran +
b|1_L\ran$ is recoverable despite errors.

The second problem is solved by realising that errors can be
digitised.  Suppose we make a measurement onto our redundant system,
and we choose our POVM to be highly degenerate, so that the only
information it yields is which subspace the system is in.  If we
choose our redundancy carefully, and make a well-chosen measurement, we
will project (and so discretise) the error and the measurement result
will tell us which error has occurred in this 
projected system.  This will be elaborated below, after we have
discussed the type of errors we are dealing with.

As mentioned previously errors are introduced through interaction
with the environment, which can be represented by the action of a
superoperator on the state we are transmitting.  Rather than using the 
operator-sum representation we will explicitly include the environment 
in our system --- and we lose no generality in representing a
superoperator if we assume that the environment starts out in a pure
state $|e_0\ran$.  Then the state of the code system which starts as
$|\phi\ran$ will evolve as
\be\label{cheerio}
|e_0\ran|\phi\ran \longrightarrow \sum_k |e_k\ran S_k|\phi\ran
\end{equation}
where the $S_k$ are unitary operators and the $|e_k\ran$ are states of 
the environment which are not necessarily orthogonal or normalised.
The $S_k$ are operators on the code system Hilbert space, and form a
complex vector space; so we choose some convenient basis for the space
consisting of operators $M_s$:
\be
S_k = \sum_s a_{ks}M_s
\end{equation}
where $a_{sk}$ are complex constants.  If we define $|e_s'\ran =
\sum_k a_{ks}|e_k\ran$, then the final state of the system+environment in
Eqn~\ref{cheerio} will be $\sum_s |e_s'\ran M_s|\phi\ran$.  Once again
the environment state vectors are not orthogonal or normalised, but we 
can now choose the operators $M_s$ to have convenient properties.

We note first of all that not all the errors can be corrected.  This
follows because not all the states $M_s|\phi\ran$ can be orthonormal,
so we cannot unambiguously distinguish between them to correct $M_s$.
So we do the next best thing: we choose a certain subset $\cM$ (called 
the {\em correctable errors}) of the
operators $M_s$ and seek to be able to correct just these errors.

What do we mean by ``correcting errors''?  Errors will be corrected by 
a {\em recovery operator}: a unitary operator $\cR$ which has the
following effect:
\be
\cR |\alpha\ran M_s|\phi\ran = |\alpha_s\ran |\phi\ran
\end{equation}
where $|\alpha\ran$ and $|\alpha_s\ran$ are states of everything else
(environment, ancilla, measuring apparatus)
and $M_s$ is any correctable error.  In general the states
$|\alpha_s\ran$ are not going to be orthogonal, but the important
feature is that the final state $|\phi\ran$ of the {\em code system}
must not depend on $|\alpha_s\ran$ --- if it did depend on the code
system state then error recovery would not work on linear combinations 
of states on which it does work, so universal error correction would
be impossible.

A {\em quantum error-correcting code} (QECC) is a subspace $V$ of the
$2^n$-dimensional space describing $n$ qubits together with a recovery 
operator $\cR$.  This QECC must satisfy the following
conditions: for every pair of code words $|u\ran, |v\ran\in V$ with
$\lan u|v\ran=0$ and every pair of correctable errors $M_s, M_t\in \cM$,
\begin{eqnarray}
\lan u|M_s^\dagger M_t|v \ran &=& 0 \label{orthog1} \\
\lan u|M_s^\dagger M_t|u \ran &=& \lan\alpha_s|\alpha_t\ran \label{orthog2}
\end{eqnarray}
with Eqn~\ref{orthog2} holding {\em independent} of the code word
$|u\ran$.  These conditions imply that errors can be corrected, as can
be seen in the following way.  For any code words the recovery
operator acts as
\be
\cR |\alpha\ran M_s|u\ran = |\alpha_s\ran|u\ran \hspace{1cm} \cR
|\alpha\ran M_t|v\ran = |\alpha_s\ran|v\ran.
\end{equation}
Taking the inner product on both sides between these we find that
\begin{eqnarray}
\lan u|M_s^\dagger\lan\alpha| \cR^\dagger \cR |\alpha\ran M_t |v\ran &=&
\lan\alpha_s|\alpha_t\ran \lan u|v\ran \\
\Rightarrow \lan u|M_s^\dagger\lan\alpha |\alpha\ran M_t |v\ran &=&
\lan\alpha_s|\alpha_t\ran \delta_{uv}
\end{eqnarray}
where $\cR^\dagger\cR = \unit$ since $\cR$ is unitary.  The
requirements listed above follow if $\lan u|v\ran=0$ or $u=v$
respectvely.

By reversing the above arguments, we find that Eqs~\ref{orthog1} and
\ref{orthog2} are also sufficient for the existence of a QECC
correcting errors $\cM$.  The aim in designing QECC's is thus to
identify the most important or likely errors acting on $n$ qubits,
identify a subspace which satisfies the requirements above, and find
the recovery operator.  This can be a complex task, but fortunately
many of the techniques of classical coding theory can be brought to
bear.  We will briefly sketch the most common ideas behind classical
error correction and indicate how these ideas may be modified for
QECC.

The simplest error model in classical coding theory is the binary
symmetric channel.  Binary refers to the fact that two signals are
used, and symmetric indicates that errors affect both signals in the
same way, by flipping $0\!\leftrightarrow\!1$; in addition, we assume
that the errors are stochastic 
i.e. affect each signal independently and with probability $p$.  In
this case we can apply the algebraic coding techniques pioneered by
Hamming \cite{MacWilliams:1978} to design parity check codes.  The
classical result is that, by using a 
$k$ bit ($2^k$-dimensional) subspace of words of length $n$ (with
$k<n$), we can design a code that corrects up to $t$ errors; the {\em
Hamming bound} on linear codes is
\be
k \leq n - \log_2 \sum_{i=0}^t \big(\!\begin{array}{c} 
n \\[-2mm] i \end{array}\!\big)
\end{equation}
where the term inside the sum is the binomial coefficient.
Essentially the sum in this expression counts the number of different
ways in which up to $t$ errors can occur, and this is a lower bound to
the number $n-k$ of redundant bits required for error correction.

Similar algebraic techniques can be applied to a quantum
generalisation of the binary symmetric channel.  In this case we take
as a basis for the errors the qubit operators $I=\unit, X=\sigma_x$,
$Y=-i\sigma_y$ and $Z=\sigma_z$ (we have introduced a factor of $-i$
into $Y$ for convenience); explicitly,
\be
I = \left( \begin{array}{cc} 1 & 0  \\ 0 & 1  \end{array} \right) \hspace{1cm}
X = \left( \begin{array}{cc} 0 & 1  \\ 1 & 0  \end{array} \right) \hspace{1cm}
Y = \left( \begin{array}{cc} 0 & -1 \\ 1 & 0  \end{array} \right) \hspace{1cm}
Z = \left( \begin{array}{cc} 1 & 0  \\ 0 & -1 \end{array} \right).
\end{equation}
These operators correspond to a bit flip $0\!\leftrightarrow\!1$ ($X$), a 
phase flip $0\!\rightarrow\!0, 1\!\rightarrow\!-1$ ($Z$) or both ($Y=XZ$)
--- a much richer set of errors than in the classical case!  We will
use a system of $n$ qubits, so a typical error will be $M_s =
I_1X_2Z_3 \ldots Y_{n-1}I_n$.  In analogy with the classical case, a
sensible choice of correctable errors will be those with {\em weight}
$\leq t$, where the weight is the number of single qubit error
operators which are not $I$ --- although if we have more sophisticated 
knowledge of the noise, such as correlations between qubits, we would
hope to include these as well.  For example, a single error-correcting 
QECC can be constructed using 5 qubits, to be compared with a
classical single error-correcting code which uses 3 bits (it only
corrects $X$ errors) \cite{Preskill:1998}.

An interesting feature to note about a $t$-error correcting QECC is
that the information is encoded highly nonlocally.  For example,
suppose we have encoded a single qubit into 5 qubits using the optimal 
1-error correcting code.  If we handed any one qubit over to an
eavesdropper Eve, then the fidelity of our quantum system after
recovery can still be arbitrarily high --- and so by the no-cloning
theorem Eve will not have been able to derive {\em any} information from her 
qubit.  Contrast this with the classical case, in which, if Eve knew
that the probability of error on any single qubit was $p\ll 1$, she
could be confident that the signal was what she measured.  In fact, if 
in the course of recovery we make a measurement (as will usually be
the case; the operator $\cR$ above includes this possibility) on the
qubits, the result cannot reveal any information about the logical
state encoded in the set of qubits, for otherwise we will have
disturbed the quantum information.  These are important features of
quantum error correction.

A pedagogical survey of QECC and coding techniques can be found in
\cite{Preskill:1998}, and a good set of references is given in the
appropriate section of \cite{Steane:1998}.

\paragraph{Quantum channel capacity}
So where does all of this leave us in terms of channel capacity?  Our
aim would be to characterise the number of qubits which can be
recovered after the action of a superoperator $\$$ with arbitrarily
good fidelity.

The quantum version of the problem of channel capacity is
mathematically a great deal more complicated than the classical
version, so we would expect an answer to this question to be more
difficult to reach.  But we are faced with even more complication: the 
capacity depends on what other resources are available.  For example,
there are channels which cannot be used to communicate quantum states
if only one-way communication is allowed, but have positive capacity
if two-way communication is allowed \cite{Bennett:1996}.  On the other 
hand, $t$-error correcting QECCs exist with an asymptotic rate $k/n =
1-2H(2t/n)$ \cite{Calderbank:1996}, where $H$ is the entropy function
--- so if the errors are uncorrelated and have low enough average
probability on each single qubit, we can transmit quantum information
with arbitrarily good fidelity.  The Shannon strategy of random coding 
is found to fail when transplanted into quantum information terms.
Altogether, there are very few answers to the general characterisation 
of quantum channel capacity.

\paragraph{Fault tolerant quantum computing}
Another, possibly fatal, criticism can be found for quantum error
correction ideas.  A QECC uses extra qubits to encode one qubit of
information, and requires us to ``smear'' the information over these
other systems.  This will require interaction between the qubits, and
such interaction will in general be noisy; and extra qubits mean that
each qubit has the potential to interact with the environment ---
noise which can spread through interactions between qubits.  So the
burning question is: Can we, by introducing more qubits and more
imperfect interactions, {\em extend} a computation?  Can we decrease
the {\em logical} error rate?

This is a serious problem, which initially seemed like a death blow to 
implementing quantum computing even in principle.  However, careful
analysis has shown that coherent quantum states can be maintained
indefinitely, despite imperfect gates and noisy qubits, as long as the 
error rate is below a certain threshhold.  Such analyses are the
subject of the enormous field of {\em fault tolerant quantum
computing}.  Further discussion of this topic can be found in
\cite{Preskill:1998b}.

\section{Measures of Entanglement}

With some idea of what entanglement can be used for and how to
interchange forms of entanglement, we can return to the problem of
quantifying an amount of entanglement.

We can immediately define two types of entanglement for any bipartite
quantum state $\rho_{AB}$:  the {\em entanglement of
formation} $E_F$, which is the minimum number of singlet states
$|\psi^-\ran$ required to prepare $\rho_{AB}$ using only local
operations and classical communication (LOCC); and the {\em
entanglement of distillation} $E_D$, which we define to be the maximum 
number of EPR pairs which can be distilled from $\rho_{AB}$ --- with
the maximum taken over all EPPs.

Bennett {\it et al} \cite{Bennett:1996} showed the very useful result
that for pure states $|\psi_{AB}\ran$ of the joint system, the
following result holds:
\be
E_F = E(|\psi_{AB}\ran) = E_D
\end{equation}
where $E(\cdot)$ is the entanglement measure proposed earlier in
Eqn~\ref{finish}.  The first equality follows from the fact that Alice 
can locally create copies of $|\psi_{AB}\ran$, compress them using
Schumacher coding, and teleport the resulting states to Bob who can
uncompress them; the second equality can be shown to hold by
demonstrating an EPP which achieves this yield.
This is the sense, alluded to previously, in which
a single system in the state $|\psi_{AB}\ran$ can be thought of as
equivalent to $E$ singlet states.  This relation immediately allows us to
simplify the definition of $E_F$ for mixed states.  If we denote by
${\cal E}_\rho = \{|\psi^i_{AB}\ran; p_i\}$ an ensemble of pure states
whose density operator is $\rho$, then the entanglement of formation
of $\rho$ is 
\be
E_F(\rho) = \min_{{\cal E}_\rho} \sum E(|\psi^i_{AB}\ran).
\end{equation}
This is justified by the fact that if we have $E_F(\rho)$ singlets,
then Alice can teleport a probabilistic mixture of pure states to Bob
to end up with the density operator $\rho$.  Wootters
\cite{Wootters:1997} has derived an exact expression for the
entanglement of formation of any state of two qubits.

Notice that in general $E_F\geq E_D$, since if this were otherwise, we 
could form an entangled state from $E_F$ EPR pairs and distill a greater
number of EPR pairs from that!  Evidence suggests \cite{Bennett:1996}
that the inequality is strict for all mixed states over the joint
system.

Recall the definition in Eqn~\ref{mixedness} of a separable bipartite
state.  We might ask whether there is any way to determine whether a
state is mixed without laboriously writing it in this form; and we
would be stuck for an answer.

Firstly, we might suppose that such states would always exhibit
nonlocal correlations such as in Bell's inequalities.  This attempt
fails because there are mixed states (in fact, the Werner states
introduced earlier) which do not violate any Bell's inequality because 
they admit hidden variables descriptions \cite{Werner:1989}.  Indeed,
there are {\em separable} states which have nonlocal properties; Bennett
{\it et al} \cite{Bennett:1999} demonstrated a set of bipartite states
of two 3-state systems which can be prepared locally but which cannot be
unambiguously discriminated by local measurements.

Secondly, we could ask whether all non-separable states can be used
for something like teleportation --- since the Werner states can
indeed be used for this \cite{Popescu:1994}.  We would once again
encounter a blind alley, since there are states which are nonlocal but 
cannot be used in quantum teleportation \cite{Horodecki:1998b}.  This
entanglement is termed ``bound'', in analogy with bound energy in
thermodynamics which cannot be used to do useful work --- this
entanglement cannot perform any useful informational work.

Note that if a state can be used for teleportation, it can be purified 
to singlet form (by Alice teleporting singlets to Bob).  Obviously, if 
a state can be distilled then it can be used for teleportation; so
there is an equivalence here.

In the spirit of trying to figure out what entanglement means,
DiVincenzo {\it et al} suggested another measure of entanglement, the
{\em entanglement of assistance} \cite{DiVincenzo:1998}.  It is
defined very similarly to entanglement of formation:
\be
E_A(\rho) = \max_{{\cal E}_\rho} \sum E(|\psi^i_{AB}\ran).
\end{equation}
Obviously this function will have properties dual to those of $E_F$.
It has some expected properties --- such as non-increase under LOCC
--- but some unexpected features too.  $E_A$ also has an
interpretation in terms of how much pure state bipartite entanglement
can be derived from a {\em tri}partite state.  And Vedral {\it et al}
\cite{Vedral:1997} have found a whole class of entanglement measures
based on quantifying the distance of a state from separability.

There is still much room for new ideas in quantifying of
entanglement, and in finding uses for it --- qualifying entanglement.
In a certain sense entanglement is a ``super'' classical correlation:
it holds the potential for classical correlation, but that potential
is only realised when a measurement is made.  On the other hand it is
also a very simple concept: it represents the fact that systems do not 
exist in classically separated states.  Entanglement is merely a
manifestation of this unity.
\chapter{Conclusion}

David Mermin \cite{Mermin:1989} once identified three post-quantum
theory generations of physicists according to their views on the
quantum.  The first, who were around when the theory developed (which
includes the ``Founding Fathers''), were forced to grapple with the
elements of the theory to give it a useful interpretation.  Once a
workable theory was developed, however, this generation came to regard
the quirks of quantum theory as due to some deeply ingrained
``classical'' modes of thought and hence to be expected.  The second
generation --- their 
students --- took this lesson further by insisting that there was {\em
nothing} unusual about quantum mechanics, and tried to make it
mundane; and indeed, this ``shut-up-and-calculate'' approach did
yield fruitful developments.  The third generation, of which Mermin
claims membership, doesn't seem to have much of an opinion one way or
the other, regarding the theory as productive and empirical and
carrying on.  But, Mermin points out, when foundational questions are
raised, the reaction of this generation varies from irritated to bored 
to plain uncomfortable.

To this we can now add another generation: those with enough
familiarity with the more bizarre parts of the theory not to be
blas\'e about it, but who take a practical approach by asking: What's
so special about {\em this} theory?  How can we characterise it?  If a 
classical ontology for this theory doesn't work --- if billiard balls
and elastic collisions don't describe the microscopic world --- then
what can be wrought from quantum theory in their place?

One famous illustration of this attitude --- albeit from a member of
the third generation --- is the EPR paradox and its resolution
by Bell.  Bohr's own response to the EPR paper had been guarded and
had appealed to his own orthodox interpretation; in a way the EPR
dilemma was likened to counting angels on the head of a pin.  The strongest
argument against EPR was that mutually contradictory experimental
setups should not be compared.  But with a more pragmatic approach,
Bell succeeded in eschewing metaphysical arguments in favour of
contemplation of the theory and reasoning of the first class.  This is 
a valuable lesson which has not gone unlearned in the field of
information physics.  Since quantum physics is an inherently
statistical theory, we cannot properly engage with it without
considering what those statistics mean in the real world --- both as
input and as output from the theory.  How do we operationally define a 
quantum state?  How much error do we introduce in our preparation, and 
how much is intrinsic according to the theory?  These are some of the
questions which information physics has addressed.

Some of the current concerns of quantum information physics have
been described in this thesis.  The compression of mixed states is
a minefield, but the curious examples cited in \cite{Barnum:2000}
illustrate that much work is still to be done in the mapping of this
minefield.  The quantification of entanglement, classification of
qualitatively different types of entanglement, their
interconvertibility and their purification are still major focuses of
work in quantum information theory.

The implications of information physics reach further than questions
of compression and capacity.  An obvious application of this study is
to high-precision measurement and to feedback control of quantum
systems.  The third generation of gravitational wave detectors for the 
LIGO experiment is expected to achieve a sensitivity beyond the
standard quantum limit for position monitoring of a free mass
\cite{Preskill:1999b}, and novel techniques for measurement will have
to be found.  For example, we could consider the Hamiltonian $H(x)$ of
the mass to be controlled by a parameter $x$, and it is our task to
determine $x$ from measurements on the mass.  In this case we can
consider the resulting state of the mass (after evolution for a time
$t$) to be one of a set of states $\rho_x$.  How much information
about $x$ can be extracted from this ensemble?  And what is the optimal
measurement?  These are questions addressed by theorems presented here 
as well as in ongoing research into techniques for optimal extraction of
information.  An illustration of the insight afforded by quantum
information theory is the following: suppose two electrons are in pure 
states with their spin polarisation vectors both pointing along the
same unknown axis but possibly in different directions.  Then more
information about this unknown direction can be extracted from the
pair of electrons if they are {\em anti}-parallel rather than parallel 
\cite{Gisin:1999}.

Indeed, ideas from quantum computing can also be of assistance in
this.  Grover's search algorithm was proposed to find some input $x$
to a ``black box'' binary-valued function $f$ such that the output is
1: $f(x)=1$.
Essentially what is achieved through this algorithm is the extraction
of global information about the black box Hamiltonian through our addition of
controlled terms to this Hamiltonian \cite{Preskill:1999b}.  Hence
through quantum algorithm analysis we can find an optimal driving to
apply to our measurement system and the measurement to apply in order to
characterise the unknown Hamiltonian affecting our LIGO III detector.

Characterisation of entanglement and information in quantum systems
is also important in studying quantum chaos \cite{Schack:1996}.  In
these cases it becomes important to specify our information about a
chaotic system and how it behaves under small perturbations.  This
leads us to introduce the concept of {\em algorithmic entropy}
\cite{Caves:1993}, which is a spin-off of classical computing theory and 
accurately captures the idea of ``complexity'' of a system.

Much current work in information theory is aimed at narrowing the gap
between the theory described in this thesis and available experimental 
techniques.  Quantum computing is a major focus of these efforts, with
a great deal of ingenuity being thrown at the problem of preserving
fragile quantum states.  Quantum error correction and fault-tolerant
quantum computing are two of the products of such efforts, and further 
analysis of particular quantum gates and particular computations is
proceeding apace.  A recent suggestion to simplify a prospective
``desk-top'' quantum computer involves supplying ready-made generic
quantum states to users who thus require slightly less sophisticated
and costly hardware \cite{Preskill:1999}; hence quantum computing may
ultimately also depend on faithful communication of quantum states.

And of course, the million-dollar question in quantum computing is:
What else can we do with a quantum computer?  Factoring large integers 
is a nice parlour trick and will enable a user to crack most
encryption protocols currently used on the internet.  But is this
enough of a reason to try and build one --- particularly considering
that in the near future quantum key distribution may become the
cryptographic protocol of choice?  Also, while Grover's search
algorithm gives a handsome speed-up over classical search techniques
(and some computer scientists are already designing quantum data
structures), it will be a long time before Telkom starts offering us
quantum phone directories.

There is something for the romantics in quantum information theory
too --- at least those romantics with a taste for foundational
physics.  The following quote (from \cite{Preskill:1999b}) is rather a
dramatic denunciation of quantum computer theorists:
\begin{quote}
It will never be possible to construct a `quantum computer' that can
factor a large number faster, and within a smaller region of space,
than a classical machine would do, if the latter could be built out of 
parts at least as large and as slow as the Planckian dimension.
\end{quote}
This statement comes from Nobel laureate Gerard 't Hooft.  Are we
missing something from quantum mechanics?  Will quantum mechanics
break down, perhaps at the Planck scale?   An interesting observation
of fault-tolerant quantum computing is that the overall behaviour of
well-designed circuit can be unitary for macroscopic periods of time
--- perhaps using concatenated coding \cite{Preskill:1998b} ---
despite almost instantaneous decoherence and non-unitary behaviour at
a lower level.  Is nature fault-tolerant?
Rather than constructing a solar system-sized particle accelerator, we
would be well-advised to assemble a quantum computer and put the
theory through its paces in the laboratory. 

On the other hand, we may be able to deduce, {\it \`a la} Wootters
\cite{Wootters:1980},
some principles governing quantum mechanics --- information-theoretic
principles, which constrain the outcomes of our proddings of the
world.  Weinberg, after failed attempts to formulate testable
alternatives to quantum mechanics, suggested \cite{Preskill:1999b}:
\begin{quote}
This theoretical failure to find a plausible alternative to quantum
mechanics suggests to me that quantum mechanics is the way it is
because any small changes in quantum mechanics would lead to
absurdities.
\end{quote}
It's been a century since the discovery of the quantum: what lies in
store in the next 100 years?




 
\specialhead{REFERENCES}
\markboth{}{}
\begin{singlespace}


\end{singlespace}

\end{document}